\documentclass[prd,aps,11pt,nofootinbib,superscriptaddress,preprintnumbers,balancelastpage,longbibliography]{revtex4-1}

\usepackage{amsmath}
\usepackage{amsfonts}
\usepackage{amssymb}
\usepackage{mathtools}
\usepackage{textcomp}
\usepackage{hyperref}
\usepackage{graphicx, rotating}
\usepackage{epstopdf}
\usepackage{epsfig}
\usepackage{latexsym}
\usepackage{graphicx}
\usepackage{color}
\usepackage{xspace}
\usepackage[dvipsnames]{xcolor}
\usepackage{amsmath,amssymb}
\usepackage{slashed}
\usepackage[vcentermath]{youngtab}
\usepackage{comment}
\usepackage{fancyhdr}
\usepackage{braket}
\usepackage{float}
\usepackage{siunitx}
\usepackage{physics}
\usepackage{longtable}
\usepackage{lmodern}

\usepackage{lipsum}
\usepackage{booktabs, tabularx, makecell, multirow, xspace}
\usepackage{listings}

\newcolumntype{C}[1]{>{\centering\let\newline\\\arraybackslash\hspace{0pt}}m{#1}}

\graphicspath{ {./Figures/} }
\newcommand{\ytab}[1]{$\includegraphics[scale=0.07]{R_#1.jpg}$}
\newcommand{\eytab}[1]{\includegraphics[scale=0.07]{R_#1.jpg}}
\newcommand{\farrar}{\includegraphics[scale=0.35]{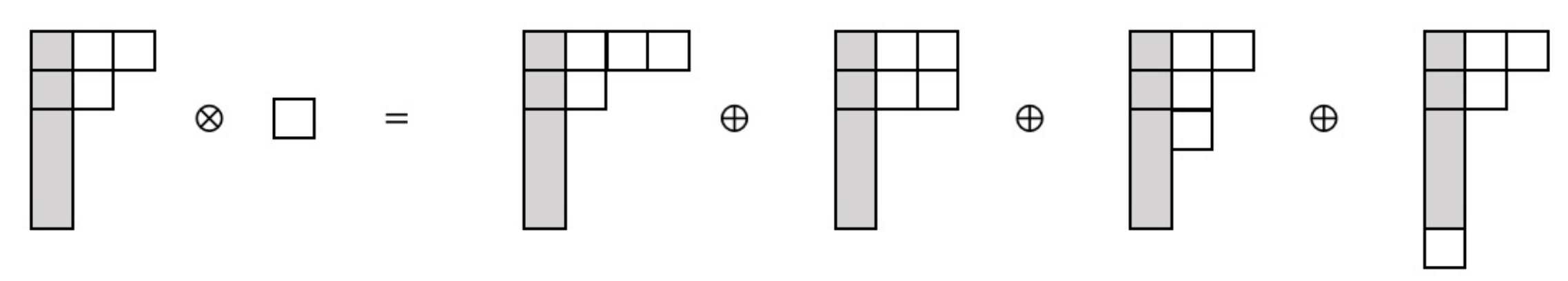}}

\definecolor{linkcolor}{rgb}{0.7752941176470588, 0.22078431372549023, 0.2262745098039215}

\hypersetup{colorlinks=true,
linkcolor=linkcolor,
citecolor=linkcolor,
urlcolor=linkcolor,
linktocpage=true,
pdfproducer=medialab,
}

\definecolor{nicered}{rgb}{0.7,0.1,0.1}
\definecolor{nicegreen}{rgb}{0.1,0.5,0.1}

\newcommand\Tstrut{\rule{0pt}{2.6ex}}         
\newcommand\Bstrut{\rule[-0.9ex]{0pt}{0pt}}   

\definecolor{codegreen}{rgb}{0,0.6,0}
\definecolor{codegray}{rgb}{0.5,0.5,0.5}
\definecolor{codepurple}{rgb}{0.58,0,0.82}
\definecolor{backcolour}{rgb}{0.95,0.95,0.92}

\newcommand{\nn}{\nonumber}

\newcommand{\be}{\begin{equation}}
\newcommand{\ee}{\end{equation}}
\newcommand{\bea}{\begin{eqnarray}}
\newcommand{\eea}{\end{eqnarray}}
\newcommand{\bc}{\begin{center}}
\newcommand{\ec}{\end{center}}

\makeatletter 
    
\renewcommand\onecolumngrid{
\do@columngrid{one}{\@ne}
\def\set@footnotewidth{\onecolumngrid}
\def\footnoterule{\kern-6pt\hrule width 1.5in\kern6pt}
}

\renewcommand\twocolumngrid{
        \def\footnoterule{
        \dimen@\skip\footins\divide\dimen@\thr@@
        \kern-\dimen@\hrule width.5in\kern\dimen@}
        \do@columngrid{mlt}{\tw@}
}

\makeatother

\begin{document}

\title{On the Proof of Chiral Symmetry Breaking through Anomaly Matching in QCD-like Theories: An Exemplification}

\author{Luca Ciambriello}
\email{luca.ciambriello@unicatt.it}
\affiliation{Interdisciplinary Laboratories for Advanced Materials Physics (i-LAMP) and Dipartimento di Matematica e Fisica, Universit\`{a} Cattolica del Sacro Cuore, Brescia, Italy}

\author{Roberto Contino}
\email{roberto.contino@uniroma1.it}
\affiliation{Dipartimento di Fisica, Sapienza Universit\`{a} di Roma, Italy}
\affiliation{Istituto Nazionale di Fisica Nucleare (INFN), Sezione di Roma, Italy}

\author{Ling-Xiao Xu}
\email{lxu@ictp.it}
\affiliation{Abdus Salam International Centre for Theoretical Physics, Trieste, Italy}


\begin{abstract}
\vspace{0.5cm}
Our recent works~\cite{Ciambriello:2022wmh, CLRX2} revisit the proof of chiral symmetry breaking in the confining phase of four-dimensional QCD-like theories, i.e. $SU(N_c)$ gauge theories with $N_f$ flavors of vectorlike quarks in the fundamental representation. The analysis relies on the structure of 't Hooft anomaly matching and persistent mass conditions for theories with same $N_c$ and different $N_f$. In this paper, we work out concrete examples with $N_c=3$ and $N_c=5$ to support and elucidate the results of~\cite{Ciambriello:2022wmh, CLRX2}.  Within the same examples, we also test some claims made in earlier works.
\end{abstract}

\maketitle

\tableofcontents

\section{Introduction}\label{sec:intro}
It is widely believed that chiral symmetry is spontaneously broken 
in the confining phase of four-dimensional QCD-like gauge theories, i.e. $SU(N_c)$ gauge theories coupled to $N_f$ flavors of vectorlike quarks in the fundamental representation. This wisdom is supported by empirical evidence of real QCD, lattice simulations (see for example~\cite{Engel:2014cka,Engel:2014eea}) and QCD inequalities~\cite{Weingarten:1983uj} (see~\cite{Nussinov:1999sx} for a comprehensive review). However, the understanding of strongly-interacting QCD-like theories is still limited. 

Among all the theoretical tools, 't Hooft anomaly matching conditions (AMC)~\cite{tHooft:1979rat} stand out as exact constraints that any theory must satisfy in its low-energy phase.
The spectrum of QCD-like gauge theories should also satisfy persistent mass conditions (PMC)~\cite{Preskill:1981sr}, which were originally proposed by 't Hooft as decoupling conditions~\cite{tHooft:1979rat} and acquired theoretical soundness from the work of Vafa and Witten~\cite{Vafa:1983tf}. Using these constraints, one would hope to derive a rigorous proof of chiral symmetry breaking ($\chi$SB), rather than simply assuming it based on empirical evidence. 
In particular, $\chi$SB is rigorously proven if one can show that there exist no integral solutions of AMC and PMC equations for any putative spectrum of massless composite fermions. Since the dynamical formation of massless bound states is not under control, it is important to consider all possible massless states in the proof. 

In a first companion paper~\cite{Ciambriello:2022wmh}, we have reconsidered the arguments put forward in the literature to prove $\chi$SB through AMC and PMC. We can distinguish three main strategies. The first makes use of a property dubbed `$N_f$-independence', according to which real solutions of AMC and PMC equations are the same for any number of flavors. This line of reasoning was indicated by 't Hooft in his seminal work~\cite{tHooft:1979rat} and later pursued in particular by Refs.~\cite{Frishman:1980dq,Farrar:1980sn,Takeshita:1981sx,Bars:1981nh,Kaul:1981fd}. A related though different argument was developed by Cohen and Frishman in~\cite{Cohen:1981iz}, where the existence of a so-called $N_f$-equation that does not depend on the number of flavors was suggested. Finally, a strategy based on the $SU(N_f|N_f)$ superalgebra was proposed by Schwimmer in~\cite{Schwimmer:1981yy}.
In Ref.~\cite{Ciambriello:2022wmh} we focused on  the argument based on $N_f$-independence. 
We provided a first proof of $N_f$-independence and clarified under which assumptions it holds. These are dynamical requirements on the putative spectrum of massless fermions. Therefore, using $N_f$-independence does not lead to a purely algebraic proof of chiral symmetry breaking, as one would hope.

In a second companion paper~\cite{CLRX2}, we presented a new proof that differs from the earlier works. Our results can be summarized as follows:
\begin{itemize}
\item The proof is based on `downlifting' solutions of AMC and PMC to lower values of~$N_f$, and applies to any spectrum of massless bound states. Specifically, given any (real) solution of the system of equations in a theory with $N_f$ flavors, we are able to provide a downlifted solution for $N_f-1$ flavors. If no integer solution exists for some specific number of flavors, this implies $\chi$SB for any larger number of flavors.
\item We proved that there is no integral solution for AMC  when $N_f$ is equal or proportional to any prime factor of $N_c$. This result is valid for any spectrum of massless bound states. Hence, $\chi$SB is rigorously proven in the confining phase for $N_f\geq p_{min}$, where $p_{min}$ is the smallest prime factor of $N_c$.
\item Finally, we presented an argument based on continuity that allows one to deduce $\chi$SB for $N_f < p_{min}$ provided no phase transition occurs when quark masses are sent to infinity.
\end{itemize}

The goal of the present paper is to provide concrete examples to elucidate and support the more abstract discussions of Refs.~\cite{Ciambriello:2022wmh,CLRX2}.
We present detailed results for QCD-like theories with $N_c =3$ and $5$ and arbitrary number of flavors. We derive explicitly the real solutions of AMC and PMC and verify that no integral solution exists. Our findings exemplify the arguments made in Refs~\cite{Ciambriello:2022wmh, CLRX2} and also test some claims made in the previous literature. 
In Section~\ref{sec:AMCPMC} we set our notation and briefly review how the AMC and PMC equations are constructed. Section~\ref{sec:discussion} contains a summary and a detailed discussion of our results, while all the AMC and PMC equations are reported in Sections~\ref{sec:resultsNc5} and~\ref{sec:resultsNc3}. There, we first discuss the example of $N_c=5$ with only baryons included in the putative spectrum of massless bound states, then we discuss the example of $N_c=3$ with massless baryons and pentaquarks in the spectrum. The dimensionalities and anomaly constants of all the representations encountered in the calculation are collected in Appendix~\ref{app:table}.

\section{Constructing AMC and PMC equations}
\label{sec:AMCPMC}
This section sets our notation and gives a brief introduction to AMC and PMC. For more details we refer the reader to Ref.~\cite{Ciambriello:2022wmh}.

A QCD-like theory with $N_f$ massless flavors  is invariant under the chiral symmetry group~\footnote{We neglect the discrete quotient that is necessary to define the symmetry group acting faithfully on quarks; see~\cite{CLRX2,Tanizaki:2018wtg} for more details. We define $U(1)_B$ so that quarks have baryon number $1/N_c$.}
\bea
\mathcal{G}[N_f]=SU(N_f)_L \times SU(N_f)_R \times U(1)_B\ . 
\label{def_G}
\eea
For $N_f$ below the conformal window, the theory is expected to confine at low energy, so that only color-singlet bound states are present in the spectrum. The quark content in the ultraviolet (UV) exhibits non-zero $[SU(N_f)_{L,R}]^2 U(1)_B$ and $[SU(N_f)_{L,R}]^3$ 't Hooft anomalies, which must be exactly reproduced in the deep infrared (IR) by massless bound states~\cite{tHooft:1979rat}.
If ${\cal G}[N_f] $ is not spontaneously broken, then the IR spectrum must contain massless spin-$1/2$ bound states~\cite{Weinberg:1980kq}.
Being color singlets, these are interpolated by local composite operators with a number of left (right) quarks, $n_L$ ($n_R$), and of left (right) antiquarks, $\bar n_L$ ($\bar n_R$), satisfying the sum rule
\bea
\label{eq:sumrule}
n_L + n_R -\bar n_L - \bar n_R=b N_c\, .
\eea
Both $N_c$ and the baryon number $b$ of the bound states must be odd integers for the latter to have spin $1/2$. For $N_c$ even, only bosonic color-singlet bound states can form, and one can immediately conclude from AMC that chiral symmetry must be spontaneously broken.

We follow closely the notation of~\cite{Ciambriello:2022wmh, CLRX2} and put massless spin-$1/2$ bound states into correspondence with irreducible representations (irreps) of ${\cal G}[N_f]$. The corresponding interpolating operators transform as traceless tensors $T$ (see for example~\cite{Tung:1985na}) possessing $n_L$ ($n_R$) left (right) upper indices and $\bar n_L$ ($\bar n_R$) left (right) lower indices. Each group of $n$ indices is appropriately (anti-)symmetrized and can be represented by a Young tableaux (YT) $\{ n\}$.
A traceless tensor is thus denoted as $T^{\{n_L\}, \{n_R\}}_{\{\bar{n}_L\}, \{\bar{n}_R\}}$ and transforms as an irrep characterized by a pair of combined Young tableaux (CYT) $\{n_L; \bar{n}_L\}$ and $\{n_R; \bar{n}_R\}$.~\footnote{See~\cite{Ciambriello:2022wmh} and Chapter 13 of~\cite{Tung:1985na} on how to construct CYT.}
Therefore, local composite operators are in correspondence with traceless tensors $T^{\{n_L\}, \{n_R\}}_{\{\bar{n}_L\}, \{\bar{n}_R\}}$ and interpolate massless bound states in irreps $r$ of ${\cal G}[N_f]$, with
\bea
\label{irreps}
r= \left(\{n_L; \bar{n}_L\}, \{n_R; \bar{n}_R\}, b\right)\, ,
\eea
where $b$ is given by the sum rule~(\ref{eq:sumrule}).
We denote the space of all possible tensors in a theory with $N_f$ flavors by ${\cal T}(N_f)$, and the space of irreps $r$
by ${\cal R}(N_f)$.
Notice that different tensors can interpolate bound states in the same irrep of ${\cal G}[N_f]$, namely tensors with different upper and lower indices can have the same pair of CYT and be associated to the same baryon number through Eq.~(\ref{eq:sumrule}). Such tensors are called equivalent tensors~\cite{Ciambriello:2022wmh}.
Following~\cite{Ciambriello:2022wmh}, it will be useful to define `class A' tensors as those satisfying
\begin{equation}
  \label{eq:classA}
n+\bar n < N_f , 
\end{equation}
where $n=n_L+n_R$ and $\bar n_L+\bar n_R$. As shown in~\cite{Ciambriello:2022wmh}, two class A tensors cannot be equivalent. Hence, there exists at most one class A tensor transforming as any given irrep $r$. Irreps corresponding to class A tensors will be dubbed class A irreps.

The AMC for a theory with $N_f$ massless flavors are of the form
\begin{equation}
  \label{eq:AMC}
\sum_{r\in {\cal R}(N_f)} \ \ell(r) A_{i}(r) = N_c \ A_{i}\!\left( \square  \right)\, ,
\end{equation}
where $i=2$ and 3 respectively for $[SU(N_f)_{L,R}]^2 U(1)_B$ and $[SU(N_f)_{L,R}]^3$.
For brevity, these equations are denoted as AMC$[N_f]$ in the following. The index $\ell(r)$ equals the number of times the bound state in the irrep $r$ appears in the spectrum with helicity $+1/2$ minus the number of times it appears with helicity $-1/2$.
The anomaly coefficients $A_{2,3}(r)$ are defined as
\begin{align}
A_2\!\left(r\right)&= b\ D_2(\{n_L; \bar{n}_L\}, N_f) \ d(\{n_R; \bar{n}_R\}, N_f)\ ,\\
A_3\!\left(r\right)&= D_3(\{n_L; \bar{n}_L\}, N_f) \ d(\{n_R; \bar{n}_R\}, N_f)\ ,
\end{align}
where~\footnote{Here, $\lambda_{a}$ are $SU(N_f)$ generators and $d(R, N_f)$ is the dimension of $R$; see~\cite{Ciambriello:2022wmh} and Chapter 13 of Ref.~\cite{Tung:1985na} for further discussions. Clearly, $D_3$ is well-defined only for $N_f\geq 3$, since all the $SU(2)$ irreps are real or pseudo-real.}
\begin{equation}
\label{eq:D2_D3}
D_2(R_i, N_f)\ \delta_{ab}=\text{tr}[\{\lambda_a, \lambda_b\}]\ , \quad\quad\quad D_3(R_i, N_f)\ d_{abc} =\text{tr}[\{\lambda_a,\lambda_b\}\lambda_c]\ .
\end{equation}
The representations $R_i$ encountered in this paper are listed in Appendix~\ref{app:table}.
The left- and right-hand sides of Eq.~(\ref{eq:AMC}) thus encode the anomalies furnished by respectively the bound states and the quarks. 
For simplicity, in this paper we consider a parity-invariant spectrum, i.e.
\bea
\label{eq:P}
\ell\left( r \right) = - \ell\left( r_P \right),
\eea
where $ r_P = \left(\{n_R; \bar{n}_R\}, \{n_L; \bar{n}_L\}, b\right)$ is the parity conjugate of $r$.
A direct consequence of this requirement is that any irrep which is identical to its parity conjugate must have vanishing index:
\begin{equation}
  \label{eq:implicationPinv}
  r = r_P \quad\Rightarrow\quad \ell(r) = 0\, .
\end{equation}
In other words: fermionic bound states transforming as parity invariant irreps of ${\cal G}[N_f]$ appear in the spectrum with an equal multiplicity of $+1/2$ and $-1/2$ helicities. Therefore, their anomaly coefficients vanish identically. An example of parity invariant irrep is of course the singlet.

Besides AMC, the spectrum of bound states must satisfy additional constraints called PMC~\cite{tHooft:1979rat,Preskill:1981sr, Vafa:1983tf}.
When one flavor is given a mass, ${\cal G}[N_f]$ is reduced to
\bea
\mathcal{G}[N_f,1]=SU(N_f-1)_L\times SU(N_f-1)_R\times U(1)_{H_1}\times U(1)_B\, ,
\label{def_G1}
\eea
where the massive quark has charge $+1$ under $U(1)_{H_1}$ while massless flavors have charge 0.
Each irrep of ${\cal G}[N_f]$ can be decomposed as a direct sum of irreps of ${\cal G} [N_f,1]$. These are characterized by a pair of YTs $\{m_L; \bar{m}_L\}$ and $\{m_R; \bar{m}_R\}$, the same baryon number $b$ as that of the parent irrep, and a $U(1)_{H_1}$ charge
\begin{equation}
  \begin{split}
H_1&=(n_L-m_L)+(n_R-m_R)-(\bar{n}_L-\bar{m}_L)-(\bar{n}_R-\bar{m}_R) \\
     &= bN_c-(m_L+m_R-\bar{m}_L-\bar{m}_R)\ .
\label{eq:Hcharge}
\end{split}
\end{equation}
We thus write
\bea
\label{eq:BS1}
r_1 = \left(\{m_L; \bar{m}_L\}, \{m_R; \bar{m}_R\}, H_1, b\right)\ .
\eea

Bound states with $H_1\neq 0$ must be massive~\cite{Ciambriello:2022wmh} and thus form vectorlike pairs with vanishing indices. This gives rise to a set of equations called PMC$[N_f,1]$:
\bea
\label{eq:PMC}
0=\ell\left(r_1\right)=\sum_{r\in \mathcal{R}(N_f)} \, \kappa\!\left(r \to r_1\right) \, \ell\left( r \right) .
\eea
Here the sum runs over all irreps $r$ in $\mathcal{R}(N_f)$, and the coefficient $\kappa\!\left( r \to r_1\right)$ denotes how many times $r_1$ appears in the decomposition of $r$.
Since tensors, in particular the ones of ${\cal G}[N_f,1]$, can become or cease to be equivalent as $N_f$ is varied, different PMC equation can collapse into a single one, or a single PMC equation can split into several different equations~\cite{Ciambriello:2022wmh}.

Irreps $r_1$ with $H_1=0$ can be decomposed further when one more flavor is given a mass, see~\cite{Ciambriello:2022wmh}. 
The corresponding PMC equations are denoted as PMC$[N_f, 2]$. In a similar way, by giving mass to $i$ flavors (with $1\leq i\leq N_f-2$), one obtains PMC$[N_f, i]$ equations. It is easy to see that PMC$[N_f, i]$ have the same form as PMC$[N_f-1, i-1]$, for $i>1$, see~\cite{Ciambriello:2022wmh}. The full set of PMC equations in a theory with $N_f$ flavors is denoted as PMC$[N_f]$.  
Notice that indices of parity invariant irreps do not enter PMC$[N_f]$ as a consequence of our assumption on the parity invariance of the spectrum. Indeed, any given PMC equation receives an equal and opposite contribution from the $+1/2$ and $-1/2$ helicity components of a parity invariant bound state.

It was shown in Ref.~\cite{CLRX2} that any solution $\{ \ell(r) \}$ of AMC$[N_f]\cup\text{PMC}[N_f]$ can be downlifted to a solution of AMC$[N_f-1]\cup\text{PMC}[N_f-1]$. The key observation behind the idea of downlifting is the following: Any tensor transforming as $r_1 = \left(\{m_L; \bar{m}_L\}, \{m_R; \bar{m}_R\}, 0, b\right)$ of ${\cal{G}}[N_f,1]$ can be identified with the tensor in $\mathcal{T}(N_f-1)$ transforming as the irrep $r'= \left(\{m_L; \bar{m}_L\}, \{m_R; \bar{m}_R\}, b\right)$ of $\mathcal{G}[N_f-1]$. In other words, the symmetry group ${\cal{G}}[N_f,1]$ acts on irreps $r_1$ with $H_1=0$ in the same fashion as $\mathcal{G}[N_f-1]$ acts on $r'$. The downlifted solution $\{ \tilde\ell(r')\}$ is defined as follows:~\footnote{\label{fot:elltilde}Here and in the following, $\tilde\ell$ denotes the index of a downlifted solution.}
\begin{equation}
  \label{eq:downliftedsol}
\tilde \ell (r') = \sum_{r\in {\cal R}(N_f)} \ell(r)\, \kappa(r\to r') ,
\end{equation}
where $\{ \ell(r) \}$ is the original solution. If one starts with a spectrum of class-A irreps $r$ of ${\cal G}(N_f)$, then for any downlifted class-A irrep $r'$ of ${\cal G}(N_f-1)$, the following property holds true~\cite{Ciambriello:2022wmh}:
\begin{equation}
  \label{eq:uplift}
\tilde \ell (r') = \ell(U[r']).
\end{equation}
Here $U[r']$ is the uplift of $r'$~\cite{Ciambriello:2022wmh}, defined as the irrep of ${\cal G}(N_f)$ which is interpolated by the same class-A tensor that interpolates $r'$.

The results summarized above are key to the proofs of chiral symmetry breaking discussed in~\cite{CLRX2} and~\cite{Ciambriello:2022wmh}. Aim of the next sections is that of illustrating them and verifying their validity by means of explicit examples.

\section{Summary and discussion of the results}
\label{sec:discussion}

\begin{figure}[t] 
\centering 
\includegraphics[width=0.7\textwidth]{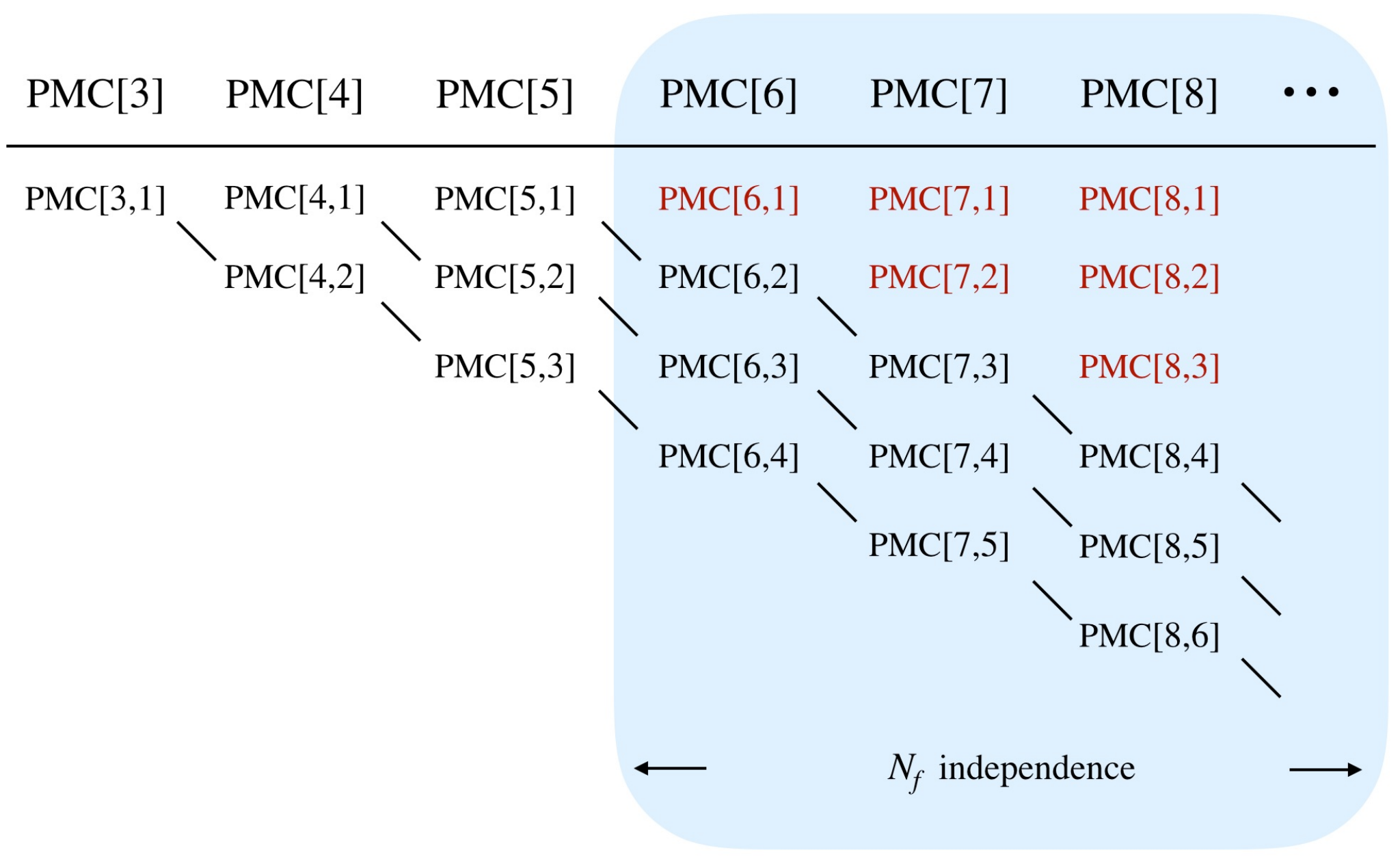} 
\caption{Structure of PMC for QCD-like theories with varying $N_f$ and fixed $N_c$~\cite{Ciambriello:2022wmh, CLRX2}.
PMC equations connected by solid lines have the same form and can be therefore identified. In both examples considered in this paper, $N_f$-independence holds only for $N_f\geq 6$ (blue region). All the equations marked in red are identical as a consequence of $N_f$-independence.} 
\label{fig:pmc} 
\end{figure}

In this paper, we work out two explicit examples of QCD-like theories.
First, we analyze theories with $N_c=5$ and $N_f\geq 2$, assuming that the putative spectrum of massless fermions consists of baryons with baryon number $b=1$. This example is simple enough, and it is already sufficient to illustrate many features of the results of~\cite{Ciambriello:2022wmh, CLRX2}, 
such as the condition under which $N_f$-independence holds, and how downlifting is implemented. 
Next, we consider theories with $N_c=3$ and $N_f\geq 2$, and assume that the spectrum of massless fermions consists of both baryons and pentaquarks with $b=1$. This example is much more involved due to the large number of possible pentaquark states in the spectrum.
As illustrated by means of Fig.~\ref{fig:pmc} and summarized below, our analysis furnishes a non-trivial test of the results of~\cite{Ciambriello:2022wmh, CLRX2}.

\begin{enumerate}
\item In both examples, we find that $N_f$-independence holds true only for $N_f\geq 6$ (see the blue region in Fig.~\ref{fig:pmc}). This confirms the general result of~\cite{Ciambriello:2022wmh}, according to which $N_f$-independence holds if the spectrum of massless fermions contains only bound states that are interpolated by class-A tensors. Given the definition of class A tensor of Eq.~(\ref{eq:classA}), it follows that baryons are interpolated by class A tensors if $N_f > bN_c$, which translates into $N_f > 5$ for $N_c =5$ and $b=1$. Similarly, pentaquarks are interpolated by class A tensors if $N_f > bN_c+2$, which also gives $N_f > 5$ for $N_c =3$ and $b=1$.
In Fig.~\ref{fig:pmc}, PMC equations connected with solid lines have the same form and can be identified. Because of $N_f$-independence, all the equations marked in red are also identical. Furthermore, we checked explicitly that all AMC have the same form for any $N_f\geq 6$.
\item Even when $N_f$-independence does not hold, it is always possible to construct a downlifted solution of AMC and PMC as indicated in~\cite{CLRX2}. In particular, we verified that AMC$[N_f-1]$ can be derived from AMC$[N_f]$ and PMC$[N_f,1]$, although these equations have different forms for different values of $N_f$.
\item We verified the validity of the `prime factor' theorem~\cite{CLRX2} (see also~\cite{Preskill:1981sr, Weinberg:1996kr}): when $N_f$ is proportional to a prime factor $p>1$ of $N_c$,  all the anomaly coefficients appearing in the $[SU(N_f)_{L,R}]^2 U(1)_B$ AMC equation are integral multiples of $p$, whereas the anomaly coefficient of quarks is normalized to 1. Hence, there exists no integral solution for the $[SU(N_f)_{L,R}]^2 U(1)_B$ AMC equation.  
\item Finally, we found no integral solutions apart from the $N_f=2$ case. The same conclusion was drawn by several earlier studies, although based on less detailed numerical calculations.
\end{enumerate}

Other arguments that appeared in previous works are also explicitly tested in our examples.
\begin{enumerate}
\item We check Farrar's argument~\cite{Farrar:1980sn} that
  the $[SU(N_f)_{L,R}]^3 $ anomaly coefficients are equal to $\pm N_c^2$ or 0 in the limit $N_f=0$. This argument is elaborated in detail within our second example, see after Eq.~(\ref{eq:farrar}). We confirm that the final conclusion is correct, although there exist some loopholes in the proof presented in~\cite{Farrar:1980sn}.
\item In both our examples, we could verify that, as observed 
  by Cohen and Frishman~\cite{Cohen:1981iz}, the two AMC$[N_f]$ equations become the same equation once evaluated on the solutions of PMC$[N_f,1]$. This was referred to as the `$N_f$-equation'. Despite the system of AMC$[N_f]$ and PMC$[N_f,1]$ depends on $N_f$ in general, Cohen and Frishman argued that the $N_f$-equation is independent of $N_f$. We find supporting evidence for this argument.
  The $N_f$-equation  can be obtained  as one linear combination of AMC$[N_f]$ and PMC$[N_f,1]$ and expressed in terms of different sets of indices. It is not guaranteed, however, that the indices chosen remain linearly independent as $N_f$ is varied. In all the cases analyzed in this paper it was possible to identify an $N_f$-equation which retains its form for any $N_f > 2$.
We do not have a proof, however, that this holds in the general case.
%
\item Our first example was partially analyzed in previous works~\cite{tHooft:1979rat,Frishman:1980dq,Schwimmer:1981yy,Cohen:1981iz}. Refs.~\cite{tHooft:1979rat,Frishman:1980dq}, in particular, found a solution in which `elbow'-shape representations have index $\pm 1/N_c^2$, while all the other indices vanish.  We confirm the validity of such solution, although it is not a generic one since it assumes that mixed representations have vanishing index.
\end{enumerate}

Table~\ref{tab:Nc_5} summarizes the total number of AMC and PMC equations, the rank of the AMC and PMC system of equations, the total number of indices, and the number of free indices that we found for various $N_f$ in our two examples. Explicit equations are reported in Sections~\ref{sec:resultsNc5} and~\ref{sec:resultsNc3}.
\vspace{0.5cm}
\begin{table}[h]
\scriptsize
\begin{center}
\begin{tabular}{C{1.8cm} | C{1.9cm}C{1.9cm}C{1.9cm}C{1.9cm}}
  \Xhline{3\arrayrulewidth}
\renewcommand{\arraystretch}{1}
   &  \multicolumn{4}{c}{\textbf{$N_c=5$, baryons with $b=1$ }} \\
$N_f$  & $\#$ of equations   &  rank  & total $\#$ of indices & $\#$ of free indices \Tstrut\Bstrut   \\   
 \hline
 2 &  1 & 1 & 6 & 5 \\
 3 &  10 & 9 & 13  & 3\\
 4  & 16 & 13  & 16  & 3\\
 5  & 18  & 14  & 18  & 3\\
 $\geq 6$  & 19 & 15  & 18  & 3\\
  \Xhline{3\arrayrulewidth}
\end{tabular}
\end{center}
\begin{center}
\begin{tabular}{C{1.8cm} | C{1.9cm}C{1.9cm}C{1.9cm}C{1.9cm}}
  \Xhline{3\arrayrulewidth}
\renewcommand{\arraystretch}{1}
   &  \multicolumn{4}{c}{\textbf{ $N_c=3$, baryons and pentaquarks with $b=1$}} \\
$N_f$  & total $\#$ of equations   &  rank  & total $\#$ of indices & $\#$ of free indices \Tstrut\Bstrut   \\   
 \hline
 2 & 1 & 1 & 6  & 5 \\
 3 & 13 & 11 & 17  & 4 \\
 4 & 37 & 18 & 22  & 4 \\
 5 & 65 & 20 & 25  & 4 \\
 $\geq 6$  & 95 & 21 & 25 & 4 \\
  \Xhline{3\arrayrulewidth}
\end{tabular}
\end{center}
\caption{Total number of AMC and PMC equations, rank of the system of AMC and PMC equations, total number of indices, and number of free indices for the two classes of QCD-like theories analyzed in this paper: $N_c=5$ theories with a spectrum of massless baryons with $b=1$ (upper panel), and $N_c=3$ theories with a spectrum of massless baryons and pentaquarks with $b=1$ (lower panel). 
In the upper panel, PMC only include PMC$[N_f,1]$, because, for a purely baryonic massless spectrum, PMC with more than one massive flavor are not independent from PMC$[N_f,1]$~\cite{CLRX2}.}
\label{tab:Nc_5}
\end{table}

\newpage
\section{$N_c=5$ QCD-like theories with massless baryons}
\label{sec:resultsNc5}

We first consider an example where, by assumption, the only massless bound states in the spectrum are baryons with $b=1$. Specifically, we discuss QCD-like theories with $N_c=5$ and generic $N_f\geq 2$.
Baryons are defined as those bound states that can be interpolated by local composite operators made of $bN_c$ quark fields and no antiquark fields (see~\cite{Ciambriello:2022wmh}). For $N_c=5$ and $b=1$, all the possible tensors of ${\cal G}[N_f]$ associated to such operators are listed in Table~\ref{var_nc_5}, together with their corresponding  indices and CYTs.
In general, only YTs with $N_f$ or fewer rows are well defined in a theory with $N_f$ flavors. This implies that some of the CYTs in Table~\ref{var_nc_5} do not exist for $N_f<5$. We assume that the spectrum is parity invariant for simplicity: each irrep listed in Table~\ref{var_nc_5} has a parity-conjugated partner with an equal and opposite index (cf. Eq.~(\ref{eq:P})).

In the following, we study the system of AMC and PMC equations for theories of different $N_f$.
Since, for a purely baryonic massless spectrum, PMC with more than one massive flavor are not independent from PMC$[N_f,1]$~\cite{CLRX2}, they will not be considered in this example.

\vspace{0.5cm} 
\begingroup
\renewcommand*{\arraystretch}{0.8} 
\begin{longtable}{|c|c|c||c|c|c|}
		\hline\hline
		Index & Irrep (CYT) & Tensor & Index & Irrep (CYT) &  Tensor \\
		\hline
		\hline
		&&&&&\\
		$\ell_{1}$& (\ytab{24}, \ytab{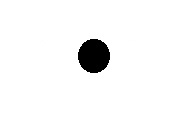}) & $T_1\sim T^{(\eytab{24}, \eytab{0} )}_{(\eytab{0},\eytab{0})}$ &$\ell_{2}$& $(\ytab{25}, \ytab{0})$ & $T_2\sim T^{(\eytab{25}, \eytab{0})}_{(\eytab{0}, \eytab{0})}$\\
		&&&&&\\
		\hline
		&&&&&\\
		$\ell_{3}$ & (\ytab{26}, \ytab{0}) & $T_3\sim T^{(\eytab{26}, \eytab{0})}_{(\eytab{0}, \eytab{0})}$ & $\ell_{4}$ & (\ytab{27}, \ytab{0}) & $T_4\sim T^{(\eytab{27}, \eytab{0})}_{(\eytab{0},\eytab{0})}$\\
		&&&&&\\
		\hline
		&&&&&\\
		$\ell_{5}$ & (\ytab{28}\ ,\  \ytab{0}) & $T_5\sim T^{(\eytab{28}, \eytab{0})}_{(\eytab{0},\eytab{0})}$ & $\ell_{6}$ & (\ytab{29}\ , \ \ytab{0}) & $T_6\sim T^{(\eytab{29} , \eytab{0})}_{(\eytab{0},\eytab{0})}$ \\
		&&&&&\\
		\hline
		&&&&&\\
		$\ell_{7}$ & (\ytab{30}, \ytab{0}) & $T_7\sim T^{(\eytab{30}, \eytab{0})}_{(\eytab{0}, \eytab{0})}$ & $\ell_{8}$ & (\ytab{7}, \ytab{1}) & $T_8\sim T^{(\eytab{7}, \eytab{1})}_{(\eytab{0}, \eytab{0})}$\\
		&&&&&\\
		\hline
		&&&&&\\
		$\ell_{9}$ & (\ytab{8}, \ytab{1}) & $T_9\sim T^{(\eytab{8}, \eytab{1})}_{(\eytab{0}, \eytab{0})}$ & $\ell_{10}$ & (\ytab{9}, \ytab{1}) & $T_{10}\sim T^{(\eytab{9}, \eytab{1}) }_{(\eytab{0}, \eytab{0})}$\\
		&&&&&\\
		\hline
		&&&&&\\
		$\ell_{11}$ & (\ytab{10}, \ytab{1}) & $T_{11}\sim T^{(\eytab{10}, \eytab{1}) }_{ (\eytab{0}, \eytab{0})}$ & $\ell_{12}$ & (\ytab{11}, \ytab{1}) & $T_{12}\sim T^{ (\eytab{11}, \eytab{1}) }_{ (\eytab{0}, \eytab{0})}$\\
		&&&&&\\
		\hline
		&&&&&\\
		$\ell_{13}$ & (\ytab{4}, \ytab{2}) & $T_{13}\sim T^{ (\eytab{4},  \eytab{2}) }_{ (\eytab{0}, \eytab{0})}$ & $\ell_{14}$ & (\ytab{5}, \ytab{2}) & $T_{14}\sim T^{ (\eytab{5}, \eytab{2}) }_{ (\eytab{0}, \eytab{0})}$\\
		&&&&&\\
		\hline
		&&&&&\\
		$\ell_{15}$ & (\ytab{6}, \ytab{2}) & $T_{15}\sim T^{ (\eytab{6}, \eytab{2}) }_{ (\eytab{0}, \eytab{0})}$ & $\ell_{16}$ & (\ytab{4}, \ytab{3}) & $T_{16}\sim T^{ (\eytab{4},  \eytab{3}) }_{ (\eytab{0}, \eytab{0})}$\\
		&&&&&\\
		\hline
		&&&&&\\
		$\ell_{17}$ & (\ytab{5}, \ytab{3}) & $T_{17}\sim T^{ (\eytab{5}, \eytab{3}) }_{ (\eytab{0}, \eytab{0})}$ & $\ell_{18}$ & (\ytab{6}, \ytab{3}) & $T_{18}\sim T^{ (\eytab{6}, \eytab{3}) }_{ (\eytab{0}, \eytab{0})}$\\
		&&&&&\\
		\hline
	\caption{Irreps for massless baryons with $b=1$ in QCD-like theories with $N_c=5$. In a parity-invariant spectrum, each irrep has a parity-conjugate partner with equal and opposite index.}
	\label{var_nc_5}
\end{longtable}
\endgroup

\subsection{Theories with $N_f\geq 6$: $N_f$-independence}

All the 18 tensors of Table~\ref{var_nc_5} are well defined for $N_f\geq 6$. Furthermore, they are all class~A and, as such, inequivalent.
Since AMC$[N_f]\cup$PMC$[N_f]$ has rank 15 (see Table~\ref{tab:Nc_5}), it admits a family of real solutions with 3 free parameters.
The anomaly contribution of each irrep $r$ can be computed following Section~\ref{sec:AMCPMC} and Table~\ref{list} of Appendix~\ref{app:table}; due to parity invariance,
it includes the contribution from the parity-conjugate irrep $r_P$.~\footnote{\label{fot:parityconj}For example, the $[SU(N_f)_{L}]^3$ anomaly coefficient for $(\ytab{7}, \ytab{1})$ is
\begin{equation}
\begin{split}
  A_3(\eytab{7},\eytab{1}) & = D_3(R_{7},N_f) d(R_1, N_f)-D_3(R_1,N_f) d(R_{7}, N_f)\\
&=\frac{(N_f+8) (N_f+4) (N_f+3)}{6} N_f-1 \frac{(N_f+3)(N_f+2)(N_f+1)N_f}{24}\\
&=\frac{ N_f (N_f+3) (N_f (N_f+15)+42) }{8}\, .
\end{split}
\end{equation}
}

The $[SU(N_f)_L]^3$ and $[SU(N_f)_L]^2 U(1)_B$ AMC$[N_f]$ equations are respectively
\begin{equation}
\begin{aligned}
\frac{(N_f+3) (N_f+4) (N_f+5)(N_f+10)}{24} \ell_{1} + \frac{(N_f+5)^2 (N_f^2+N_f-6)}{6}  \ell_{2} \\
+\frac{ N_f (N_f+5) (N_f (5 N_f-3)-50) }{24}  \ell_{3} +\frac{(N_f^4-17N_f^2+100)}{4}  \ell_{4}\\
+\frac{ N_f (N_f (5N_f+3)-50) (N_f-5) }{24}  \ell_{5} +\frac{(N_f-3) (N_f+2) (N_f-5)^2}{6}   \ell_{6} \\
+\frac{(N_f-10) (N_f-4) (N_f-3)(N_f-5)}{24}  \ell_{7} +\frac{N_f (N_f+3) (N_f (N_f+15)+42)}{8}   \ell_{8} \\
+\frac{3 N_f (N_f+3) (N_f(N_f+3)-14) }{8}   \ell_{9} + \frac{N_f^2 (N_f^2-21)}{4} \ell_{10} \\
+\frac{3 (N_f-3) N_f ((N_f-3) N_f-14) }{8}   \ell_{11} +\frac{(N_f-3) N_f ((N_f-15) N_f+42)}{8}  \ell_{12} \\
+\frac{N_f (N_f+1)(N_f (N_f+15)+38)}{12} \ell_{13} +\frac{N_f (N_f+1)((N_f-6) N_f-19)}{6}  \ell_{14} \\
+\frac{N_f (N_f((N_f-26) N_f+47)+38)}{12} \ell_{15} +\frac{N_f (N_f(N_f (N_f+26)+47)-38)}{12}   \ell_{16} \\
+\frac{(N_f-1) N_f(N_f (N_f+6)-19) }{6}   \ell_{17} + \frac{(N_f-1) N_f ((N_f-15) N_f+38)}{12}   \ell_{18} &=5\, ,
\end{aligned}
\end{equation}
and
\begin{equation}
\begin{aligned}
\frac{(N_f+2) (N_f+3) (N_f+4)(N_f+5)}{24}   \ell_{1} +\frac{(N_f+2) (N_f+3) (N_f(N_f+2)-5)}{6}   \ell_{2} \\
+\frac{(N_f+2) (N_f (5 N_f+4)-25)N_f}{24}   \ell_{3} +\frac{ (N_f^4-9N_f^2+20)}{4}   \ell_{4} \\
+\frac{(N_f-2) (N_f (5N_f-4)-25) N_f }{24}   \ell_{5} +\frac{(N_f-3) (N_f-2)((N_f-2) N_f-5)}{6}  \ell_{6} \\
+\frac{(N_f-5) (N_f-4) (N_f-3) (N_f-2) }{24}   \ell_{7} +\frac{(N_f+2) (N_f+3) (N_f+5) N_f }{8}  \ell_{8} \\
+\frac{(N_f+2) (N_f+3) (3N_f-5) N_f }{8}  \ell_{9} +\frac{ (N_f^2-5)N_f^2 }{4}   \ell_{10} \\
+\frac{(N_f-3) (N_f-2) (3 N_f+5)N_f }{8}  \ell_{11} +\frac{(N_f-5) (N_f-3) (N_f-2) N_f }{8}   \ell_{12} \\
+\frac{(N_f+1) (N_f+2) (N_f+5)N_f }{12}   \ell_{13} + \frac{(N_f+1) ((N_f-2) N_f-5)N_f }{6}   \ell_{14} \\
+\frac{(N_f-2) ((N_f-8) N_f-5) N_f }{12}   \ell_{15} +\frac{(N_f+2) (N_f (N_f+8)-5)N_f }{12}   \ell_{16} \\
+\frac{ (N_f (N_f^2+N_f-7)+5) N_f }{6}   \ell_{17} +\frac{(N_f-5) (N_f-2) (N_f-1) N_f }{12}  \ell_{18} &=1\, .
\end{aligned}
\end{equation}
Since all the anomaly constants explicitly depend on $N_f$, the solutions of AMC$[N_f]$ in general depend on the value of $N_f$. 

The PMC$[N_f,1]$ equations are
\begin{equation}
\begin{split}
0=\ell\left(\eytab{0},\eytab{1} \,; 4\right)=&-\ell_{1}-\ell_{2}-\ell_{9}-\ell_{14}+\ell_{16}\ , \\
0=\ell\left(\eytab{0},\eytab{2} \,; 3\right)=& -\ell_{1}-\ell_{2}-\ell_{3}-\ell_{8}-\ell_{9}-\ell_{10}-\ell_{14}\ , \\
0=\ell\left(\eytab{0},\eytab{3} \,; 3\right)=& -\ell_{2}-\ell_{4}-\ell_{9}-\ell_{11}-\ell_{14}-\ell_{15}+\ell_{16}\ , \\
0=\ell\left(\eytab{0},\eytab{4} \,; 2\right)=& -\ell_{1}-\ell_{2}-\ell_{3}-\ell_{8}-\ell_{9}-\ell_{13}\ , \\
0=\ell\left(\eytab{0},\eytab{5} \,; 2\right)=& -\ell_{2}-\ell_{3}-\ell_{4}-\ell_{5}-\ell_{9}-\ell_{10}-\ell_{11}-\ell_{14}\ , \\
0=\ell\left(\eytab{0},\eytab{6} \,; 2\right)=& -\ell_{4}-\ell_{6}-\ell_{11}-\ell_{12}-\ell_{15}\ , \\
0=\ell\left(\eytab{0},\eytab{7} \,; 1\right)=& -\ell_{1}-\ell_{2}-\ell_{8}\ , \\
0=\ell\left(\eytab{0},\eytab{8} \,; 1\right)=& -\ell_{2}-\ell_{3}-\ell_{4}-\ell_{9}\ , \\
0=\ell\left(\eytab{0},\eytab{9} \,; 1\right)=& -\ell_{3}-\ell_{5}-\ell_{10}\ , \\
0=\ell\left(\eytab{0},\eytab{10} \,; 1\right)=& -\ell_{4}-\ell_{5}-\ell_{6}-\ell_{11}\ , \\
0=\ell\left(\eytab{0},\eytab{11} \,; 1\right)=& -\ell_{6}-\ell_{7}-\ell_{12}\ , \\
0=\ell\left(\eytab{1},\eytab{2} \,; 2\right)=& -\ell_{8}-\ell_{9}-\ell_{10}-\ell_{16}-\ell_{17}\ , \\
0=\ell\left(\eytab{1},\eytab{3} \,; 2\right)=& -\ell_{9}-\ell_{11}-\ell_{14}-\ell_{15}+\ell_{16}-\ell_{18}\ , \\
0=\ell\left(\eytab{1},\eytab{4} \,; 1\right)=& -\ell_{8}-\ell_{9}-\ell_{13}-\ell_{16}\ , \\
0=\ell\left(\eytab{1},\eytab{5} \,; 1\right)=& -\ell_{9}-\ell_{10}-\ell_{11}-\ell_{14}-\ell_{17}\ , \\
0=\ell\left(\eytab{1},\eytab{6} \,; 1\right)=& -\ell_{11}-\ell_{12}-\ell_{15}-\ell_{18}\ , \\
0=\ell\left(\eytab{2},\eytab{3} \,; 1\right)=& -\ell_{14}-\ell_{15}+\ell_{16}+\ell_{17}\ .\\ 
\end{split}
\end{equation}
There is one equation for each irrep of ${\cal G}[N_f,1]$. We denote irreps of ${\cal G}[N_f,1]$ by their CYTs and $U(1)_{H_1}$ charge $H_1$ (we omit for simplicity the baryon number which is always equal to 1 in this example). For baryons, $H_1$ simply equals the number of boxes removed in the decomposition, see Eq.~(\ref{eq:Hcharge}).
Each PMC equation has a parity-conjugate equation. For example, the parity-conjugate of the first equation is
\bea
0=\ell\left(\eytab{1},\eytab{0} ; 4\right)=\ell_{1}+\ell_{2}+\ell_{9}+\ell_{14}-\ell_{16}\ .
\eea
As it can be seen from this example, parity-conjugate equations are not truly new equations and will not be considered in the following.

It is simple to see that the two AMC equations become the same once evaluated on the solutions of PMC$[N_f,1]$. By solving PMC$[N_f,1]$ and plugging the solution into AMC$[N_f]$, we obtain
\begin{align}
\label{Nfeq_Nc=5}
25 (4\ell_1 -\ell_2 -2\ell_{16}+\ell_{17}) = 5\ ,\\[0.2cm]
  \label{Nfeq2_Nc=5}
5 (4\ell_1 -\ell_2 -2\ell_{16}+\ell_{17}) = 1\ ,
\end{align}
for respectively the $[SU(N_f)_L]^3$ and $[SU(N_f)_L]^2 U(1)_B$ AMC equations. Remarkably, these two equations become independent of $N_f$, even though the solutions of AMC$[N_f]$ themselves are $N_f$ dependent.   

We find the following solution of AMC$[N_f]$ and PMC$[N_f]$:
\begin{equation}
\label{eq:Nf>=6}
\text{SOL}[N_c=5, N_f\geq 6]=
\left(
\begin{array}{c}
 \ell_{1}= \frac{1}{50} (30 \ell_{16}-15 \ell_{17}+5 \ell_{18}+2) \\
 \ell_{2}= \frac{1}{25} (10 \ell_{16}-5 \ell_{17}+10 \ell_{18}-1) \\
 \ell_{3}= \frac{1}{2}(\ell_{17}-\ell_{18}) \\
 \ell_{4}= \frac{1}{25} (-10 \ell_{16}+5 \ell_{17}+15 \ell_{18}+1) \\
 \ell_{5}= \frac{1}{2}(\ell_{17}-\ell_{18}) \\
 \ell_{6}= \frac{1}{25} (10 \ell_{16}-5 \ell_{17}+10 \ell_{18}-1) \\
 \ell_{7}= \frac{1}{50} (30 \ell_{16}-15 \ell_{17}+5 \ell_{18}+2) \\
 \ell_{8}= \frac{1}{2} (-2 \ell_{16}+\ell_{17}-\ell_{18}) \\
 \ell_{9}= \frac{1}{2} (-\ell_{17}-\ell_{18}) \\
 \ell_{10}= \ell_{18}-\ell_{17} \\
 \ell_{11}= \frac{1}{2} (-\ell_{17}-\ell_{18}) \\
 \ell_{12}= \frac{1}{2} (-2 \ell_{16}+\ell_{17}-\ell_{18}) \\
 \ell_{13}= \ell_{18} \\
 \ell_{14}= \ell_{17} \\
 \ell_{15}= \ell_{16} \\
\end{array}
\right)\ ,
\end{equation}
where we have chosen the free indices to be $\{\ell_{16}, \ell_{17}, \ell_{18}\}$. Since Eq.~(\ref{eq:Nf>=6}) solves AMC$[N_f]\cup$PMC$[N_f]$ for any $N_f\geq 6$, $N_f$-independence is explicitly checked for $N_f\geq 6$.

\subsection{Theory with $N_f=5$}

In a theory with 5 flavors, all the 18 tensors of Tab.~\ref{var_nc_5} are well defined and inequivalent. However, $T_7$ transforms as a singlet of ${\cal G}[5]$; this means that the index $\ell_7$ drops from AMC and PMC and identically vanishes (see Eq.~(\ref{eq:implicationPinv})).
The rank of AMC$[5]\cup$PMC$[5]$ is 14 (see Table~\ref{tab:Nc_5}), and we conclude that the family of real solutions has 3 free parameters.

The $[SU(5)_L]^3$ and $[SU(5)_L]^2 U(1)_B$ AMC$[5]$ equations are respectively
\begin{equation}
  \begin{split}
450 \ell_{1}+400 \ell_{2}+125 \ell_{3}+75 \ell_{4}+710 \ell_{8}+390\ell_{9}+25 \ell_{10} \\
-15 \ell_{11}-10 \ell_{12}+345 \ell_{13}-120 \ell_{14}-105\ell_{15}+405 \ell_{16}+120 \ell_{17}-20 \ell_{18}&=5\ , 
\end{split}
\end{equation}
and
\begin{equation}
  \label{eq:AMC2Nf5}
\begin{split}
210 \ell_{1}+280 \ell_{2}+175 \ell_{3}+105 \ell_{4}+50 \ell_{5}+10\ell_{6}+350 \ell_{8}+350 \ell_{9}+125 \ell_{10} \\
+75 \ell_{11}+175 \ell_{13}+50 \ell_{14}-25 \ell_{15}+175 \ell_{16}+100 \ell_{17}&=1\ .
\end{split}
\end{equation}
One can check the `prime factor' theorem~\cite{Preskill:1981sr, Weinberg:1996kr, CLRX2} explicitly. All the anomaly coefficients in Eq.~(\ref{eq:AMC2Nf5}) are multiples of $5$, so there is no integral solution for this equation and more in general for the entire system. This is enough to prove $\chi$SB in this case.

The PMC$[5,1]$ equations are
\begin{equation}
\begin{split}
0=\ell\left(\eytab{0},\eytab{1} \,; 4\right) &=-\ell_{1}-\ell_{2}-\ell_{9}-\ell_{14}+\ell_{16}\ , \\
0=\ell\left(\eytab{0},\eytab{2} \,; 3\right) &=-\ell_{1}-\ell_{2}-\ell_{3}-\ell_{8}-\ell_{9}-\ell_{10}-\ell_{14}\ , \\
0=\ell\left(\eytab{0},\eytab{3} \,; 3\right) &= -\ell_{2}-\ell_{4}-\ell_{9}-\ell_{11}-\ell_{14}-\ell_{15}+\ell_{16}\ , \\
0=\ell\left(\eytab{0},\eytab{4} \,; 2\right) &= -\ell_{1}-\ell_{2}-\ell_{3}-\ell_{8}-\ell_{9}-\ell_{13}\ , \\
0=\ell\left(\eytab{0},\eytab{5} \,; 2\right) &= -\ell_{2}-\ell_{3}-\ell_{4}-\ell_{5}-\ell_{9}-\ell_{10}-\ell_{11}-\ell_{14}\ ,\\
0=\ell\left(\eytab{0},\eytab{6} \,; 2\right) &= -\ell_{4}-\ell_{6}-\ell_{11}-\ell_{12}-\ell_{15} \ ,\\
0=\ell\left(\eytab{0},\eytab{7} \,; 1\right) &= -\ell_{1}-\ell_{2}-\ell_{8} \ ,\\
0=\ell\left(\eytab{0},\eytab{8} \,; 1\right) &= -\ell_{2}-\ell_{3}-\ell_{4}-\ell_{9} \ ,\\
0=\ell\left(\eytab{0},\eytab{9} \,; 1\right) &= -\ell_{3}-\ell_{5}-\ell_{10} \ ,\\
0=\ell\left(\eytab{0},\eytab{10} \,; 1\right) &= -\ell_{4}-\ell_{5}-\ell_{6}-\ell_{11} \ ,\\
0=\ell\left(\eytab{1},\eytab{2} \,; 2\right) &= -\ell_{8}-\ell_{9}-\ell_{10}-\ell_{16}-\ell_{17} \ ,\\
0=\ell\left(\eytab{1},\eytab{3} \,; 2\right) &= -\ell_{9}-\ell_{11}-\ell_{14}-\ell_{15}+\ell_{16}-\ell_{18} \ ,\\
0=\ell\left(\eytab{1},\eytab{4} \,; 1 \right) &= -\ell_{8}-\ell_{9}-\ell_{13}-\ell_{16} \ ,\\
0=\ell\left(\eytab{1},\eytab{5} \,; 1 \right) &= -\ell_{9}-\ell_{10}-\ell_{11}-\ell_{14}-\ell_{17} \ ,\\
0=\ell\left(\eytab{1},\eytab{6} \,; 1 \right) &=-\ell_{11}-\ell_{12}-\ell_{15}-\ell_{18} \ ,\\
0=\ell\left(\eytab{2},\eytab{3} \,; 1 \right) &= -\ell_{14}-\ell_{15}+\ell_{16}+\ell_{17}\ , 
\end{split}
\end{equation}
We see that PMC$[5,1]$ are not exactly the same as PMC$[6,1]$. In particular, the parity-conjugated equations
\begin{equation}
\begin{split}
0 & =\ell \left(\ {\tiny\yng(1,1,1,1)}\ ,\ \cdot\ \right) =\ell_{6}+\ell_{7}+\ell_{12}\\[0.3cm]
0 & =\ell \left(\ \cdot\ ,\ {\tiny\yng(1,1,1,1)}\  \right)=-\ell_{6}-\ell_{7}-\ell_{12}
\end{split}
\end{equation}
exist in PMC$[6,1]$, but they collapse into a single identically-vanishing equation in PMC$[5,1]$.
Since AMC$[5]$ and PMC$[5,1]$ are not the same as AMC$[N_f]$ and PMC$[N_f,1]$ for $N_f\geq 6$, it is clear that $N_f$-independence does not hold anymore for $N_f =5$. Despite this fact, we verified that the same $N_f$-equation that we found for $N_f\geq 6$ (see Eqs.~(\ref{Nfeq_Nc=5}) and (\ref{Nfeq2_Nc=5})) is obtained also in this case once AMC$[5]$ are evaluated on the solutions of PMC$[5,1]$. Hence, the $N_f$-independence of the $N_f$-equation seems to hold for $N_f=5$.

The solution of AMC$[5]$ and PMC$[5]$ is
\begin{equation}
\label{eq:Nf=5}
\text{SOL}[N_c=5,N_f=5]=
\left(
\begin{array}{c}
 \ell_{1}= \frac{1}{50} (30 \ell_{16}-15 \ell_{17}+5 \ell_{18}+2) \\
 \ell_{2}= \frac{1}{25} (10 \ell_{16}-5 \ell_{17}+10 \ell_{18}-1) \\
 \ell_{3}= \frac{1}{2} (\ell_{17}-\ell_{18}) \\
 \ell_{4}= \frac{1}{25} (-10 \ell_{16}+5 \ell_{17}+15 \ell_{18}+1) \\
 \ell_{5}= \frac{1}{2} (\ell_{17}-\ell_{18}) \\
  \ell_{6}= \frac{1}{25} (10 \ell_{16}-5 \ell_{17}+10 \ell_{18}-1) \\
  \ell_7 = 0 \\
 \ell_{8}= \frac{1}{2} (-2 \ell_{16}+\ell_{17}-\ell_{18}) \\
 \ell_{9}= \frac{1}{2} (-\ell_{17}-\ell_{18}) \\
 \ell_{10}= \ell_{18}-\ell_{17} \\
 \ell_{11}= \frac{1}{2} (-\ell_{17}-\ell_{18}) \\
 \ell_{12}= \frac{1}{2} (-2 \ell_{16}+\ell_{17}-\ell_{18}) \\
 \ell_{13}= \ell_{18} \\
 \ell_{14}= \ell_{17} \\
  \ell_{15}= \ell_{16}
\end{array}
\right)\, ,
\end{equation}
where we have chosen the free indices to be $\{\ell_{16}, \ell_{17}, \ell_{18}\}$. Notice that $\ell_{7}$ vanishes due to parity invariance of the spectrum.

\subsection{Theory with $N_f=4$}

The form of AMC and PMC equations changes substantially for $N_f=4$.
First, we notice that the tensor $T_7$ is not well defined in this case, since one cannot have 5 fully antisymmetrized indices if $N_f=4$.
Furthermore, $T_6$ and the parity conjugate of $T_{12}$ are equivalent tensors, and thus transform as the same irrep:
\bea
{\left(\ {\tiny\yng(2,1,1,1)}\ ,\ \cdot\ \right)} \sim \left(\ {\tiny\yng(1)}\ ,\ {\tiny\yng(1,1,1,1)} \ \right)  \ \xleftrightarrow{\text{parity}}\  \left(\ {\tiny\yng(1,1,1,1)}\ ,\ {\tiny\yng(1)} \ \right)\ .
\eea
We denote the index of such irrep as $\ell_6$;~\footnote{In the following, whenever two or more tensors $T_i$ are equivalent, we denote the index of the corresponding irrep as $\ell_{i_{min}}$, where $i_{min}$ is the smallest among the tensors' labels.}
consequently, there is no $\ell_{12}$ index in this case. There are in total $16$ irreps for $N_f=4$, each with a corresponding index.
In this case, AMC$[N_f]\cup$PMC$[N_f]$ has rank 13 (see Table~\ref{tab:Nc_5}), and its family of solutions has 3 free parameters.

The $[SU(4)_L]^3$ and $[SU(4)_L]^2 U(1)_B$ AMC$[4]$ equations are respectively
\begin{align}
  \begin{aligned}
    294 \ell_{1}+189 \ell_{2}+27 \ell_{3}+21 \ell_{4}-7 \ell_{5}+\ell_{6}+413 \ell_{8}+147 \ell_{9}-20 \ell_{10} \\
-15 \ell_{11}+190 \ell_{13}-90\ell_{14}-42 \ell_{15}+210 \ell_{16}+42 \ell_{17}-6 \ell_{18}&=5\ , \\
\end{aligned} 
  \intertext{and}
\begin{aligned}
126 \ell_{1}+133 \ell_{2}+71 \ell_{3}+33 \ell_{4}+13 \ell_{5}+\ell_{6}+189 \ell_{8}+147 \ell_{9}+44 \ell_{10} \\
+17 \ell_{11}+90 \ell_{13}+10 \ell_{14}-14 \ell_{15}+86 \ell_{16}+38 \ell_{17}-2 \ell_{18}&=1\, . 
\end{aligned}
\end{align}

The PMC$[4,1]$ equations are
\begin{equation}
\begin{split}
0=\ell\left(\eytab{0},\eytab{1} \,; 4\right) &=-\ell_{1}-\ell_{2}-\ell_{9}-\ell_{14}+\ell_{16}\ , \\
0=\ell\left(\eytab{0},\eytab{2} \,; 3\right) &=-\ell_{1}-\ell_{2}-\ell_{3}-\ell_{8}-\ell_{9}-\ell_{10}-\ell_{14}\ , \\
0=\ell\left(\eytab{0},\eytab{3} \,;3\right) &= -\ell_{2}-\ell_{4}-\ell_{9}-\ell_{11}-\ell_{14}-\ell_{15}+\ell_{16}\ , \\
0=\ell\left(\eytab{0},\eytab{4} \,;2\right) &= -\ell_{1}-\ell_{2}-\ell_{3}-\ell_{8}-\ell_{9}-\ell_{13}\ , \\
0=\ell\left(\eytab{0},\eytab{5} \,;2\right) &= -\ell_{2}-\ell_{3}-\ell_{4}-\ell_{5}-\ell_{9}-\ell_{10}-\ell_{11}-\ell_{14}\ , \\
0=\ell\left(\eytab{0},\eytab{7} \,;1\right) &= -\ell_{1}-\ell_{2}-\ell_{8}\ , \\
0=\ell\left(\eytab{0},\eytab{8} \,;1\right) &= -\ell_{2}-\ell_{3}-\ell_{4}-\ell_{9}\ , \\
0=\ell\left(\eytab{0},\eytab{9} \,;1\right) &=-\ell_{3}-\ell_{5}-\ell_{10}\ , \\
0=\ell\left(\eytab{1},\eytab{2} \,;2\right) &= -\ell_{8}-\ell_{9}-\ell_{10}-\ell_{16}-\ell_{17}\ , \\
0=\ell\left(\eytab{1},\eytab{3} \,;2\right) &= -\ell_{9}-\ell_{11}-\ell_{14}-\ell_{15}+\ell_{16}-\ell_{18}\ , \\
0=\ell\left(\eytab{1},\eytab{4} \,;1\right) &= -\ell_{8}-\ell_{9}-\ell_{13}-\ell_{16}\ , \\
0=\ell\left(\eytab{1},\eytab{5} \,;1\right) &= -\ell_{9}-\ell_{10}-\ell_{11}-\ell_{14}-\ell_{17}\ , \\
0=\ell\left(\eytab{1},\eytab{6} \,;1\right)+\ell\left(\eytab{10},\eytab{0}\,;1 \right)&= \ell_{4}+\ell_{5}+\ell_{6}-\ell_{15}-\ell_{18}\ , \\
0=\ell\left(\eytab{2},\eytab{3}\,; 1 \right) &= -\ell_{14}-\ell_{15}+\ell_{16}+\ell_{17}\ .
\end{split}
\end{equation}

Let us compare PMC$[4,1]$ with PMC$[5,1]$.
\begin{itemize}
\item The following two equations in PMC$[5,1]$
\bea
0 &=\ell\left(\eytab{0},\eytab{10}\,; 1 \right) = -\ell_{4}-\ell_{5}-\ell_{6}-\ell_{11} \ ,\\
0 &=\ell\left(\eytab{1},\eytab{6}\,; 1 \right)  =-\ell_{11}-\ell_{12}-\ell_{15}-\ell_{18} \ ,
\eea
collapse into a single equation in PMC$[4,1]$:
\bea
0=\ell\left(\eytab{1},\eytab{6}\,; 1\right)+\tilde\ell\left(\eytab{10},\eytab{0}\,;1\right) = \ell_{4}+\ell_{5}+\ell_{6}-\ell_{15}-\ell_{18}\ .
\eea
This is because the two CYTs involved are inequivalent under $\mathcal{G}[5,1]$, but become equivalent under $\mathcal{G}[4,1]$.
\item The following equation 
\bea
0=\ell\left(\eytab{0},\eytab{6}\,; 2\right) &= -\ell_{4}-\ell_{6}-\ell_{11}-\ell_{12}-\ell_{15}\ ,
\eea
appears in PMC$[5,1]$ but not in PMC$[4,1]$.
This is because the CYT involved is equivalent to its parity conjugate under $\mathcal{G}[4,1]$, hence the corresponding equation identically vanishes in PMC$[4,1]$:
\begin{equation}
  \begin{split}
0=&\ell\left(\eytab{0},\eytab{6}\,; 2\right) + \ell\left(\eytab{6}, \eytab{0} \,; 2\right) \\
  =& (-\ell_{4}-\ell_{6}-\ell_{11}-\ell_{15})+(\ell_{4}+\ell_{6}+\ell_{11}+\ell_{15}) \ .
\end{split}
\end{equation}
\end{itemize}

The differences between PMC$[4,1]$ and PMC$[5,1]$ show very clearly that PMC$[N_f,1]$ do depend on the value of $N_f$.
Hence $N_f$-independence cannot be a general feature of PMC$[N_f,1]$. 
On the other hand, we find that once the two AMC$[4]$ equations are evaluated on the solutions of PMC$[4,1]$, they are linearly dependent and identical to Eqs.~(\ref{Nfeq_Nc=5}),~(\ref{Nfeq2_Nc=5}). Therefore, the $N_f$-independence of the $N_f$-equations holds true also for $N_f=4$.

The solution of AMC$[4]\cup$PMC$[4]$ is:
\begin{equation}
\label{eq:Nf=4}
\text{SOL}[N_c=5,N_f=4]=
\left(
\begin{array}{c}
 \ell_{1}= \frac{1}{50} (30 \ell_{16}-15 \ell_{17}+5 \ell_{18}+2) \\
 \ell_{2}= \frac{1}{25} (10 \ell_{16}-5 \ell_{17}+10 \ell_{18}-1) \\
 \ell_{3}= \frac{1}{2}\left(\ell_{17}-\ell_{18}\right)\\
 \ell_{4}= \frac{1}{25} (-10 \ell_{16}+5 \ell_{17}+15 \ell_{18}+1) \\
  \ell_{5}= \frac{1}{2}\left(\ell_{17}-\ell_{18}\right) \\
  \ell_{6}= \frac{1}{50} (70 \ell_{16}-35 \ell_{17}+45 \ell_{18}-2) \\
 \ell_{8}= \frac{1}{2} (-2 \ell_{16}+\ell_{17}-\ell_{18}) \\
 \ell_{9}= \frac{1}{2} (-\ell_{17}-\ell_{18}) \\
 \ell_{10}= \ell_{18}-\ell_{17} \\
 \ell_{11}= \frac{1}{2} (-\ell_{17}-\ell_{18}) \\
 \ell_{13}= \ell_{18} \\
 \ell_{14}= \ell_{17} \\
 \ell_{15}= \ell_{16} \\
\end{array}
\right)\ ,
\end{equation}
where we have chosen the free indices to be $\{\ell_{16}, \ell_{17}, \ell_{18}\}$.

\subsection{Theory with $N_f=3$}

Tensors $T_{6}, T_{7}, T_{12}$ do not exist in this theory, because one cannot have CYTs with columns of more than 3 boxes if $N_f=3$. 
Furthermore, tensor $T_4$ and the parity conjugate of $T_{15}$ are equivalent:
\bea
\left(\ {\tiny\yng(3,1,1)}\ , \ \cdot\ \right) \sim \left(\ {\tiny\yng(2)}\ , \ {\tiny\yng(1,1,1)}\ \right) \ \xleftrightarrow{\text{parity}}\  \left(\ {\tiny\yng(1,1,1)}\ ,\ {\tiny\yng(2)}\ \right)\ ;
\eea
the index of the corresponding representation will be denoted as $\ell_4$ (while $\ell_{15}$ is not defined in this case).
Similarly, $T_5$ and the parity conjugate of $T_{18}$ are equivalent:
\bea
\left(\ {\tiny\yng(2,2,1)}\ , \ \cdot\ \right) \sim \left(\ {\tiny\yng(1,1)}\ , \ {\tiny\yng(1,1,1)}\ \right) \ \xleftrightarrow{\text{parity}}\  \left(\ {\tiny\yng(1,1,1)}\ ,\ {\tiny\yng(1,1)}\ \right)\ ;
\eea
the index of the corresponding representation will be denoted as $\ell_5$ (while $\ell_{18}$ is not defined in this case).
Therefore, there are $13$ irreps in total, each with a corresponding index. 
Notice, however, that tensor $T_{11}$ is equivalent to its parity conjugate:
\bea
\left(\ {\tiny\yng(2,1,1)}\ , \ {\tiny\yng(1)}\ \right) \sim \left(\ {\tiny\yng(1)}\ ,\ {\tiny\yng(2,1,1)}\ \right)\ .
\eea
This implies that the index $\ell_{11}$ drops from AMC or PMC and identically vanishes (see Eq.~(\ref{eq:implicationPinv})).
The rank of AMC$[3]\cup$PMC$[3]$ is 9 (see Table~\ref{tab:Nc_5}), and we conclude that the family of solutions has 3 free parameters.

The $[SU(3)_L]^3$ and $[SU(3)_L]^2 U(1)_B$ AMC$[3]$ equations are respectively
\begin{align}
182 \ell_{1}+64 \ell_{2}-14 \ell_{3}+7 \ell_{4}-\ell_{5}+216 \ell_{8}+27\ell_{9}-27 \ell_{10}
+92 \ell_{13}-56 \ell_{14}+91 \ell_{16}+8\ell_{17}&=5\ ,
\\[0.3cm]
  70 \ell_{1}+50 \ell_{2}+20 \ell_{3}+5 \ell_{4}+\ell_{5}+90 \ell_{8}+45 \ell_{9}+9 \ell_{10} 
+40 \ell_{13}-4 \ell_{14}+35 \ell_{16}+10\ell_{17}&=1 \ .
\end{align}
The PMC$[3,1]$ equations are
\begin{equation}
\begin{split}
0=\ell\left(\eytab{0},\eytab{1} \,; 4\right)&=-\ell_{1}-\ell_{2}-\ell_{9}-\ell_{14}+\ell_{16}\ , \\
0=\ell\left(\eytab{0},\eytab{5}\,;2\right)+\ell\left(\eytab{3},\eytab{1}\,;2\right)&=-\ell_{2}-\ell_{3}-\ell_{4}-\ell_{5}-\ell_{10}-\ell_{16}\ ,\\
0=\ell\left(\eytab{0},\eytab{2}\,;3\right)&= -\ell_{1}-\ell_{2}-\ell_{3}-\ell_{8}-\ell_{9}-\ell_{10}-\ell_{14}\ , \\
0=\ell\left(\eytab{0},\eytab{8}\,;1\right)+\ell\left(\eytab{3}, \eytab{2}\,;1\right) &= -\ell_{2}-\ell_{3}-\ell_{4}-\ell_{9}+\ell_{14}-\ell_{16}-\ell_{17}\ ,\\
0=\ell\left(\eytab{0},\eytab{4}\,;2\right)&= -\ell_{1}-\ell_{2}-\ell_{3}-\ell_{8}-\ell_{9}-\ell_{13}\ ,\\
0=\ell\left(\eytab{0},\eytab{7}\,;1\right)&= -\ell_{1}-\ell_{2}-\ell_{8}\ , \\
0=\ell\left(\eytab{1},\eytab{2}\,;2\right)&= -\ell_{8}-\ell_{9}-\ell_{10}-\ell_{16}-\ell_{17}\ , \\
0=\ell\left(\eytab{1},\eytab{4}\,;1\right)&= -\ell_{8}-\ell_{9}-\ell_{13}-\ell_{16} \ .
\end{split}
\end{equation}

Let us compare PMC$[3,1]$ with PMC$[4,1]$. 
\begin{itemize}
\item The following two equations in PMC$[4,1]$
\begin{align}
0=\ell\left(\eytab{0},\eytab{5}\,;2\right) =& -\ell_{2}-\ell_{3}-\ell_{4}-\ell_{5}-\ell_{9}-\ell_{10}-\ell_{11}-\ell_{14}\ , \\
0=\ell\left(\eytab{1},\eytab{3}\,;2\right) =& -\ell_{9}-\ell_{11}-\ell_{14}-\ell_{15}+\ell_{16}-\ell_{18}\ ,
\end{align}
collapse into a single equation in PMC$[3,1]$
\begin{equation}
0=\ell\left(\eytab{0},\eytab{5}\,;2\right)+\ell\left(\eytab{3},\eytab{1}\,;2\right)=-\ell_{2}-\ell_{3}-\ell_{4}-\ell_{5}-\ell_{10}-\ell_{16}\ ,
\end{equation}
because $(\eytab{0},\eytab{5})$ and $(\eytab{3},\eytab{1})$ correspond to the same irrep of $SU(2)_L\times SU(2)_R$.
Similarly, the following two equations in PMC$[4,1]$
\begin{align}
0=\ell\left(\eytab{0},\eytab{8}\,;1\right) =& -\ell_{2}-\ell_{3}-\ell_{4}-\ell_{9}\ , \\
0=\ell\left(\eytab{2},\eytab{3}\,;1\right) =& -\ell_{14}-\ell_{15}+\ell_{16}+\ell_{17}\ ,
\end{align}
collapse into a single equation in PMC$[3,1]$
\begin{equation}
0=\ell\left(\eytab{0},\eytab{8}\,; 1\right)+\ell\left(\eytab{3}, \eytab{2}\,;1\right) = -\ell_{2}-\ell_{3}-\ell_{4}-\ell_{9}+\ell_{14}-\ell_{16}-\ell_{17}\ ,
\end{equation}
because $(\eytab{0},\eytab{8})$ and $(\eytab{3}, \eytab{2})$ correspond to the same irrep of $SU(2)_L\times SU(2)_R$.
\item The following equations appear only in PMC$[4,1]$ but not in PMC$[3,1]$:
\begin{align}
0=\ell\left(\eytab{0},\eytab{3}\,; 3\right) =& -\ell_{2}-\ell_{4}-\ell_{9}-\ell_{11}-\ell_{14}-\ell_{15}+\ell_{16}\ , \\
0=\ell\left(\eytab{0},\eytab{9}\,; 1\right) =& -\ell_{3}-\ell_{5}-\ell_{10}\ , \\
0=\ell\left(\eytab{1},\eytab{5}\,; 1\right) =& -\ell_{9}-\ell_{10}-\ell_{11}-\ell_{14}-\ell_{17}\ , \\
0=\ell\left(\eytab{1},\eytab{6}\,; 1\right)+\ell\left(\eytab{10},\eytab{0}\,; 1\right)=& \ell_{4}+\ell_{5}+\ell_{6}-\ell_{15}-\ell_{18}\ .
\end{align}
The first three equations vanish identically in PMC$[3,1]$ because they correspond to CYTs that are equivalent to their parity conjugates under $\mathcal{G}[3,1]$. The fourth equation does not exist in PMC$[3,1]$ simply because its CYT is not a valid irrep of  $\mathcal{G}[3,1]$.
\end{itemize}
This shows that $N_f$-independence does not hold as $N_f$ decreases from 4 to 3. 

As before, we find that, once evaluated on the solutions of PMC$[3,1]$, the two AMC equations become linearly dependent and identical respectively to Eq.~(\ref{Nfeq_Nc=5}), and~(\ref{Nfeq2_Nc=5}). This proves the $N_f$-independence of the $N_f$-equation also for $N_f=3$.

The solution of AMC$[3]\cup$PMC$[3]$ reads
\begin{equation}
\label{eq:Nf=3}
\text{SOL}[N_c=5,N_f=3]=
\left(
\begin{array}{c}
 \ell_{2}= 4 \ell_{1}-2 \ell_{16}+\ell_{17}-\frac{1}{5} \\
  \ell_{3}= -5 \ell_{1}+3 \ell_{16}-\ell_{17}+\frac{1}{5} \\
  \ell_{4}= 6 \ell_{1}-5 \ell_{16}+2 \ell_{17}-\frac{1}{5} \\
 \ell_{5}= -15 \ell_{1}+9 \ell_{16}-4 \ell_{17}+\frac{3}{5} \\
 \ell_{8}= -5 \ell_{1}+2 \ell_{16}-\ell_{17}+\frac{1}{5} \\
 \ell_{9}= -5 \ell_{1}+3 \ell_{16}-2 \ell_{17}+\frac{1}{5} \\
  \ell_{10}= 10 \ell_{1}-6 \ell_{16}+2 \ell_{17}-\frac{2}{5} \\
  \ell_{11} = 0 \\
 \ell_{13}= 10 \ell_{1}-6 \ell_{16}+3 \ell_{17}-\frac{2}{5} \\
 \ell_{14}= \ell_{17}  
\end{array}
\right)\ ,
\end{equation}
where we have chosen the free indices to be $\{\ell_{1}, \ell_{16}, \ell_{17}\}$. Notice that $\ell_{11}$ vanishes as a consequence of parity invariance of the spectrum.

\subsection{Theory with $N_f=2$}

Tensors $T_{4}, T_{5}, T_{6}, T_{7}, T_{11}, T_{12}, T_{15}$, and $T_{18}$ are not well defined in the theory with $N_f=2$.
Furthermore, we identify the following groups of equivalent tensors: $T_3$, $T_{17}$ and the parity conjugate of $T_{10}$; $T_2$ and the parity conjugate of $T_{16}$; $T_9$ and the parity conugate of $T_{14}$. The indices of the corresponding irreps are denoted respectively by $\ell_3$, $\ell_2$ and $\ell_9$. There are in total 6 irreps. 
The $[SU(2)_L]^2 U(1)_B$ AMC equation is
\begin{equation}
\label{eq:Nf=2}
35 \ell_{1}+10 \ell_{2}+\ell_{3}+35 \ell_{8}+5 \ell_{9}+14 \ell_{13}=1\ ,
\end{equation}
while there is no $[SU(2)_L]^3$ AMC equation, since all the representations of $SU(2)_L$ are real or pseudo-real. There are no PMC equations either. Indeed, when one flavor is given a mass, the unbroken global symmetry group ${\cal G}[2,1]=U(1)_{H_1}\times U(1)_B$ is vectorial, and all the states after decomposition are automatically paired for a parity-invariant spectrum. 
It is easy to see that Eq.~(\ref{eq:Nf=2}) admits integral solutions.

\subsection{Downlifting: from $N_f=6$ to $N_f=2$}
\label{downlifting_baryons}

As we demonstrated in the previous sections, solutions of AMC$[N_f]\cup$PMC$[N_f]$ are $N_f$-independent in our first example only for $N_f\geq 6$. This result agrees with the general condition stated in~\cite{Ciambriello:2022wmh}. Even when 
$N_f$-independence does not hold, one can always downlift solutions by means of Eq.~(\ref{eq:downliftedsol}), i.e. one can obtain a solution for a theory with $N_f-1$ massless flavors from the solution of a theory with $N_f$ massless flavors. 
By iteration, one can construct downlifted solutions down to $N_f=2$.

For $N_f \geq 6$, all massless baryons in the spectrum are class A. Equation~(\ref{eq:uplift}) then holds (i.e. downlifting is the inverse of uplifting~\cite{Ciambriello:2022wmh}), and the correspondence between irreps of ${\cal G}[N_f+1]$ and ${\cal G}[N_f]$ is one-to-one. Hence, the downlifted solution implied by Eqs.~(\ref{eq:downliftedsol}) and~(\ref{eq:uplift}) is simply~\footnote{We recall that the symbol $\tilde \ell$ denotes an index of the downlifted solution, c.f. footnote~\ref{fot:elltilde}. In this section, indices are equipped with an upper label to indicate that they refer to the theory with the corresponding number of flavors.}
\begin{equation}
  \label{eq:downliftNfp1toNf}
\tilde\ell^{(N_f)}_i\equiv \ell^{(N_f+1)}_i ,\quad i=1,\cdots,18\ .
\end{equation}

When downlifting from $N_f = 6$ to $N_f=5$, one finds the following downlifted solution of AMC$[5]\cup$PMC$[5]$:
\begin{equation}
  \label{eq:downlift6to5}
  \begin{split}
    \tilde\ell^{(5)}_7 & \equiv 0 \, , \\
\tilde\ell^{(5)}_i & \equiv \ell^{(6)}_i ,\quad i=1,\cdots, 6, 8, \cdots ,18\ .
\end{split}
\end{equation}
Notice that $\tilde\ell^{(5)}_7$ vanishes consistently with the fact that $\ell_7$ also vanishes in Eq.~(\ref{eq:Nf=5}) and as a consequence of parity invariance of the spectrum ($T_7$ transforms as a singlet of ${\cal G}[5]$).


Starting from $N_f=5$, the downlifted solution of AMC$[4]\cup$PMC$[4]$ is the following:
\begin{equation}
  \label{eq:downlift5to4}
  \begin{split}
  \tilde\ell^{(4)}_i  & \equiv \ell^{(5)}_i ,\quad i=1,\cdots, 5, 8,\cdots,11,13 \cdots, 18 \, , \\
  \tilde\ell^{(4)}_6 &\equiv \ell^{(5)}_6-\ell^{(5)}_{12}\, .
  \end{split}
\end{equation}
The last line follows from Eq.~(\ref{eq:downliftedsol}) because $T_6$ and the parity conjugate of $T_{12}$ become equivalent for $N_f=4$.

By downlifting from $N_f=4$ to $N_f=3$, one instead finds:
\begin{equation}
  \label{eq:downlift4to3}
  \begin{split}
\tilde\ell^{(3)}_{4} & \equiv  \ell^{(4)}_4 - \ell^{(4)}_{15} \, ,\\
\tilde\ell^{(3)}_{5} & \equiv  \ell^{(4)}_5 - \ell^{(4)}_{18} \, , \\
\tilde\ell^{(3)}_{11} & \equiv 0 \, , \\
\tilde\ell^{(3)}_i    &\equiv \ell^{(4)}_i ,\quad i=1, 2, 3, 8, 9, 10, 13, 14, 16, 17 \ . \\
\end{split}
\end{equation}
The first two lines follow from Eq.~(\ref{eq:downliftedsol}) because, for $N_f=3$, $T_4$ is equivalent to the parity conjugate of $T_{15}$, and $T_5$ is equivalent to the parity conjugate of $T_{18}$. Furthermore, $\tilde\ell^{(3)}_{11} $ vanishes consistently with Eq.~(\ref{eq:Nf=3}) and as a consequence of parity invariance of the spectrum ($T_{11}$ transforms as a singlet of ${\cal G}[3]$).

Finally, by downlifting from $N_f=3$ to $N_f=2$ one obtains:
\begin{equation}
  \label{eq:downlift3to2}
  \begin{split}
\tilde\ell^{(2)}_i & \equiv \ell^{(3)}_i ,\quad i=1, 8, 13\, ,\\
\tilde\ell^{(2)}_2 & \equiv \tilde\ell^{(3)}_2 - \tilde\ell^{(3)}_{16}\, ,\\
\tilde\ell^{(2)}_3 & \equiv \tilde\ell^{(3)}_3 - \tilde\ell^{(3)}_{10} + \tilde\ell^{(3)}_{17} \, ,\\
\tilde\ell^{(2)}_9 & \equiv \tilde\ell^{(3)}_9 - \tilde\ell^{(3)}_{14}\, .\\
\end{split}
\end{equation}

In all the cases except for $N_f=2$, the downlifted solution is the most general solution of AMC and PMC equations.


\section{$N_c=3$ QCD-like theories with massless baryons and pentaquarks}
\label{sec:resultsNc3}

As our second example, we consider QCD-like theories with $N_c=3$ and $N_f\geq 2$, and assume a spectrum of massless fermions consisting of baryons and pentaquarks with $b=1$. Pentaquarks are defined as those bound states that can be interpolated by local composite operators made of $bN_c+1$ quarks and 1 antiquark, and that cannot be interpolated by pure quark operators. For $N_c=3$ and $b=1$, all the possible tensors of ${\cal G}[N_f]$ associated to the interpolating operators are listed in Table~\ref{var_pentaq}, together with their corresponding indices and CYTs. Notice that some of the CYTs do not exist for $N_f <5$. Similarly to our first example, also in this case we assume for simplicity that the spectrum is parity invariant, that is: each irrep  in Table~\ref{var_pentaq} has a parity-conjugated partner with equal and opposite index. 
In the following, we study the system of AMC and PMC equations for theories of different $N_f$.

\vspace{1.2cm}
\begingroup
\renewcommand*{\arraystretch}{1.8}
\begin{longtable}{|c|c|c||c|c|c|}
	\hline\hline
		Index & Irrep (CYT) & Tensor & Index & Irrep (CYT) & Tensor\\
		\hline
		\hline
		$\ell_{a}$ & (\ytab{4}, \ytab{0}) & $T_a\sim T^{(\eytab{4}, \eytab{0})}_{(\eytab{0}, \eytab{0})}$ & $\ell_{b}$ & (\ytab{6}, \ytab{0}) & $T_b\sim T^{(\eytab{6}, \eytab{0})}_{(\eytab{0},\eytab{0})}$\\
		\hline
		$\ell_{c}$ & (\ytab{5}, \ytab{0}) & $T_c\sim T^{(\eytab{5}, \eytab{0})}_{(\eytab{0}, \eytab{0})}$ & $\ell_{d}$ & (\ytab{2}, \ytab{1}) & $T_d\sim T^{(\eytab{2}, \eytab{1})}_{(\eytab{0}, \eytab{0})}$\\
		\hline
		$\ell_{e}$ & (\ytab{3}, \ytab{1}) & $ T_e\sim T^{(\eytab{3}, \eytab{1})}_{(\eytab{0}, \eytab{0})} $ &&&\\
		\hline
  $\ell_{1}$ & (\ytab{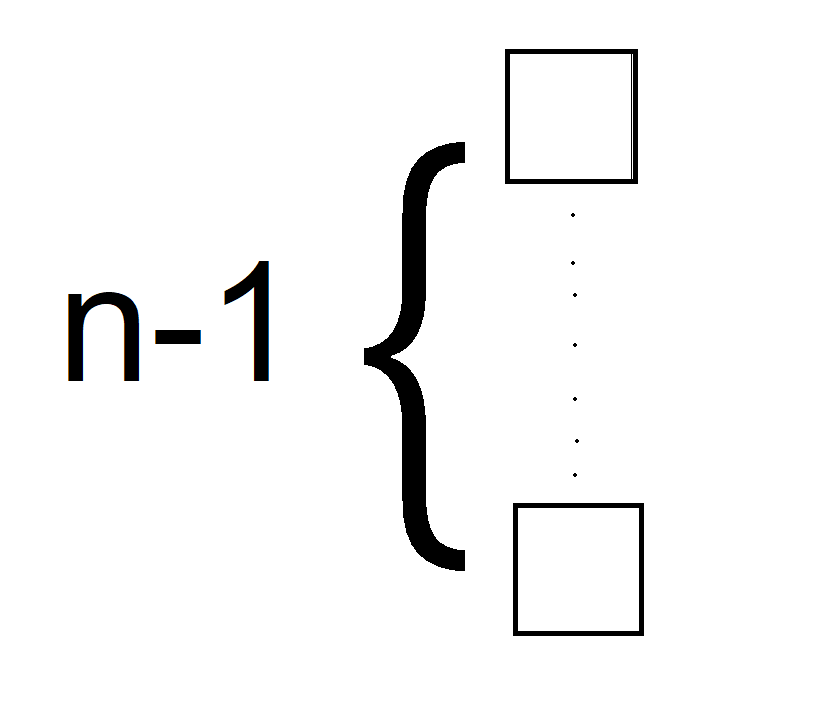}, \ytab{7}) & $T_1\sim T^{(\eytab{0}, \eytab{7})}_{(\eytab{1}, \eytab{0})}$ & $\ell_{2}$ & (\ytab{12}, \ytab{8}) & $T_2\sim T^{(\eytab{0}, \eytab{8})}_{(\eytab{1}, \eytab{0})} $\\
		\hline
		$\ell_{3}$ & (\ytab{12}, \ytab{9}) & $T_3\sim T^{(\eytab{0}, \eytab{9})}_{(\eytab{1}, \eytab{0})}$ & $\ell_{4}$ & (\ytab{12}, \ytab{10}) & $T_4\sim T^{(\eytab{0}, \eytab{10})}_{(\eytab{1}, \eytab{0})}$ \\
		\hline
		$\ell_{5}$ & (\ytab{12}, \ytab{11}) & $T_5\sim T^{(\eytab{0}, \eytab{11})}_{(\eytab{1}, \eytab{0})}$ & $\ell_{6}$ & (\ytab{13}, \ytab{4}) & $T_6\sim T^{(\eytab{1}, \eytab{4})}_{(\eytab{1}, \eytab{0})}$\\
		\hline
		$\ell_{7}$& (\ytab{13}, \ytab{5}) & $T_7\sim T^{(\eytab{1}, \eytab{5})}_{(\eytab{1}, \eytab{0})}$ & $\ell_{8}$ &  (\ytab{13}, \ytab{6}) & $T_8\sim T^{(\eytab{1}, \eytab{6})}_{(\eytab{1}, \eytab{0})}$\\
		\hline
		$\ell_{9}$& (\ytab{14}, \ytab{2}) & $T_9\sim T^{(\eytab{3}, \eytab{2})}_{(\eytab{1}, \eytab{0})}$ & $\ell_{10}$ & (\ytab{14}, \ytab{3}) & $T_{10}\sim T^{(\eytab{3}, \eytab{3})}_{(\eytab{1}, \eytab{0})}$\\
		\hline
		$\ell_{11}$ & (\ytab{15}, \ytab{2}) & $T_{11}\sim T^{(\eytab{2}, \eytab{2})}_{(\eytab{1}, \eytab{0})}$ & $\ell_{12}$ & (\ytab{15}, \ytab{3}) & $T_{12}\sim T^{(\eytab{2}, \eytab{3})}_{(\eytab{1}, \eytab{0})}$\\
		\hline
		$\ell_{13}$ & (\ytab{16}, \ytab{1}) & $T_{13}\sim T^{(\eytab{6}, \eytab{1})}_{(\eytab{1}, \eytab{0})}$ & $\ell_{14}$ & (\ytab{17}, \ytab{1}) & $T_{14}\sim T^{(\eytab{5}, \eytab{1})}_{(\eytab{1}, \eytab{0})}$\\
		\hline
		$\ell_{15}$& (\ytab{18}, \ytab{1}) & $T_{15}\sim T^{(\eytab{4}, \eytab{1})}_{(\eytab{1}, \eytab{0})}$ & $\ell_{16}$ & (\ytab{19}, \ytab{0}) & $T_{16}\sim T^{(\eytab{11}, \eytab{0})}_{(\eytab{1}, \eytab{0})}$\\
		\hline
		$\ell_{17}$ & (\ytab{20}, \ytab{0}) & $T_{17}\sim T^{(\eytab{10}, \eytab{0})}_{(\eytab{1}, \eytab{0})}$ & $\ell_{18}$ & (\ytab{21}, \ytab{0}) & $T_{18}\sim T^{(\eytab{9}, \eytab{0})}_{(\eytab{1}, \eytab{0})}$\\
		\hline
		$\ell_{19}$ & (\ytab{22}, \ytab{0}) & $T_{19}\sim T^{(\eytab{8}, \eytab{0})}_{(\eytab{1}, \eytab{0})}$ & $\ell_{20}$ & (\ytab{23}, \ytab{0}) & $T_{20}\sim T^{(\eytab{7}, \eytab{0})}_{(\eytab{1}, \eytab{0})}$\\
  \hline
  \caption{Irreps for massless baryons and pentaquarks with $b=1$ in QCD-like theories with $N_c=3$. In a parity-invariant spectrum, each irrep has a parity-conjugate partner with equal and opposite index.}
	\label{var_pentaq}
\end{longtable}
\endgroup

\subsection{Theories with $N_f\geq 6$: $N_f$-independence}

All the 25 tensors of Table~\ref{var_pentaq} are well defined for $N_f\geq 6$. Furthermore, they are all class~A and, as such, inequivalent.
Since AMC$[N_f]\cup$PMC$[N_f]$ has rank 21 (see Table~\ref{tab:Nc_5}), it admits a family of real solutions with 4 free parameters.

The anomaly coefficients can be computed following Section~\ref{sec:AMCPMC} and Table~\ref{list} of Appendix~\ref{app:table}; they encode the contribution from each irrep and its parity conjugate (see footnote~\ref{fot:parityconj}).
The $[SU(N_f)_{L}]^3$ AMC$[N_f]$ equation reads
\scriptsize
\begin{align}
\label{eq:am_nf>6}
\frac{(N_f+3) (N_f+6)}{2}  \ell_{a} +\frac{(N_f-6) (N_f-3)}{2}   \ell_{b} + \left(N_f^2-9\right) \ell_{c} +\frac{(N_f+7) N_f}{2}   \ell_{d}  +\frac{(N_f-7) N_f }{2}  \ell_{e}  \nn\\
-\frac{(N_f+3) (N_f+5) (5 N_f+26) N_f}{24} \ell_{1}+\frac{(N_f [17-N_f (5 N_f+22)]+130) N_f}{8}   \ell_{2} \nn\\
-\frac{5 \left(N_f^2-13\right) N_f^2}{12}   \ell_{3} +\frac{(N_f ((22-5N_f) N_f+17)-130) N_f}{8}   \ell_{4} -\frac{(N_f-5) (N_f-3) (5 N_f-26) N_f}{24}   \ell_{5} \nn\\
-\frac{(N_f+3) (N_f+6) \left(N_f^2-1\right)}{2}   \ell_{6} - \left(N_f^2-9\right) \left(N_f^2-1\right) \ell_{7}  -\frac{(N_f-6) (N_f-3)\left(N_f^2-1\right)}{2}   \ell_{8}  \nn\\
-\frac{(N_f+1) (N_f(N_f+11)-14) N_f}{4}   \ell_{9} -\frac{(N_f+2) ((N_f-4) N_f+7)N_f }{4}  \ell_{10} -\frac{(N_f-2) (N_f (N_f+4)+7) N_f}{4}   \ell_{11}  \nn\\
-\frac{(N_f-1) ((N_f-11) N_f-14) N_f}{4}   \ell_{12} +\frac{(N_f-3) ((N_f-18) N_f-7) N_f}{6}   \ell_{13} +\frac{\left(N_f^2-28\right) N_f^2}{3} \ell_{14}   \nn\\
+\frac{(N_f+3) (N_f (N_f+18)-7) N_f }{6}  \ell_{15} +\frac{(N_f-4) (N_f-3) ((N_f-11) N_f-6)}{8}   \ell_{16} +\frac{3 (N_f-6) (N_f-3) (N_f+3)N_f  }{8}  \ell_{17}  \nn\\
+\frac{\left(N_f^4-25 N_f^2+36\right) }{4} \ell_{18} +\frac{3 (N_f-3) (N_f+3) (N_f+6) N_f}{8}   \ell_{19} +\frac{(N_f+3) (N_f+4)(N_f (N_f+11)-6)}{8}   \ell_{20}  &=3\ ,
\end{align}
\normalsize
while the $[SU(N_f)_{L}]^2 U(1)_B$ AMC$[N_f]$ equation is
\scriptsize
\begin{align}
\label{eq:am_nf>6_2}
\frac{(N_f+2) (N_f+3)}{2}   \ell_{a} +\frac{(N_f-3) (N_f-2)}{2}   \ell_{b} + \left(N_f^2-3\right) \ell_{c} +\frac{(N_f+3) N_f}{2}  \ell_{d} +\frac{(N_f-3) N_f}{2} \ell_{e} \nn\\
-\frac{(N_f+2) (N_f+3) (N_f+5) N_f }{8}  \ell_{1} - \frac{(N_f+2) (N_f+3) (3 N_f-5) N_f}{8}   \ell_{2} \nn\\
-\frac{\left(N_f^2-5\right) N_f^2}{4}   \ell_{3} -\frac{(N_f-3) (N_f-2) (3 N_f+5) N_f}{8}   \ell_{4} -\frac{(N_f-5) (N_f-3) (N_f-2) N_f}{8}   \ell_{5} \nn\\
-\frac{(N_f+1) (N_f+2) (N_f (N_f+6)-9)}{6}   \ell_{6} +\frac{\left(-N_f^4+10 N_f^2-9\right)}{3}   \ell_{7} - \frac{(N_f-2) (N_f-1) ((N_f-6) N_f-9)}{6}  \ell_{8} \nn\\
+\frac{(N_f-3) (N_f-2) (N_f+1) N_f}{4}   \ell_{9} +\frac{(N_f-2) \left(N_f^2+3\right) N_f}{4}   \ell_{10}+\frac{(N_f+2) \left(N_f^2+3\right) N_f}{4}  \ell_{11} \nn\\
+\frac{(N_f-1) (N_f+2) (N_f+3) N_f}{4}   \ell_{12} +\frac{(N_f ((N_f-5) N_f+5)+3) N_f}{2}  \ell_{13}+ \left(N_f^2-4\right) N_f^2 \ell_{14} \nn\\
+\frac{(N_f+3) (N_f (N_f+2)-1) N_f}{2} \ell_{15} +\frac{(N_f-4) (N_f-3) (N_f-2) (5 N_f+3)}{24}  \ell_{16}+\frac{(N_f-3) (N_f+2) (5 N_f-9) N_f}{8}  \ell_{17} \nn\\
+\frac{\left(5 N_f^4-29 N_f^2+36\right)}{12}  \ell_{18} +\frac{(N_f-2) (N_f+3) (5 N_f+9) N_f}{8}  \ell_{19}+\frac{(N_f+2) (N_f+3) (N_f+4) (5 N_f-3)}{24}  \ell_{20}&=1\ .
\end{align}
\normalsize

By setting $N_f= 0$ in the $[SU(N_f)_{L}]^3$ AMC equation we obtain:
\begin{equation}
\label{eq:farrar}
  9 \ell_{a}+9 \ell_{b}-9 \ell_{c}+9 \ell_{6}-9 \ell_{7}+9 \ell_{8}-9 \ell_{16}+9 \ell_{18}-9 \ell_{20}=3 \ .
\end{equation}
We see that all the non-vanishing anomaly coefficients are equal to $\pm N_c^2 =\pm 9$, as suggested by Farrar in~\cite{Farrar:1980sn}. However, differently from the claim made in~\cite{Farrar:1980sn}, not all the CYTs associated to the indices appearing in Eq.~(\ref{eq:farrar}) are of `non-mixing', `elbow' type.~\footnote{Here, mixed CYTs are defined to be those which transform non-trivially under both $SU(N_f)_L$ and $SU(N_f)_R$. Elbow YTs are those with at most one column and one row.} More in detail, while $\ell_{a}, \ell_{b}, \ell_{c}, \ell_{20}$ are associated to non-mixed elbow CYTs, the remaining indices are associated to mixed CYTs ($\ell_{6}, \ell_{7}, \ell_{8}$) or non-mixed non-elbow ones ($\ell_{16}, \ell_{18}$), see Table~\ref{var_pentaq}.
One can notice the following:
\begin{itemize}
\item The CYT associated to $\ell_{6}, \ell_{7}, \ell_{8}$ involve the adjoint representation (see Table~\ref{var_pentaq}), which is the only non-trivial irrep of Table~\ref{list} in Appendix~\ref{app:table} whose dimension does not vanish for $N_f\to 0$. This is an exception to the argument of~\cite{Farrar:1980sn}. Since however the dimension of the adjoint equals~$-1$ in the limit $N_f\to 0$, we still obtain an anomaly coefficient equal to $\pm N_c^2$.
\item Although the CYT associated to $\ell_{16}$ is not an elbow one, its conjugate is.  In this sense, the CYT associated to $\ell_{16}$ is still related to a non-mixing elbow-shape CYT, and this explains why its anomaly constant does not vanish in the limit $N_f\to 0$.
\item The index $\ell_{18}$ is related to an elbow-shape representation in a less trivial manner. Let us denote the CYTs associated to $\ell_{17}, \ell_{18}, \ell_{19}$ and $\ell_{c}$ respectively by $r_{17}, r_{18}, r_{19}$ and $r_c$. These CYTs are related by the following identity
  \begin{equation}
  \label{eq:product}
\farrar\ ,
\end{equation}
where the column marked in gray has $N_f-1$ boxes.
Since (as pointed out in~\cite{Farrar:1980sn}) the anomaly of the left-hand side vanishes in the limit $N_f\to 0$, one obtains that
\begin{equation}
\lim_{N_f\to 0} \, A_3(r_{18})= -\lim_{N_f\to 0} \, \left[A_3(r_{17})+A_3(r_{19})+A_3(r_{c})\right]\ ,
\end{equation}
where we used the fact that $r_c$ is equivalent to the last factor on the right-hand side of Eq.~(\ref{eq:product}), since they differ by a singlet column with $N_f$ boxes. Now, the anomaly coefficients of $r_{17}$ and $r_{19}$ vanish in the $N_f\to 0$ limit, but that of $r_{c}$ does not (see Table~\ref{list}).
Therefore, the anomaly of $r_{18}$ is equal and opposite to that of $r_c$, which has an elbow shape.
This is another exception to the argument of~\cite{Farrar:1980sn}.
\end{itemize}
In conclusion, there exist exceptions to the argument of~\cite{Farrar:1980sn} that arise from the presence of adjoint representations, conjugates representations, and singlet columns of $N_f$ boxes in YTs. The conclusions drawn by Farrar, however, are correct: all the non-vanishing anomaly coefficients are equal to $\pm N_c^2$ in the limit $N_f\to 0$.

Let us now turn to PMC. Differently from the case of a purely baryonic massless spectrum, the PMC$[N_f, i]$ with $1< i\leq N_f-2$, are in general linearly independent from each other and from PMC$[N_f, 1]$, and must be therefore included. We find that:
\begin{enumerate}
\item All the PMC$[N_f,i]$ equations are identical to PMC$[6,1]$ when $1\leq i \leq N_f-5$; these are the sets of equations marked in red in Fig.~\ref{fig:pmc}.
\item PMC$[N_f, N_f-4]$, PMC$[N_f, N_f-3]$, and PMC$[N_f, N_f-2]$ are different sets of equations, but they are identical to PMC$[6, 2]$, PMC$[6, 3]$, and PMC$[6, 4]$, respectively.
\end{enumerate}
In the following, we write explicitly all the relevant sets of PMC.

When one flavor is given a mass, we obtain the following PMC$[N_f,1]$ equations:
\begin{itemize}
\small
\item \textbf{PMC$[N_f,1]$ with $H_1>0$}
\begin{equation}
\begin{split}
0&=\ell\left(T^{(\eytab{2},\eytab{0})}_{(\eytab{0},\eytab{0})} \,; 1\right) =\ell_{a}+\ell_{c}+\ell_{d} -\ell_{1}-\ell_{2}-\ell_{3}-\ell_{6}-\ell_{7}+\ell_{14}+\ell_{15}+\ell_{18}+\ell_{19}+\ell_{20}\ ,\\
0&=\ell\left(T^{(\eytab{3},\eytab{0})}_{(\eytab{0},\eytab{0})} \,; 1\right) =\ell_{b}+\ell_{c}+\ell_{e}-\ell_{2}-\ell_{4}-\ell_{7}-\ell_{8}+\ell_{9}-\ell_{12}+\ell_{13}+\ell_{14}+\ell_{17}+\ell_{19} \ ,\\
0&=\ell\left(T^{(\eytab{1},\eytab{0})}_{(\eytab{0},\eytab{0})}\,; 2\right)=\ell_{a}+\ell_{c}+\ell_{e}-\ell_{1}-\ell_{2}-\ell_{7}+\ell_{9}-\ell_{12}+\ell_{14}+\ell_{19}+\ell_{20} \ ,\\
0&= \ell\left(T^{(\eytab{4},\eytab{0})}_{(\eytab{1},\eytab{0})}\,; 1 \right)=\ell_{15}+\ell_{19}+\ell_{20} \ , \quad\quad
0= \ell\left(T^{(\eytab{4},\eytab{0})}_{(\eytab{0},\eytab{1})}\,; 1 \right)=-\ell_{1}-\ell_{2}-\ell_{6} \ ,\\
0&= \ell\left(T^{(\eytab{6},\eytab{0})}_{(\eytab{1},\eytab{0})}\,; 1 \right)= \ell_{13}+\ell_{16}+\ell_{17} \ , \quad\quad
0= \ell\left(T^{(\eytab{6},\eytab{0})}_{(\eytab{0},\eytab{1})}\,; 1\right)= -\ell_{4}-\ell_{5}-\ell_{8} \ ,\\
0&= \ell\left(T^{(\eytab{5},\eytab{0})}_{(\eytab{1},\eytab{0})}\,; 1\right)=\ell_{14}+\ell_{17}+\ell_{18}+\ell_{19} \ ,\quad\quad
0=\ell\left(T^{(\eytab{5},\eytab{0})}_{(\eytab{0},\eytab{1})}\,; 1 \right)= -\ell_{2}-\ell_{3}-\ell_{4}-\ell_{7} \ ,\\
0&=\ell\left(T^{(\eytab{2},\eytab{1})}_{(\eytab{1},\eytab{0})}\,; 1 \right)= \ell_{11}+\ell_{12}+\ell_{14}+\ell_{15} \ ,\quad\quad
0=\ell\left(T^{(\eytab{2},\eytab{1})}_{(\eytab{0},\eytab{1})}\,; 1 \right)= -\ell_{6}-\ell_{7}-\ell_{9}-\ell_{11} \ , \\
0&= \ell\left(T^{(\eytab{3},\eytab{1})}_{(\eytab{1},\eytab{0})}\,; 1 \right)= \ell_{9}+\ell_{10}+\ell_{13}+\ell_{14}\ , \quad\quad
0= \ell\left(T^{(\eytab{3},\eytab{1})}_{(\eytab{0},\eytab{1})}\,; 1 \right)= -\ell_{7}-\ell_{8}-\ell_{10}-\ell_{12} \ ,\\
0&= \ell\left(T^{(\eytab{2},\eytab{0})}_{(\eytab{1},\eytab{0})}\,; 2 \right)=\ell_{11}+\ell_{14}+\ell_{15}+\ell_{18}+\ell_{19}+\ell_{20} \ , \\
0&= \ell\left(T^{(\eytab{2},\eytab{0})}_{(\eytab{0},\eytab{1})}\,; 2 \right)=-\ell_{1}-\ell_{2}-\ell_{3}-\ell_{6}-\ell_{7}-\ell_{11} \ ,\\
0&=\ell\left(T^{(\eytab{3},\eytab{0})}_{(\eytab{1},\eytab{0})}\,; 2\right)= \ell_{9}+\ell_{13}+\ell_{14}+\ell_{17}+\ell_{19} \ ,\\
0&= \ell\left(T^{(\eytab{3},\eytab{0})}_{(\eytab{0},\eytab{1})}\,; 2\right)= -\ell_{2}-\ell_{4}-\ell_{7}-\ell_{8}-\ell_{12} \ ,\\
0&= \ell\left(T^{(\eytab{1},\eytab{1})}_{(\eytab{1},\eytab{0})}\,; 2\right)=\ell_{6}+\ell_{7}+\ell_{9}+\ell_{10}+\ell_{11}+\ell_{12}+\ell_{14}+\ell_{15} \ ,\\
0&=\ell\left(T^{(\eytab{1},\eytab{0})}_{(\eytab{1},\eytab{0})}\,; 3\right)= \ell_{6}+\ell_{9}+\ell_{11}+\ell_{14}+\ell_{15}+\ell_{19}+\ell_{20} \ , \\
0&= \ell\left(T^{(\eytab{1},\eytab{0})}_{(\eytab{0},\eytab{1})}\,; 3\right)=-\ell_{1}-\ell_{2}-\ell_{6}-\ell_{7}-\ell_{11}-\ell_{12}-\ell_{15} \ ,\\
0&= \ell\left(T^{(\eytab{0},\eytab{0})}_{(\eytab{1},\eytab{0})}\,; 4\right)=\ell_{1}+\ell_{6}+\ell_{11}+\ell_{15}+\ell_{20} \ .\\
\end{split}
\label{eq:pmc61_H>0}
\end{equation}
\item \textbf{PMC$[N_f,1]$ with $H_1<0$}
\begin{equation}
\begin{split}
0&= \ell\left(T^{(\eytab{0},\eytab{7})}_{(\eytab{0},\eytab{0})}\,; -1\right)=\ell_{1}-\ell_{20} \ , \quad\quad
0= \ell\left(T^{(\eytab{0},\eytab{8})}_{(\eytab{0},\eytab{0})}\,; -1\right)=\ell_{2}-\ell_{19} \ ,\\
0&= \ell\left(T^{(\eytab{0},\eytab{9})}_{(\eytab{0},\eytab{0})}\,; -1 \right)=\ell_{3}-\ell_{18} \ ,\quad\quad
0=\ell\left(T^{(\eytab{0},\eytab{10})}_{(\eytab{0},\eytab{0})}\,; -1\right)=\ell_{4}-\ell_{17} \ ,\\
0&=\ell\left(T^{(\eytab{0},\eytab{11})}_{(\eytab{0},\eytab{0})}\,; -1\right)=\ell_{5}-\ell_{16} \ , \quad\quad
0= \ell\left(T^{(\eytab{1},\eytab{4})}_{(\eytab{0},\eytab{0})}\,; -1\right)=\ell_{6}-\ell_{15} \ , \\ 
0&= \ell\left(T^{(\eytab{1},\eytab{5})}_{(\eytab{0},\eytab{0})}\,; -1\right)=\ell_{7}-\ell_{14} \ , \quad\quad
0= \ell\left(T^{(\eytab{1},\eytab{6})}_{(\eytab{0},\eytab{0})}\,; -1\right)=\ell_{8}-\ell_{13} \ , \\
0&=\ell\left(T^{(\eytab{3},\eytab{2})}_{(\eytab{0},\eytab{0})}\,; -1\right)=\ell_{9}-\ell_{12} \ .
\end{split}
\label{eq:pmc61_H<0}
\end{equation}
\normalsize
\end{itemize}
There is one equation for each irrep of ${\cal G}[N_f,1]$. For convenience, we have denoted irreps of ${\cal G}[N_f,1]$ by the tensor transforming as their CYT and by their $U(1)_{H_1}$ charge $H_1$ (we omit the baryon number since it is always equal to 1). Using tensors instead of CYTs is convenient because the dual YT from antiquarks has a number of boxes which depends on $N_f$.

We notice that:
\begin{itemize}
\item The form of PMC$[N_f,1]$, i.e. Eqs.~(\ref{eq:pmc61_H>0}) and~(\ref{eq:pmc61_H<0}), is the same for any $N_f\geq 6$, despite CYTs do have different shapes for different values of $N_f$ (the number of boxes in the antiquark column varies with $N_f$).
This follows because, for $N_f\geq 6$, there are no equivalent tensors of ${\cal G}[N_f,1]$ and the factor $\kappa(r \to r_1)$ appearing in the decomposition of Eq.~(\ref{eq:PMC}) does not depend on $N_f$.
\item The anomaly coefficients in AMC$[N_f]$ depend explicitly on the value of $N_f$. On the other hand, once evaluated on the solutions of PMC$[N_f,1]$, Eqs.~(\ref{eq:am_nf>6}) and~(\ref{eq:am_nf>6_2})  can be rewritten as 
\begin{align}
\label{Nfeq_Nc=3}
  9\left(-4 \ell_{1}+2 \ell_6+2 \ell_9 + 3 \ell_a+\ell_d \right)&=3\ ,\\
  \label{Nfeq_Nc=3_2}
  3\left(-4 \ell_{1}+2 \ell_6+2 \ell_9 + 3 \ell_a+\ell_d \right)&=1\ ,
\end{align}
Therefore, on the solutions of PMC$[N_f,1]$, the two AMC equations become identical and independent of $N_f$. This confirms the existence of the $N_f$-equation for $N_f\geq 6$, as argued by Cohen and Frishman in~\cite{Cohen:1981iz}.
\end{itemize}

By giving mass to a second flavor, one can derive PMC$[N_f,2]$. These are constraints on the indices of the irreps $r_1$ of ${\cal G}[N_f,1]$ with charge $H_1=0$. Let us denote by $\ell^{(N_f-1)}_\alpha$ these indices (with $\alpha = a,\cdots, e, 1, \cdots, 20$).~\footnote{\label{fot:upperlable}We introduce the upper label $(N_f-1)$ to signify that $\ell^{(N_f-1)}_\alpha$ is associated to a bound state in a theory with $N_f-1$ massless flavors.} They are computed in terms of the indices of the irreps $r$ of ${\cal G}[N_f]$ by means of Eq.~(\ref{eq:downliftedsol}). In this way we find, for $N_f>6$:
\begin{equation}
\label{eq:ident_0}
\begin{split}
\ell^{(N_f-1)}_a&= \ell_a-\ell_{1}-\ell_{2}-\ell_{6}+\ell_{15}+\ell_{19}+\ell_{20}\ ,\\
\ell^{(N_f-1)}_b&= \ell_b-\ell_{4}-\ell_{5}-\ell_{8}+\ell_{13}+\ell_{16}+\ell_{17} \ ,\\
\ell^{(N_f-1)}_c&= \ell_c-\ell_{2}-\ell_{3}-\ell_{4}-\ell_{7}+\ell_{14}+\ell_{17}+\ell_{18}+\ell_{19} \ ,\\
\ell^{(N_f-1)}_d&= \ell_d-\ell_{6}-\ell_{7}-\ell_{9}+\ell_{12}+\ell_{14}+\ell_{15} \ ,\\
\ell^{(N_f-1)}_e&= \ell_e-\ell_{7}-\ell_{8}+\ell_{9}-\ell_{12}+\ell_{13}+\ell_{14} \ ,\\
\ell^{(N_f-1)}_\alpha&=\ell_\alpha ,\quad \text{for}\  \alpha=1,\cdots, 20\ .
\end{split}
\end{equation}
These expressions can be simplified by making use of PMC$[N_f,1]$, Eqs.~(\ref{eq:pmc61_H>0}) and~(\ref{eq:pmc61_H<0}). In this way we obtain
\begin{equation}
\label{eq:ident_1}
\ell^{(N_f-1)}_\alpha=\ell_\alpha\quad  \text{for }  \alpha=a,\cdots,e, 1, \cdots, 20\ .
\end{equation}
This result was actually expected based on Eq.~(\ref{eq:uplift}). Indeed, each irrep $r_1$ of ${\cal G}[N_f,1]$ with charge $H_1=0$ is in one-to-one correspondence with an irrep $r'$ of ${\cal G}[N_f-1]$. For $N_f>6$, all such irreps $r'$ are class A, hence Eq.~(\ref{eq:uplift}) holds true. Therefore, Eq.~(\ref{eq:ident_1}) is an explicit verification of the validity of Eq.~(\ref{eq:uplift}). At this point it is easy to see that
\begin{itemize}
\item PMC$[N_f,2]$ have the same form as PMC$[N_f-1,1]$; this is a direct consequence of the correspondence between irreps $r_1$ of ${\cal G}[N_f,1]$ with charge $H_1=0$ and irreps $r'$ of ${\cal G}[N_f-1]$.
\item PMC$[N_f,2]$ have the same form as PMC$[N_f,1]$. Indeed, irreps $r_1$ of ${\cal G}[N_f,1]$ with charge $H_1=0$ are also in one-to-one correspondence with irreps $r$ of ${\cal G}[N_f]$. The correspondence is between irreps $r$ and $r_1$  interpolated by the same tensor, and is one-to-one thanks to the absence of equivalent tensors: all tensors are class A.
It follows that PMC$[N_f,2]$ coincide with PMC$[N_f,1]$ as a consequence of Eq.~(\ref{eq:ident_1}):
\begin{equation}
\text{PMC}[N_f,2]=\text{PMC}[N_f,1]\lvert_{\ell_\alpha \to \ell^{(N_f-1)}_\alpha}\ .
\end{equation}
\end{itemize}

By repeating the same steps, one can show that for, $N_f > 7$, PMC$[N_f,3]$ coincide with PMC$[N_f,2]$, hence with  PMC$[N_f,1]$, and also with PMC$[N_f-1,2]$, hence with PMC$[N_f-2,1]$. In general, one can show that all the PMC$[N_f,i]$ are identical for any $N_f\geq 6$ and $1\leq i \leq N_f-5$, and that they coincide with PMC$[6,1]$. These sets of PMC are marked in red in Fig.~\ref{fig:pmc}.

Let us discuss at this point PMC$[N_f, N_f-4]$. These are obtained by giving (unequal) masses to $N_f-4$ flavors, and constrain the indices $\ell^{(5)}_\alpha$ of irreps of ${\cal G}[N_f, N_f-5]$ with $H_{N_f-5} = 0$. Tensors interpolating such irreps are not class A, and there is no one-to-one correspondence between irreps of ${\cal G}[N_f, N_f-5]$ and irreps of ${\cal G}[N_f, N_f-6]$.
In particular, the CYT associated to the index $\ell^{(5)}_5$, 
\begin{equation}
\left(\ {\tiny\yng(1,1,1,1)}\ , \ {\tiny\yng(1,1,1,1)}\ \right)\ ,
\end{equation}
coincides with its parity conjugate, i.e. they correspond to the same irrep of $\mathcal{G}[N_f, N_f-5]$. Due to parity invariance of the spectrum, the index $\ell^{(5)}_5$ disappears from PMC$[N_f, N_f-4]$. This implies that PMC$[N_f, N_f-4]$ do not coincide with PMC$[N_f, N_f-5]$.
It is easy to verify that by inspecting the explicit expression of PMC$[N_f, N_f-4]$. For the latter, we find:
%
\begin{itemize}
%
\item \textbf{PMC$[N_f,N_f-4]$ with $H_{N_f-4}>0$}
\scriptsize
 \begin{equation}
\begin{split}
0&=\ell\left(T^{(\eytab{2},\eytab{0})}_{(\eytab{0},\eytab{0})} \,; 1\right) =\ell^{(5)}_{a}+\ell^{(5)}_{c}+\ell^{(5)}_{d} -\ell^{(5)}_{1}-\ell^{(5)}_{2}-\ell^{(5)}_{3}-\ell^{(5)}_{6}-\ell^{(5)}_{7}+\ell^{(5)}_{14}+\ell^{(5)}_{15}+\ell^{(5)}_{18}+\ell^{(5)}_{19}+\ell^{(5)}_{20}\ ,\\
0&=\ell\left(T^{(\eytab{3},\eytab{0})}_{(\eytab{0},\eytab{0})}\,; 1\right) = \ell^{(5)}_{b}+\ell^{(5)}_{c}+\ell^{(5)}_{e}-\ell^{(5)}_{2}-\ell^{(5)}_{4}-\ell^{(5)}_{7} -\ell^{(5)}_{8}+\ell^{(5)}_{9}-\ell^{(5)}_{12}+\ell^{(5)}_{13}+\ell^{(5)}_{14}+\ell^{(5)}_{17}+\ell^{(5)}_{19} \ ,\\
0&=\ell\left(T^{(\eytab{1},\eytab{0})}_{(\eytab{0},\eytab{0})} \,; 2\right) = \ell^{(5)}_{a}+\ell^{(5)}_{c}+\ell^{(5)}_{e}-\ell^{(5)}_{1}-\ell^{(5)}_{2}-\ell^{(5)}_{7} +\ell^{(5)}_{9}-\ell^{(5)}_{12}+\ell^{(5)}_{14}+\ell^{(5)}_{19}+\ell^{(5)}_{20} \ ,\\
0& = \ell\left(T^{(\eytab{4},\eytab{0})}_{(\eytab{1},\eytab{0})}\,; 1\right) =  \ell^{(5)}_{15}+ \ell^{(5)}_{19}+ \ell^{(5)}_{20} \ , \quad\quad
0= \ell\left(T^{(\eytab{4},\eytab{0})}_{(\eytab{0},\eytab{1})} \,; 1 \right) = - \ell^{(5)}_{1}- \ell^{(5)}_{2}- \ell^{(5)}_{6} \ , \\
0&=\ell\left(T^{(\eytab{6},\eytab{0})}_{(\eytab{1},\eytab{0})} \,; 1\right) =  \ell^{(5)}_{13}+ \ell^{(5)}_{16}+ \ell^{(5)}_{17} \ ,\\
0&= \ell\left(T^{(\eytab{5},\eytab{0})}_{(\eytab{1},\eytab{0})}\,; 1\right) =  \ell^{(5)}_{14}+ \ell^{(5)}_{17}+ \ell^{(5)}_{18}+ \ell^{(5)}_{19} \ , \quad\quad
0 = \ell\left(T^{(\eytab{5},\eytab{0})}_{(\eytab{0},\eytab{1})} \,; 1\right) = - \ell^{(5)}_{2}- \ell^{(5)}_{3}- \ell^{(5)}_{4}- \ell^{(5)}_{7} \ ,\\
0 &= \ell\left(T^{(\eytab{2},\eytab{1})}_{(\eytab{1},\eytab{0})} \,; 1\right) =  \ell^{(5)}_{11}+ \ell^{(5)}_{12}+ \ell^{(5)}_{14}+ \ell^{(5)}_{15} \ , \quad\quad
0 = \ell\left(T^{(\eytab{2},\eytab{1})}_{(\eytab{0},\eytab{1})}\,; 1 \right) = - \ell^{(5)}_{6}- \ell^{(5)}_{7}- \ell^{(5)}_{9}- \ell^{(5)}_{11} \ , \\
0 &=  \ell\left(T^{(\eytab{3},\eytab{1})}_{(\eytab{1},\eytab{0})}\,; 1\right) =  \ell^{(5)}_{9}+ \ell^{(5)}_{10}+ \ell^{(5)}_{13}+ \ell^{(5)}_{14}\ , \quad\quad
0 =  \ell\left(T^{(\eytab{3},\eytab{1})}_{(\eytab{0},\eytab{1})} \,; 1 \right) = - \ell^{(5)}_{7}- \ell^{(5)}_{8}- \ell^{(5)}_{10}- \ell^{(5)}_{12} \ ,\\
0 &=  \ell\left(T^{(\eytab{2},\eytab{0})}_{(\eytab{1},\eytab{0})} \,; 2\right) =  \ell^{(5)}_{11}+ \ell^{(5)}_{14}+ \ell^{(5)}_{15}+ \ell^{(5)}_{18}+ \ell^{(5)}_{19}+ \ell^{(5)}_{20} \ , \\
0 &=  \ell\left(T^{(\eytab{2},\eytab{0})}_{(\eytab{0},\eytab{1})} \,; 2 \right) = - \ell^{(5)}_{1}- \ell^{(5)}_{2}- \ell^{(5)}_{3}- \ell^{(5)}_{6}- \ell^{(5)}_{7}- \ell^{(5)}_{11} \ ,\\
0 &= \ell\left(T^{(\eytab{3},\eytab{0})}_{(\eytab{1},\eytab{0})}\,; 2 \right) =  \ell^{(5)}_{9}+ \ell^{(5)}_{13}+ \ell^{(5)}_{14}+ \ell^{(5)}_{17}+ \ell^{(5)}_{19} \ ,\\
0 &=  \ell\left(T^{(\eytab{3},\eytab{0})}_{(\eytab{0},\eytab{1})}\,; 2 \right) = - \ell^{(5)}_{2}- \ell^{(5)}_{4}- \ell^{(5)}_{7}- \ell^{(5)}_{8}- \ell^{(5)}_{12} \ ,\\
0 &=  \ell\left(T^{(\eytab{1},\eytab{1})}_{(\eytab{1},\eytab{0})}\,; 2 \right) =  \ell^{(5)}_{6}+ \ell^{(5)}_{7}+ \ell^{(5)}_{9}+ \ell^{(5)}_{10}+ \ell^{(5)}_{11}+ \ell^{(5)}_{12}+ \ell^{(5)}_{14}+ \ell^{(5)}_{15} \ ,\\
0 &= \ell\left(T^{(\eytab{1},\eytab{0})}_{(\eytab{1},\eytab{0})}\,; 3\right) =  \ell^{(5)}_{6}+ \ell^{(5)}_{9}+ \ell^{(5)}_{11}+ \ell^{(5)}_{14}+ \ell^{(5)}_{15}+ \ell^{(5)}_{19}+ \ell^{(5)}_{20} \ , \\
0 &=  \ell\left(T^{(\eytab{1},\eytab{0})}_{(\eytab{0},\eytab{1})}\,; 3\right) = - \ell^{(5)}_{1}- \ell^{(5)}_{2}- \ell^{(5)}_{6}- \ell^{(5)}_{7}- \ell^{(5)}_{11}- \ell^{(5)}_{12}- \ell^{(5)}_{15} \ ,\\
0 &=  \ell\left(T^{(\eytab{0},\eytab{0})}_{(\eytab{1},\eytab{0})}\,; 4\right) =  \ell^{(5)}_{1}+ \ell^{(5)}_{6}+ \ell^{(5)}_{11}+ \ell^{(5)}_{15}+ \ell^{(5)}_{20} \ ,
\end{split}
\label{eq:pmc62_H>0}
\end{equation}
\normalsize
\item \textbf{PMC$[N_f,N_f-4]$ with $H_{N_f-4}<0$}
  \scriptsize
\begin{equation}
\begin{split}
0&=  \ell\left(T^{(\eytab{0},\eytab{7})}_{(\eytab{0},\eytab{0})}\,; -1\right)= \ell^{(5)}_{1}- \ell^{(5)}_{20} \ ,\quad\quad
0=  \ell\left(T^{(\eytab{0},\eytab{8})}_{(\eytab{0},\eytab{0})}\,;-1 \right)= \ell^{(5)}_{2}- \ell^{(5)}_{19} \ , \\
0&=  \ell\left(T^{(\eytab{0},\eytab{9})}_{(\eytab{0},\eytab{0})}\,; -1\right)= \ell^{(5)}_{3}- \ell^{(5)}_{18} \ ,\quad\quad
0= \ell\left(T^{(\eytab{0},\eytab{10})}_{(\eytab{0},\eytab{0})}\,; -1\right)= \ell^{(5)}_{4}- \ell^{(5)}_{17} \ ,\\
0&=  \ell\left(T^{(\eytab{1},\eytab{4})}_{(\eytab{0},\eytab{0})}\,; -1\right)= \ell^{(5)}_{6}- \ell^{(5)}_{15} \ , \quad\quad
0=  \ell\left(T^{(\eytab{1},\eytab{5})}_{(\eytab{0},\eytab{0})}\,; -1\right)= \ell^{(5)}_{7}- \ell^{(5)}_{14} \ ,\\
0&=  \ell\left(T^{(\eytab{1},\eytab{6})}_{(\eytab{0},\eytab{0})}\,; -1\right)= \ell^{(5)}_{8}- \ell^{(5)}_{13} \ , \quad\quad
0=  \ell\left(T^{(\eytab{3},\eytab{2})}_{(\eytab{0},\eytab{0})}\,; -1\right)= \ell^{(5)}_{9}- \ell^{(5)}_{12} \ .
\end{split}
\label{eq:pmc62_H<0}
\end{equation}
\end{itemize}
\normalsize

PMC$[N_f,N_f-3]$ are obtained by giving (unequal) mass to $N_f-3$ flavors, and constrain the indices $  \ell^{(4)}_\alpha$ of irreps of ${\cal G}[N_f, N_f-4]$ with $H_{N_f-4} = 0$. Such indices can be computed by means of Eq.~(\ref{eq:downliftedsol}). We find:
\begin{equation}
\label{eq:ident_5to4}
\begin{split}
  \ell^{(4)}_a&=  \ell^{(5)}_a- \ell^{(5)}_{1}- \ell^{(5)}_{2}- \ell^{(5)}_{6}+ \ell^{(5)}_{15}+ \ell^{(5)}_{19}+ \ell^{(5)}_{20}\ ,\\
  \ell^{(4)}_b&=  \ell^{(5)}_b- \ell^{(5)}_{4}- \ell^{(5)}_{8}+ \ell^{(5)}_{13}+ \ell^{(5)}_{16}+ \ell^{(5)}_{17} \ ,\\
  \ell^{(4)}_c&=  \ell^{(5)}_c- \ell^{(5)}_{2}- \ell^{(5)}_{3}- \ell^{(5)}_{4}- \ell^{(5)}_{7}+ \ell^{(5)}_{14}+ \ell^{(5)}_{17}+ \ell^{(5)}_{18}+ \ell^{(5)}_{19} \ ,\\
  \ell^{(4)}_d&=  \ell^{(5)}_d- \ell^{(5)}_{6}- \ell^{(5)}_{7}- \ell^{(5)}_{9}+ \ell^{(5)}_{12}+ \ell^{(5)}_{14}+ \ell^{(5)}_{15} \ ,\\
  \ell^{(4)}_e&=  \ell^{(5)}_e- \ell^{(5)}_{7}- \ell^{(5)}_{8}+ \ell^{(5)}_{9}- \ell^{(5)}_{12}+ \ell^{(5)}_{13}+ \ell^{(5)}_{14} \ ,\\
  \ell^{(4)}_4 &=   \ell^{(5)}_4-  \ell^{(5)}_8\ ,\\
  \ell^{(4)}_\alpha&= \ell^{(5)}_\alpha ,\quad \text{for }  \alpha=1,2,3,6,7,9,\cdots ,15, 17,\cdots, 20\ .
\end{split}
\end{equation}
Tensors $T_{b}$,  $T_{5}$ and $T_{16}$ transform as the same irrep of $\mathcal{G}[N_f, N_f-4]$, whose index is denoted by~$  \ell^{(4)}_b$. Similarly, $T_{4}$ and the parity conjugate of $T_{8}$ transform as the same irrep of $\mathcal{G}[N_f, N_f-4]$, whose index is denoted by~$  \ell^{(4)}_4$
We obtain the following expression for PMC$[N_f, N_f-3]$:

\begin{itemize}
\item \textbf{PMC$[N_f,N_f-3]$ with $H_{N_f-3}>0$}
\small
\begin{equation}
\begin{split}
0&= \ell\left(T^{(\eytab{2},\eytab{0})}_{(\eytab{0},\eytab{0})} \,; 1\right) = \ell^{(4)}_{a}+ \ell^{(4)}_{c}+ \ell^{(4)}_{d}- \ell^{(4)}_{1}- \ell^{(4)}_{2}- \ell^{(4)}_{3}- \ell^{(4)}_{6}- \ell^{(4)}_{7}+ \ell^{(4)}_{14}+ \ell^{(4)}_{15}+ \ell^{(4)}_{18}+ \ell^{(4)}_{19}+ \ell^{(4)}_{20} \ ,\\
0&= \ell\left(T^{(\eytab{3},\eytab{0})}_{(\eytab{0},\eytab{0})}\,; 1\right)+ \ell\left(T^{(\eytab{6},\eytab{0})}_{(\eytab{1},\eytab{0})}\,; 1\right)+ \ell\left(T^{(\eytab{0},\eytab{6})}_{(\eytab{1},\eytab{0})}\,; 1\right) \\
&=  \ell^{(4)}_{b}+ \ell^{(4)}_{c}+ \ell^{(4)}_{e} - \ell^{(4)}_{2}- \ell^{(4)}_{7}+ \ell^{(4)}_{9}- \ell^{(4)}_{12}+ \ell^{(4)}_{13}+ \ell^{(4)}_{14}+ \ell^{(4)}_{17}+ \ell^{(4)}_{19} \ ,\\
0&= \ell\left(T^{(\eytab{1},\eytab{0})}_{(\eytab{0},\eytab{0})}\,; 2\right)=  \ell^{(4)}_{a}+ \ell^{(4)}_{c}+ \ell^{(4)}_{e} - \ell^{(4)}_{1}- \ell^{(4)}_{2}- \ell^{(4)}_{7}+ \ell^{(4)}_{9}- \ell^{(4)}_{12}+ \ell^{(4)}_{14}+ \ell^{(4)}_{19}+ \ell^{(4)}_{20}\ ,\\
0&= \ell\left(T^{(\eytab{4},\eytab{0})}_{(\eytab{1},\eytab{0})}\,; 1\right)=  \ell^{(4)}_{15}+ \ell^{(4)}_{19}+ \ell^{(4)}_{20} \ , \quad\quad\quad\quad\quad
0= \ell\left(T^{(\eytab{4},\eytab{0})}_{(\eytab{0},\eytab{1})}\,; 1\right)= - \ell^{(4)}_{1}- \ell^{(4)}_{2}- \ell^{(4)}_{6} \ , \nonumber
\end{split}
\end{equation}
\begin{equation}
\begin{split}
0&= \ell\left(T^{(\eytab{5},\eytab{0})}_{(\eytab{1},\eytab{0})}\,; 1\right)=  \ell^{(4)}_{14}+ \ell^{(4)}_{17}+ \ell^{(4)}_{18}+ \ell^{(4)}_{19} \ ,\\
0&= \ell\left(T^{(\eytab{5},\eytab{0})}_{(\eytab{0},\eytab{1})}\,; 1\right)+ \ell\left(T^{(\eytab{1},\eytab{3})}_{(\eytab{1},\eytab{0})}\,; 1\right)=- \ell^{(4)}_{2}- \ell^{(4)}_{3}- \ell^{(4)}_{4}+  \ell^{(4)}_{10}+ \ell^{(4)}_{12} \ ,\\
0&= \ell\left(T^{(\eytab{2},\eytab{1})}_{(\eytab{1},\eytab{0})}\,; 1\right)=  \ell^{(4)}_{11}+ \ell^{(4)}_{12}+ \ell^{(4)}_{14}+ \ell^{(4)}_{15} \ , \quad\quad\quad
0= \ell\left(T^{(\eytab{2},\eytab{1})}_{(\eytab{0},\eytab{1})}\,; 1\right)=- \ell^{(4)}_{6}- \ell^{(4)}_{7}- \ell^{(4)}_{9}- \ell^{(4)}_{11} \ ,\\
0&= \ell\left(T^{(\eytab{3},\eytab{1})}_{(\eytab{1},\eytab{0})}\,;1 \right)=  \ell^{(4)}_{9}+ \ell^{(4)}_{10}+ \ell^{(4)}_{13}+ \ell^{(4)}_{14} \ , \\
0&= \ell\left(T^{(\eytab{2},\eytab{0})}_{(\eytab{1},\eytab{0})}\,;2 \right)=  \ell^{(4)}_{11}+ \ell^{(4)}_{14}+ \ell^{(4)}_{15}+ \ell^{(4)}_{18}+ \ell^{(4)}_{19}+ \ell^{(4)}_{20} \ ,\\
0&= \ell\left(T^{(\eytab{2},\eytab{0})}_{(\eytab{0},\eytab{1})}\,; 2 \right)= - \ell^{(4)}_{1}- \ell^{(4)}_{2}- \ell^{(4)}_{3}- \ell^{(4)}_{6}- \ell^{(4)}_{7}- \ell^{(4)}_{11} \ , \\
0&= \ell\left(T^{(\eytab{3},\eytab{0})}_{(\eytab{1},\eytab{0})}\,; 2\right)=  \ell^{(4)}_{9}+ \ell^{(4)}_{13}+ \ell^{(4)}_{14}+ \ell^{(4)}_{17}+ \ell^{(4)}_{19} \ ,\\
0&= \ell\left(T^{(\eytab{1},\eytab{1})}_{(\eytab{1},\eytab{0})}\,; 2\right)=  \ell^{(4)}_{6}+ \ell^{(4)}_{7}+ \ell^{(4)}_{9}+ \ell^{(4)}_{10}+ \ell^{(4)}_{11}+ \ell^{(4)}_{12}+ \ell^{(4)}_{14}+ \ell^{(4)}_{15} \ ,\\
0&= \ell\left(T^{(\eytab{1},\eytab{0})}_{(\eytab{1},\eytab{0})}\,; 3\right)=  \ell^{(4)}_{6}+ \ell^{(4)}_{9}+ \ell^{(4)}_{11}+ \ell^{(4)}_{14}+ \ell^{(4)}_{15}+ \ell^{(4)}_{19}+ \ell^{(4)}_{20} \ ,\\
0&= \ell\left(T^{(\eytab{1},\eytab{0})}_{(\eytab{0},\eytab{1})}\,; 3\right)= - \ell^{(4)}_{1}- \ell^{(4)}_{2}- \ell^{(4)}_{6}- \ell^{(4)}_{7}- \ell^{(4)}_{11}- \ell^{(4)}_{12}- \ell^{(4)}_{15} \ ,\\
0&= \ell\left(T^{(\eytab{0},\eytab{0})}_{(\eytab{1},\eytab{0})}\,; 4\right)=  \ell^{(4)}_{1}+ \ell^{(4)}_{6}+ \ell^{(4)}_{11}+ \ell^{(4)}_{15}+ \ell^{(4)}_{20} \ .
\end{split}
\label{eq:pmc63_H>0}
\end{equation}
\normalsize
\item \textbf{PMC$[N_f,N_f-3]$ with $H_{N_f-3}<0$}
\small
  \begin{equation}
\begin{split}
0&= \ell\left(T^{(\eytab{0},\eytab{7})}_{(\eytab{0},\eytab{0})}\,; -1\right)=  \ell^{(4)}_{1}- \ell^{(4)}_{20} \ , \quad\quad\quad\quad\quad
0= \ell\left(T^{(\eytab{0},\eytab{8})}_{(\eytab{0},\eytab{0})}\,; -1\right)=  \ell^{(4)}_{2}- \ell^{(4)}_{19} \ , \\
 0&= \ell\left(T^{(\eytab{0},\eytab{9})}_{(\eytab{0},\eytab{0})}\,; -1\right)=  \ell^{(4)}_{3}- \ell^{(4)}_{18} \ ,  \quad\quad\quad\quad\quad
 0= \ell\left(T^{(\eytab{1},\eytab{4})}_{(\eytab{0},\eytab{0})}\,; -1\right)= \ell^{(4)}_{6}- \ell^{(4)}_{15} \ ,\\
 0&= \ell\left(T^{(\eytab{0},\eytab{10})}_{(\eytab{0},\eytab{0})}\,; -1\right)+ \ell\left(T^{(\eytab{6},\eytab{1})}_{(\eytab{0},\eytab{0})}\,; -1\right)=  \ell^{(4)}_{4}+ \ell^{(4)}_{13}- \ell^{(4)}_{17} \ ,\\
0&= \ell\left(T^{(\eytab{1},\eytab{5})}_{(\eytab{0},\eytab{0})}\,; -1\right)= \ell^{(4)}_{7}- \ell^{(4)}_{14} \ ,  \quad\quad\quad\quad\quad
 0= \ell\left(T^{(\eytab{3},\eytab{2})}_{(\eytab{0},\eytab{0})}\,; -1\right)= \ell^{(4)}_{9}- \ell^{(4)}_{12} \ .
 \end{split}
 \label{eq:pmc63_H<0}
 \end{equation}
 \normalsize
\end{itemize}

Finally, PMC$[N_f,N_f-2]$ are obtained by giving  mass to $N_f-2$ flavors, and constrain the indices $  \ell^{(3)}_\alpha$ of irreps of ${\cal G}[N_f, N_f-3]$ with $H_{N_f-3} = 0$. Computing such indices by means of Eq.~(\ref{eq:downliftedsol}), we find:
\small
\begin{equation}
\label{eq:ident_4to3}
\begin{split}
  \ell^{(3)}_a &=  \ell^{(4)}_a- \ell^{(4)}_{1}- \ell^{(4)}_{2}- \ell^{(4)}_{6}+ \ell^{(4)}_{15}+ \ell^{(4)}_{19}+ \ell^{(4)}_{20}\ ,\\
  \ell^{(3)}_b &= 0 \ , \\
  \ell^{(3)}_c&=  \ell^{(4)}_c- \ell^{(4)}_{2}- \ell^{(4)}_{3}- \ell^{(4)}_{4}- \ell^{(4)}_{7}+ \ell^{(4)}_{14}+ \ell^{(4)}_{17}+ \ell^{(4)}_{18}+ \ell^{(4)}_{19} \ ,\\
  \ell^{(3)}_d &=  \ell^{(4)}_d- \ell^{(4)}_{6}- \ell^{(4)}_{7}- \ell^{(4)}_{9}+ \ell^{(4)}_{12}+ \ell^{(4)}_{14}+ \ell^{(4)}_{15} \ ,\\
  \ell^{(3)}_e&=  \ell^{(4)}_e+ \ell^{(4)}_{4}- \ell^{(4)}_{7}+ \ell^{(4)}_{9}- \ell^{(4)}_{12}+ \ell^{(4)}_{13}+ \ell^{(4)}_{14} \ ,\\
  \ell^{(3)}_{2}&=   \ell^{(4)}_{2}-  \ell^{(4)}_{12}\ ,\\
  \ell^{(3)}_{3} &=   \ell^{(4)}_{3}-  \ell^{(4)}_{10}\ ,\\
  \ell^{(3)}_7 &= 0 \ , \\
  \ell^{(3)}_\alpha&= \ell^{(4)}_\alpha ,\quad \text{for}\  \alpha=1,6,9,11,14,15,18,19,20\ .\\
\end{split}
\end{equation}
\normalsize
We took into account that, as tensors of $\mathcal{G}[N_f, N_f-3]$: $T_{5}, T_{16}$ are not well defined; $T_b$ and $T_7$ transform as the same irreps as their parity conjugate, their indices $  \ell^{(3)}_{b}$, $  \ell^{(3)}_{7}$ thus vanish identically and disappear from PMC; $T_{c}$,  $T_{8}$ and $T_{17}$ are equivalent, the index of the corresponding irrep is denoted by $  \ell^{(3)}_{c}$; $T_{e}$,  $T_{4}$ and $T_{13}$ are equivalent, the index of the corresponding irrep is denoted by $  \ell^{(3)}_{e}$; $T_{2}$ and the parity conjugate of $T_{12}$ are equivalent, the index of the corresponding irrep is denoted by $  \ell^{(3)}_{2}$; $T_{3}$ and the parity conjugate of $T_{10}$ are equivalent, the index of the corresponding irrep is denoted by $  \ell^{(3)}_{3}$.

The PMC$[N_f, N_f-2]$ equations are: 
\begin{itemize}
%
\item \textbf{PMC$[N_f,N_f-2]$ with $H_{N_f-2}>0$}
\begin{equation}
\begin{split}
0&= \ell\left(T^{(\eytab{1},\eytab{0})}_{(\eytab{0},\eytab{0})}\,; 2\right)+ \ell\left(T^{(\eytab{3},\eytab{0})}_{(\eytab{1},\eytab{0})}\,;2\right)+ \ell\left(T^{(\eytab{0},\eytab{3})}_{(\eytab{1},\eytab{0})}\,;2\right) \\
&= \ell^{(3)}_{a}+ \ell^{(3)}_{c}+ \ell^{(3)}_{e}- \ell^{(3)}_{1}+ \ell^{(3)}_{9}+ \ell^{(3)}_{14}+ \ell^{(3)}_{19}+ \ell^{(3)}_{20}\ ,\\
0&= \ell\left(T^{(\eytab{2},\eytab{0})}_{(\eytab{0},\eytab{0})}\,; 1\right)+ \ell\left(T^{(\eytab{5},\eytab{0})}_{(\eytab{1},\eytab{0})}\,; 1\right)+ \ell\left(T^{(\eytab{1},\eytab{3})}_{(\eytab{1},\eytab{0})}\,; 1\right)\\
&= \ell^{(3)}_{a}+ \ell^{(3)}_{c}+ \ell^{(3)}_{d}- \ell^{(3)}_{1}- \ell^{(3)}_{2}- \ell^{(3)}_{3}- \ell^{(3)}_{6}+ \ell^{(3)}_{14}+ \ell^{(3)}_{15}+ \ell^{(3)}_{18}+ \ell^{(3)}_{19}+ \ell^{(3)}_{20} \ ,\\
0&= \ell\left(T^{(\eytab{4},\eytab{0})}_{(\eytab{1},\eytab{0})}\,; 1\right)=  \ell^{(3)}_{15}+ \ell^{(3)}_{19}+ \ell^{(3)}_{20}\ , \\
0&=  \ell\left(T^{(\eytab{4},\eytab{0})}_{(\eytab{0},\eytab{1})}\,; 1\right)+ \ell\left(T^{(\eytab{2},\eytab{1})}_{(\eytab{1},\eytab{0})}\,; 1\right)= - \ell^{(3)}_{1}- \ell^{(3)}_{2}- \ell^{(3)}_{6}+ \ell^{(3)}_{11}+ \ell^{(3)}_{14}+ \ell^{(3)}_{15} \ , \\
0&=  \ell\left(T^{(\eytab{2},\eytab{0})}_{(\eytab{1},\eytab{0})}\,;2\right)=  \ell^{(3)}_{11}+ \ell^{(3)}_{14}+ \ell^{(3)}_{15}+ \ell^{(3)}_{18}+ \ell^{(3)}_{19}+ \ell^{(3)}_{20}\ , \\
0&=  \ell\left(T^{(\eytab{2},\eytab{0})}_{(\eytab{0},\eytab{1})}\,; 2\right)+ \ell\left(T^{(\eytab{1},\eytab{1})}_{(\eytab{1},\eytab{0})}\,; 2\right)= - \ell^{(3)}_{1}- \ell^{(3)}_{2}- \ell^{(3)}_{3}+ \ell^{(3)}_{9}+ \ell^{(3)}_{14}+ \ell^{(3)}_{15}\ , \\
0&=  \ell\left(T^{(\eytab{1},\eytab{0})}_{(\eytab{1},\eytab{0})}\,; 3\right)=  \ell^{(3)}_{6}+ \ell^{(3)}_{9}+ \ell^{(3)}_{11}+ \ell^{(3)}_{14}+ \ell^{(3)}_{15}+ \ell^{(3)}_{19}+ \ell^{(3)}_{20}\ , \\
0&=  \ell\left(T^{(\eytab{0},\eytab{0})}_{(\eytab{1},\eytab{0})}\,; 4\right) =  \ell^{(3)}_{1}+ \ell^{(3)}_{6}+ \ell^{(3)}_{11}+ \ell^{(3)}_{15}+ \ell^{(3)}_{20} \ .
\end{split}
\label{eq:pmc64_H>0}
\end{equation}
\item \textbf{PMC$[N_f,N_f-2]$ with $H_{N_f-2}<0$}
\begin{equation}
\begin{split}
0&= \ell\left(T^{(\eytab{0},\eytab{7})}_{(\eytab{0},\eytab{0})}\,; -1\right)= \ell^{(3)}_{1}- \ell^{(3)}_{20} \ ,\\
0&= \ell\left(T^{(\eytab{0},\eytab{8})}_{(\eytab{0},\eytab{0})}\,; -1\right)+ \ell\left(T^{(\eytab{3},\eytab{2})}_{(\eytab{0},\eytab{0})}\,; -1\right)= \ell^{(3)}_{2}+ \ell^{(3)}_{9} - \ell^{(3)}_{19}\ ,\\
0&= \ell\left(T^{(\eytab{1},\eytab{4})}_{(\eytab{0},\eytab{0})}\,; -1\right)= \ell^{(3)}_{6}- \ell^{(3)}_{15} \ .
\end{split}
\label{eq:pmc64_H<0}
\end{equation}
\end{itemize}
\normalsize

The solution of AMC$[N_f]$ (Eqs.~(\ref{eq:am_nf>6}),~(\ref{eq:am_nf>6_2})) and PMC$[N_f]$ (Eqs.~(\ref{eq:pmc61_H>0}),~(\ref{eq:pmc61_H<0}),~(\ref{eq:pmc62_H>0}),~(\ref{eq:pmc62_H<0})~(\ref{eq:pmc63_H>0}),~(\ref{eq:pmc63_H<0}),~(\ref{eq:pmc64_H>0}) and~(\ref{eq:pmc64_H<0})) is the following:
\begin{equation}
\label{eq:Nc=3_Nf>=6}
\text{SOL}[N_c=3, N_f\geq 6]=
\left(
\begin{array}{c}
 \ell_{b}= \ell_{a}  \\
 \ell_{d}= -\ell_{a}-\ell_{c}  \\
 \ell_{e}= -\ell_{a}-\ell_{c} \\
 \ell_{3}= 2 \ell_{1}-\ell_{a}+\frac{1}{2}\ell_{c}+\frac{1}{6} \\
 \ell_{4}= \ell_{2} \\
 \ell_{5}= \ell_{1} \\
 \ell_{6}= -\ell_{1}-\ell_{2} \\
 \ell_{7}= -2\ell_{1}-2\ell_{2}+\ell_{a}-\frac{1}{2}\ell_{c}-\frac{1}{6} \\
 \ell_{8}= -\ell_{1}-\ell_{2} \\
 \ell_{9}= 3\ell_{1}+\ell_{2}-\ell_{a}+\frac{1}{2}\ell_{c}+\frac{1}{6} \\
 \ell_{10}= 2 \ell_{2} \\
 \ell_{11}=  2 \ell_{2} \\
 \ell_{12}= 3\ell_{1}+\ell_{2} -\ell_{a}+\frac{1}{2}\ell_{c}+\frac{1}{6} \\
 \ell_{13}=  -\ell_{1}-\ell_{2} \\
 \ell_{14}= -2\ell_{1} - 2 \ell_{2} +\ell_{a}-\frac{1}{2}\ell_{c}-\frac{1}{6} \\
 \ell_{15}=  -\ell_{1}-\ell_{2} \\
 \ell_{16}= \ell_{1} \\
 \ell_{17}=  \ell_{2}  \\
 \ell_{18}= 2 \ell_{1}-\ell_{a}+\frac{1}{2}\ell_{c}+\frac{1}{6} \\
 \ell_{19}=  \ell_{2} \\
 \ell_{20}= \ell_{1} 
\end{array}
\right)\ ,
\end{equation}
\normalsize
where we have chosen the free indices to be $\{ \ell_{a}, \ell_{c}, \ell_{1},\ell_{2}\}$. Since Eq.~(\ref{eq:Nc=3_Nf>=6}) solves AMC$[N_f]\cup$PMC$[N_f]$ for $N_c=3$ and $N_f\geq 6$, $N_f$-independence is explicitly checked for $N_f\geq 6$.

\subsection{Theory with $N_f=5$}

In a theory with 5 flavors, the 25 tensors of~Table~\ref{var_pentaq} are all well defined. Tensor $T_5$ transforms as the representation
\bea
\left(\ {\tiny\yng(1,1,1,1)}\ , \ {\tiny\yng(1,1,1,1)}\ \right)\ ,
\eea
which is equivalent to its parity conjugate. Parity invariance of the spectrum then implies that the index $\ell_5$ vanishes identically and disappears from AMC and PMC equations.
The rank of AMC$[5]\cup$PMC$[5]$ is 20 (see Table~\ref{tab:Nc_5}), and we conclude that the family of real solutions has 4 free parameters.

The $[SU(5)_{L}]^3$ and $[SU(5)_{L}]^2U(1)_B$ AMC$[5]$ equations are respectively
\begin{align}
\label{eq:am_nf=5}
\begin{aligned}
  44 \ell_{a}-\ell_{b}+16 \ell_{c}+30 \ell_{d}-5 \ell_{e} & \\
-850 \ell_{1}-600 \ell_{2}-125 \ell_{3}-75 \ell_{4}-1056 \ell_{6}-384 \ell_{7}+24 \ell_{8}-495 \ell_{9}-105 \ell_{10} & \\
-195 \ell_{11}+220 \ell_{12}-120 \ell_{13}-25 \ell_{14}+720 \ell_{15}-9 \ell_{16}-30 \ell_{17}+9 \ell_{18}+330\ell_{19}+666 \ell_{20}&=3 \ , 
\end{aligned}
\intertext{and}
\label{eq:am_nf=5_2}
\begin{aligned}
28\ell_{a}+3 \ell_{b}+22 \ell_{c}+20 \ell_{d}+5 \ell_{e} & \\
  -350 \ell_{1}-350 \ell_{2}-125 \ell_{3}-75 \ell_{4}-322 \ell_{6}-128 \ell_{7}+28 \ell_{8}+45 \ell_{9} +105 \ell_{10} & \\
+245 \ell_{11}+280 \ell_{12}+70 \ell_{13}+525 \ell_{14}+680 \ell_{15}+7 \ell_{16}+140 \ell_{17}+203\ell_{18}+510 \ell_{19}+462 \ell_{20}&=1 \ .
\end{aligned}
\end{align}

Next, we consider PMC equations.
By giving mass to one flavor, one can derive PMC$[5,1]$. These are:
%
\begin{itemize}
\item \textbf{PMC$[5,1]$ with $H_{1}>0$}
  \scriptsize
\begin{equation}
\begin{split}
0&= \ell\left(T^{(\eytab{2},\eytab{0})}_{(\eytab{0},\eytab{0})}\,; 1\right) = \ell_{a}+ \ell_{c}+\ell_{d} -\ell_{1}-\ell_{2}-\ell_{3}-\ell_{6}-\ell_{7}+\ell_{14}+\ell_{15}+\ell_{18}+\ell_{19}+\ell_{20}\ ,\\
0&= \ell\left(T^{(\eytab{3},\eytab{0})}_{(\eytab{0},\eytab{0})}\,; 1\right) = \ell_{b}+\ell_{c}+\ell_{e}-\ell_{2}-\ell_{4}-\ell_{7} -\ell_{8}+\ell_{9}-\ell_{12}+\ell_{13}+\ell_{14}+\ell_{17}+\ell_{19} \ ,\\
0&= \ell\left(T^{(\eytab{1},\eytab{0})}_{(\eytab{0},\eytab{0})}\,; 2\right) = \ell_{a}+\ell_{c}+\ell_{e}-\ell_{1}-\ell_{2}-\ell_{7} +\ell_{9}-\ell_{12}+\ell_{14}+\ell_{19}+\ell_{20} \ ,\\
0& =  \ell\left(T^{(\eytab{4},\eytab{0})}_{(\eytab{1},\eytab{0})}\,;1\right) = \ell_{15} + \ell_{19}+ \ell_{20} \ , \quad\quad
0=  \ell\left(T^{(\eytab{4},\eytab{0})}_{(\eytab{0},\eytab{1})}\,; 1\right) = -\ell_{1}- \ell_{2}-\ell_{6} \ , \\
0&= \ell\left(T^{(\eytab{6},\eytab{0})}_{(\eytab{1},\eytab{0})}\,; 1\right) = \ell_{13}+ \ell_{16}+ \ell_{17} \ ,\\
0&=  \ell\left(T^{(\eytab{5},\eytab{0})}_{(\eytab{1},\eytab{0})}\,; 1\right) = \ell_{14}+ \ell_{17}+ \ell_{18}+ \ell_{19} \ , \quad\quad
0 = \ell\left(T^{(\eytab{5},\eytab{0})}_{(\eytab{0},\eytab{1})}\,; 1\right) = -\ell_{2}-\ell_{3}-\ell_{4}-\ell_{7} \ ,\\
0 &= \ell\left(T^{(\eytab{2},\eytab{1})}_{(\eytab{1},\eytab{0})}\,; 1\right) = \ell_{11}+\ell_{12}+\ell_{14}+\ell_{15} \ , \quad\quad
0 = \ell\left(T^{(\eytab{2},\eytab{1})}_{(\eytab{0},\eytab{1})}\,; 1\right) = -\ell_{6}-\ell_{7}- \ell_{9}- \ell_{11} \ , \\
0 &=  \ell\left(T^{(\eytab{3},\eytab{1})}_{(\eytab{1},\eytab{0})}\,; 1\right) = \ell_{9}+ \ell_{10}+ \ell_{13}+ \ell_{14}\ , \quad\quad
0 =  \ell\left(T^{(\eytab{3},\eytab{1})}_{(\eytab{0},\eytab{1})}\,; 1\right) = -\ell_{7}-\ell_{8}- \ell_{10}- \ell_{12} \ ,\\
0 &=  \ell\left(T^{(\eytab{2},\eytab{0})}_{(\eytab{1},\eytab{0})}\,; 2 \right) = \ell_{11}+ \ell_{14}+ \ell_{15}+ \ell_{18}+ \ell_{19}+ \ell_{20} \ , \\
0 &=  \ell\left(T^{(\eytab{2},\eytab{0})}_{(\eytab{0},\eytab{1})}\,; 2\right) = -\ell_{1}- \ell_{2}- \ell_{3}-\ell_{6}-\ell_{7}-\ell_{11} \ ,\\
0 &= \ell\left(T^{(\eytab{3},\eytab{0})}_{(\eytab{1},\eytab{0})}\,; 2\right) = \ell_{9}+ \ell_{13}+ \ell_{14}+ \ell_{17}+ \ell_{19} \ ,\\
0 &=  \ell\left(T^{(\eytab{3},\eytab{0})}_{(\eytab{0},\eytab{1})}\,; 2\right) = -\ell_{2}-\ell_{4}-\ell_{7}-\ell_{8}-\ell_{12} \ ,\\
0 &=  \ell\left(T^{(\eytab{1},\eytab{1})}_{(\eytab{1},\eytab{0})}\,; 2\right) = \ell_{6}+\ell_{7}+\ell_{9}+\ell_{10}+\ell_{11}+\ell_{12}+\ell_{14}+\ell_{15} \ ,\\
0 &= \ell\left(T^{(\eytab{1},\eytab{0})}_{(\eytab{1},\eytab{0})}\,; 3\right) = \ell_{6}+\ell_{9}+\ell_{11}+\ell_{14}+\ell_{15}+\ell_{19}+\ell_{20} \ , \\
0 &=  \ell\left(T^{(\eytab{1},\eytab{0})}_{(\eytab{0},\eytab{1})}\,; 3\right) = -\ell_{1}-\ell_{2}-\ell_{6}-\ell_{7}-\ell_{11}-\ell_{12}-\ell_{15} \ ,\\
0 &=  \ell\left(T^{(\eytab{0},\eytab{0})}_{(\eytab{1},\eytab{0})}\,; 4\right) = \ell_{1}+\ell_{6}+\ell_{11}+\ell_{15}+\ell_{20} \ ,\\
\end{split}
\label{eq:pmc51_H>0}
\end{equation}
\normalsize
\item \textbf{PMC$[5,1]$ with $H_{1}<0$}
  \scriptsize
\begin{equation}
\begin{split}
0&=  \ell\left(T^{(\eytab{0},\eytab{7})}_{(\eytab{0},\eytab{0})}\,; -1\right)=\ell_{1}-\ell_{20} \ ,\quad\quad
0=  \ell\left(T^{(\eytab{0},\eytab{8})}_{(\eytab{0},\eytab{0})}\,; -1\right)=\ell_{2}-\ell_{19} \ , \\
0&=  \ell\left(T^{(\eytab{0},\eytab{9})}_{(\eytab{0},\eytab{0})}\,; -1\right)=\ell_{3}-\ell_{18} \ ,\quad\quad
0= \ell\left(T^{(\eytab{0},\eytab{10})}_{(\eytab{0},\eytab{0})}\,; -1\right)=\ell_{4}-\ell_{17} \ ,\\
0&=  \ell\left(T^{(\eytab{1},\eytab{4})}_{(\eytab{0},\eytab{0})}\,; -1\right)=\ell_{6}-\ell_{15} \ , \quad\quad
0=  \ell\left(T^{(\eytab{1},\eytab{5})}_{(\eytab{0},\eytab{0})}\,; -1\right)=\ell_{7}-\ell_{14} \ ,\\
0&=  \ell\left(T^{(\eytab{1},\eytab{6})}_{(\eytab{0},\eytab{0})}\,; -1\right)=\ell_{8}-\ell_{13} \ , \quad\quad
0=  \ell\left(T^{(\eytab{3},\eytab{2})}_{(\eytab{0},\eytab{0})}\,; -1\right)=\ell_{9}-\ell_{12} \ ,\\
\end{split}
\label{eq:pmc51_H<0}
\end{equation}
\end{itemize}
\normalsize
By comparing with Eqs.~(\ref{eq:pmc61_H>0}) and (\ref{eq:pmc61_H<0}), we see that PMC$[5,1]$ are not exactly the same as PMC$[6,1]$, hence $N_f$-independence does not hold for $N_f=5$. This is expected based on~\cite{Ciambriello:2022wmh}, since pentaquark states are not interpolated by class A tensors for $N_f=5$.

Once evaluated on the solutions of  PMC$[5,1]$, the two AMC equations become linearly dependent and identical to respectively Eq.~(\ref{Nfeq_Nc=3}) and~(\ref{Nfeq_Nc=3_2}). This proves the $N_f$-independence of the $N_f$-equation for $N_f=5$.

PMC$[5,2]$ equations are obtained by turning on different masses for two flavors, and constrain the indices $ \ell^{(4)}_\alpha$ of irreps of $\mathcal{G}[5,1]$ with $H_1 =0$. The indices $ \ell^{(4)}_\alpha$ are computed by means of Eq.~(\ref{eq:downliftedsol}); we find:
\begin{equation}
\label{eq:ident_5to4_Nf5}
\begin{split}
  \ell^{(4)}_a&= \ell_a- \ell_{1}-\ell_{2}- \ell_{6}+ \ell_{15}+ \ell_{19}+ \ell_{20}\ ,\\
  \ell^{(4)}_b&= \ell_b- \ell_{4}-\ell_{8}+\ell_{13}+ \ell_{16}+ \ell_{17} \ ,\\
  \ell^{(4)}_c&= \ell_c-\ell_{2}-\ell_{3}-\ell_{4}-\ell_{7}+ \ell_{14}+ \ell_{17}+ \ell_{18}+ \ell_{19} \ ,\\
  \ell^{(4)}_d&= \ell_d- \ell_{6}- \ell_{7}- \ell_{9}+ \ell_{12}+ \ell_{14}+ \ell_{15} \ ,\\
  \ell^{(4)}_e&= \ell_e- \ell_{7}- \ell_{8}+ \ell_{9}- \ell_{12}+ \ell_{13}+ \ell_{14} \ ,\\
  \ell^{(4)}_4 &=  \ell_4- \ell_8\ ,\\
  \ell^{(4)}_\alpha&= \ell_\alpha ,\quad \text{for }  \alpha=1,2,3,6,7,9,\cdots ,15, 17,\cdots, 20\ .
\end{split}
\end{equation}
Notice that tensors $T_{b}$,  $T_{5}$ and $T_{16}$ transform as the same irrep of $\mathcal{G}[5,1]$, whose index is denoted by $  \ell^{(4)}_b$. Similarly, $T_4$ is equivalent to the parity conjugate of $T_8$, the corresponding index is $  \ell^{(4)}_4$.

The PMC$[5, 2]$ equations are:
\begin{itemize}
\item \textbf{PMC$[5,2]$ with $H_{2}>0$}
\scriptsize
  \begin{equation}
\begin{split}
0&= \ell\left(T^{(\eytab{2},\eytab{0})}_{(\eytab{0},\eytab{0})}\,; 1\right) = \ell^{(4)}_{a}+ \ell^{(4)}_{c}+ \ell^{(4)}_{d}- \ell^{(4)}_{1}- \ell^{(4)}_{2}- \ell^{(4)}_{3}- \ell^{(4)}_{6}- \ell^{(4)}_{7}+ \ell^{(4)}_{14}+ \ell^{(4)}_{15}+ \ell^{(4)}_{18}+ \ell^{(4)}_{19}+ \ell^{(4)}_{20} \ ,\\
0&= \ell\left(T^{(\eytab{3},\eytab{0})}_{(\eytab{0},\eytab{0})}\,; 1\right)+ \ell\left(T^{(\eytab{6},\eytab{0})}_{(\eytab{1},\eytab{0})}\,; 1\right)+ \ell\left(T^{(\eytab{0},\eytab{6})}_{(\eytab{1},\eytab{0})}\,; 1\right) \\
&=  \ell^{(4)}_{b}+ \ell^{(4)}_{c}+ \ell^{(4)}_{e} - \ell^{(4)}_{2}- \ell^{(4)}_{7}+ \ell^{(4)}_{9}- \ell^{(4)}_{12}+ \ell^{(4)}_{13}+ \ell^{(4)}_{14}+ \ell^{(4)}_{17}+ \ell^{(4)}_{19} \ ,\\
0&= \ell\left(T^{(\eytab{1},\eytab{0})}_{(\eytab{0},\eytab{0})}\,; 2\right)=  \ell^{(4)}_{a}+ \ell^{(4)}_{c}+ \ell^{(4)}_{e} - \ell^{(4)}_{1}- \ell^{(4)}_{2}- \ell^{(4)}_{7}+ \ell^{(4)}_{9}- \ell^{(4)}_{12}+ \ell^{(4)}_{14}+ \ell^{(4)}_{19}+ \ell^{(4)}_{20}\ ,\\
0&= \ell\left(T^{(\eytab{4},\eytab{0})}_{(\eytab{1},\eytab{0})}\,; 1\right)=  \ell^{(4)}_{15}+ \ell^{(4)}_{19}+ \ell^{(4)}_{20} \ , \quad\quad\quad\quad\quad
0= \ell\left(T^{(\eytab{4},\eytab{0})}_{(\eytab{0},\eytab{1})}\,; 1\right)= - \ell^{(4)}_{1}- \ell^{(4)}_{2}- \ell^{(4)}_{6} \ ,\\
0&= \ell\left(T^{(\eytab{5},\eytab{0})}_{(\eytab{1},\eytab{0})}\,; 1\right)=  \ell^{(4)}_{14}+ \ell^{(4)}_{17}+ \ell^{(4)}_{18}+ \ell^{(4)}_{19} \ ,\\
0&= \ell\left(T^{(\eytab{5},\eytab{0})}_{(\eytab{0},\eytab{1})}\,; 1\right)+ \ell\left(T^{(\eytab{1},\eytab{3})}_{(\eytab{1},\eytab{0})}\,; 1\right)=- \ell^{(4)}_{2}- \ell^{(4)}_{3}- \ell^{(4)}_{4}+  \ell^{(4)}_{10}+ \ell^{(4)}_{12} \ ,\\
0&= \ell\left(T^{(\eytab{2},\eytab{1})}_{(\eytab{1},\eytab{0})}\,; 1\right)=  \ell^{(4)}_{11}+ \ell^{(4)}_{12}+ \ell^{(4)}_{14}+ \ell^{(4)}_{15} \ , \quad\quad\quad
0= \ell\left(T^{(\eytab{2},\eytab{1})}_{(\eytab{0},\eytab{1})}\,; 1\right)=- \ell^{(4)}_{6}- \ell^{(4)}_{7}- \ell^{(4)}_{9}- \ell^{(4)}_{11} \ ,\nonumber
\end{split}
\end{equation}
\begin{equation}
\begin{split}
0&= \ell\left(T^{(\eytab{3},\eytab{1})}_{(\eytab{1},\eytab{0})}\,; 1\right)=  \ell^{(4)}_{9}+ \ell^{(4)}_{10}+ \ell^{(4)}_{13}+ \ell^{(4)}_{14} \ , \\
0&= \ell\left(T^{(\eytab{2},\eytab{0})}_{(\eytab{1},\eytab{0})}\,; 2\right)=  \ell^{(4)}_{11}+ \ell^{(4)}_{14}+ \ell^{(4)}_{15}+ \ell^{(4)}_{18}+ \ell^{(4)}_{19}+ \ell^{(4)}_{20} \ ,\\
0&= \ell\left(T^{(\eytab{2},\eytab{0})}_{(\eytab{0},\eytab{1})}\,; 2\right)= - \ell^{(4)}_{1}- \ell^{(4)}_{2}- \ell^{(4)}_{3}- \ell^{(4)}_{6}- \ell^{(4)}_{7}- \ell^{(4)}_{11} \ , \\
0&= \ell\left(T^{(\eytab{3},\eytab{0})}_{(\eytab{1},\eytab{0})}\,; 2\right)=  \ell^{(4)}_{9}+ \ell^{(4)}_{13}+ \ell^{(4)}_{14}+ \ell^{(4)}_{17}+ \ell^{(4)}_{19} \ ,\\
0&= \ell\left(T^{(\eytab{1},\eytab{1})}_{(\eytab{1},\eytab{0})}\,; 2\right)=  \ell^{(4)}_{6}+ \ell^{(4)}_{7}+ \ell^{(4)}_{9}+ \ell^{(4)}_{10}+ \ell^{(4)}_{11}+ \ell^{(4)}_{12}+ \ell^{(4)}_{14}+ \ell^{(4)}_{15} \ ,\\
0&= \ell\left(T^{(\eytab{1},\eytab{0})}_{(\eytab{1},\eytab{0})}\,; 3\right)=  \ell^{(4)}_{6}+ \ell^{(4)}_{9}+ \ell^{(4)}_{11}+ \ell^{(4)}_{14}+ \ell^{(4)}_{15}+ \ell^{(4)}_{19}+ \ell^{(4)}_{20} \ ,\\
0&= \ell\left(T^{(\eytab{1},\eytab{0})}_{(\eytab{0},\eytab{1})}\,; 3\right)= - \ell^{(4)}_{1}- \ell^{(4)}_{2}- \ell^{(4)}_{6}- \ell^{(4)}_{7}- \ell^{(4)}_{11}- \ell^{(4)}_{12}- \ell^{(4)}_{15} \ ,\\
0&= \ell\left(T^{(\eytab{0},\eytab{0})}_{(\eytab{1},\eytab{0})}\,; 4\right)=  \ell^{(4)}_{1}+ \ell^{(4)}_{6}+ \ell^{(4)}_{11}+ \ell^{(4)}_{15}+ \ell^{(4)}_{20} \ .
\end{split}
\label{eq:pmc52_H>0}
\end{equation}
\normalsize
\item \textbf{PMC$[5,2]$ with $H_{2}<0$}
\scriptsize
  \begin{equation}
\begin{split}
0&= \ell\left(T^{(\eytab{0},\eytab{7})}_{(\eytab{0},\eytab{0})}\,; -1\right)=  \ell^{(4)}_{1}- \ell^{(4)}_{20} \ , \quad\quad\quad\quad\quad
0= \ell\left(T^{(\eytab{0},\eytab{8})}_{(\eytab{0},\eytab{0})}\,; -1\right)=  \ell^{(4)}_{2}- \ell^{(4)}_{19} \ , \\
 0&= \ell\left(T^{(\eytab{0},\eytab{9})}_{(\eytab{0},\eytab{0})}\,; -1\right)=  \ell^{(4)}_{3}- \ell^{(4)}_{18} \ ,  \quad\quad\quad\quad\quad
 0= \ell\left(T^{(\eytab{1},\eytab{4})}_{(\eytab{0},\eytab{0})}\,; -1\right)= \ell^{(4)}_{6}- \ell^{(4)}_{15} \ ,\\
 0&= \ell\left(T^{(\eytab{0},\eytab{10})}_{(\eytab{0},\eytab{0})}\,; -1\right)+ \ell\left(T^{(\eytab{6},\eytab{1})}_{(\eytab{0},\eytab{0})}\,; -1\right)=  \ell^{(4)}_{4}+ \ell^{(4)}_{13}- \ell^{(4)}_{17} \ ,\\
0&= \ell\left(T^{(\eytab{1},\eytab{5})}_{(\eytab{0},\eytab{0})}\,; -1\right)= \ell^{(4)}_{7}- \ell^{(4)}_{14} \ ,  \quad\quad\quad\quad\quad
 0= \ell\left(T^{(\eytab{3},\eytab{2})}_{(\eytab{0},\eytab{0})}\,; -1\right)= \ell^{(4)}_{9}- \ell^{(4)}_{12} \ .
 \end{split}
 \label{eq:pmc52_H<0}
 \end{equation}
 \normalsize
\end{itemize}

PMC$[5,3]$ equations are obtained by turning on different masses for three flavors, and constrain the indices $ \ell^{(3)}_\alpha$ of irreps of $\mathcal{G}[5,2]$ with $H_1=H_2 =0$. The indices $ \ell^{(3)}_\alpha$ are computed by means of Eq.~(\ref{eq:downliftedsol}); we find:
\begin{equation}
\label{eq:ident_4to3_Nf5}
\begin{split}
  \ell^{(3)}_a &=  \ell^{(4)}_a- \ell^{(4)}_{1}- \ell^{(4)}_{2}- \ell^{(4)}_{6}+ \ell^{(4)}_{15}+ \ell^{(4)}_{19}+ \ell^{(4)}_{20}\ ,\\
  \ell^{(3)}_b &= 0 \ , \\
  \ell^{(3)}_c&=  \ell^{(4)}_c- \ell^{(4)}_{2}- \ell^{(4)}_{3}- \ell^{(4)}_{4}- \ell^{(4)}_{7}+ \ell^{(4)}_{14}+ \ell^{(4)}_{17}+ \ell^{(4)}_{18}+ \ell^{(4)}_{19} \ ,\\
  \ell^{(3)}_d &=  \ell^{(4)}_d- \ell^{(4)}_{6}- \ell^{(4)}_{7}- \ell^{(4)}_{9}+ \ell^{(4)}_{12}+ \ell^{(4)}_{14}+ \ell^{(4)}_{15} \ ,\\
  \ell^{(3)}_e&=  \ell^{(4)}_e+ \ell^{(4)}_{4}- \ell^{(4)}_{7}+ \ell^{(4)}_{9}- \ell^{(4)}_{12}+ \ell^{(4)}_{13}+ \ell^{(4)}_{14} \ ,\\
  \ell^{(3)}_{2} &=   \ell^{(4)}_{2}-  \ell^{(4)}_{12}\ ,\\
  \ell^{(3)}_{3} &=   \ell^{(4)}_{3}-  \ell^{(4)}_{10}\ ,\\
  \ell^{(3)}_7 &= 0 \ , \\
  \ell^{(3)}_\alpha&= \ell^{(4)}_\alpha ,\quad \text{for }  \alpha=1,6,9,11,14,15,18,19,20\ .
\end{split}
\end{equation}
Notice that: $T_{5}$ and $T_{16}$ are not well defined tensors of $\mathcal{G}[5,2]$, hence the indices $  \ell^{(3)}_{5,16}$ do not appear in Eq.~(\ref{eq:ident_4to3_Nf5}); $T_b$ and $T_7$ transform as the same irreps of $\mathcal{G}[5,2]$ as their parity conjugate, hence their indices vanish identically. Furthermore, there exist the following groups of equivalent tensors: $T_c$, $T_8$ and $T_{17}$; $T_e$, $T_4$ and $T_{13}$; $T_2$ and the parity conjugate of $T_{12}$; $T_3$ and the parity conjugate of $T_{10}$. The indices of the corresponding irreps are denoted respectively by $  \ell^{(3)}_c$, $  \ell^{(3)}_e$, $  \ell^{(3)}_2$ and $  \ell^{(3)}_3$.

The PMC$[5, 3]$ equations are: 
\begin{itemize}
%
\item \textbf{PMC$[5,3]$ with $H_{3}>0$}
\begin{equation}
\begin{split}
0&= \ell\left(T^{(\eytab{1},\eytab{0})}_{(\eytab{0},\eytab{0})}\,; 2\right)+ \ell\left(T^{(\eytab{3},\eytab{0})}_{(\eytab{1},\eytab{0})}\,; 2\right)+ \ell\left(T^{(\eytab{0},\eytab{3})}_{(\eytab{1},\eytab{0})}\,; 2\right) \\
&= \ell^{(3)}_{a}+ \ell^{(3)}_{c}+ \ell^{(3)}_{e}- \ell^{(3)}_{1}+ \ell^{(3)}_{9}+ \ell^{(3)}_{14}+ \ell^{(3)}_{19}+ \ell^{(3)}_{20}\ ,\\
0&= \ell\left(T^{(\eytab{2},\eytab{0})}_{(\eytab{0},\eytab{0})}\,; 1\right)+ \ell\left(T^{(\eytab{5},\eytab{0})}_{(\eytab{1},\eytab{0})}\,; 1\right)+ \ell\left(T^{(\eytab{1},\eytab{3})}_{(\eytab{1},\eytab{0})}\,; 1\right)\\
&= \ell^{(3)}_{a}+ \ell^{(3)}_{c}+ \ell^{(3)}_{d}- \ell^{(3)}_{1}- \ell^{(3)}_{2}- \ell^{(3)}_{3}- \ell^{(3)}_{6}+ \ell^{(3)}_{14}+ \ell^{(3)}_{15}+ \ell^{(3)}_{18}+ \ell^{(3)}_{19}+ \ell^{(3)}_{20} \ ,\\
0&= \ell\left(T^{(\eytab{4},\eytab{0})}_{(\eytab{1},\eytab{0})}\,; 1\right)=  \ell^{(3)}_{15}+ \ell^{(3)}_{19}+ \ell^{(3)}_{20}\ , \\
0&=  \ell\left(T^{(\eytab{4},\eytab{0})}_{(\eytab{0},\eytab{1})}\,; 1\right)+ \ell\left(T^{(\eytab{2},\eytab{1})}_{(\eytab{1},\eytab{0})}\,; 1\right)= - \ell^{(3)}_{1}- \ell^{(3)}_{2}- \ell^{(3)}_{6}+ \ell^{(3)}_{11}+ \ell^{(3)}_{14}+ \ell^{(3)}_{15} \ , \\
0&=  \ell\left(T^{(\eytab{2},\eytab{0})}_{(\eytab{1},\eytab{0})}\,; 2\right)=  \ell^{(3)}_{11}+ \ell^{(3)}_{14}+ \ell^{(3)}_{15}+ \ell^{(3)}_{18}+ \ell^{(3)}_{19}+ \ell^{(3)}_{20}\ , \\
0&=  \ell\left(T^{(\eytab{2},\eytab{0})}_{(\eytab{0},\eytab{1})}\,; 2\right)+ \ell\left(T^{(\eytab{1},\eytab{1})}_{(\eytab{1},\eytab{0})}\,; 2\right)= - \ell^{(3)}_{1}- \ell^{(3)}_{2}- \ell^{(3)}_{3}+ \ell^{(3)}_{9}+ \ell^{(3)}_{14}+ \ell^{(3)}_{15}\ , \\
0&=  \ell\left(T^{(\eytab{1},\eytab{0})}_{(\eytab{1},\eytab{0})}\,; 3\right)=  \ell^{(3)}_{6}+ \ell^{(3)}_{9}+ \ell^{(3)}_{11}+ \ell^{(3)}_{14}+ \ell^{(3)}_{15}+ \ell^{(3)}_{19}+ \ell^{(3)}_{20}\ , \\
0&=  \ell\left(T^{(\eytab{0},\eytab{0})}_{(\eytab{1},\eytab{0})}\,; 4\right) =  \ell^{(3)}_{1}+ \ell^{(3)}_{6}+ \ell^{(3)}_{11}+ \ell^{(3)}_{15}+ \ell^{(3)}_{20} \ ,
\end{split}
\label{eq:pmc53_H>0}
\end{equation}
\item \textbf{PMC$[5,3]$ with $H_{3}<0$}
\begin{equation}
\begin{split}
0&= \ell\left(T^{(\eytab{0},\eytab{7})}_{(\eytab{0},\eytab{0})}\,; -1\right)= \ell^{(3)}_{1}- \ell^{(3)}_{20} \ ,\\
0&= \ell\left(T^{(\eytab{0},\eytab{8})}_{(\eytab{0},\eytab{0})}\,; -1\right)+ \ell\left(T^{(\eytab{3},\eytab{2})}_{(\eytab{0},\eytab{0})}\,; -1\right)= \ell^{(3)}_{2}+ \ell^{(3)}_{9} - \ell^{(3)}_{19}\ ,\\
0&= \ell\left(T^{(\eytab{1},\eytab{4})}_{(\eytab{0},\eytab{0})}\,; -1\right)= \ell^{(3)}_{6}- \ell^{(3)}_{15} \ ,\\
\end{split}
\label{eq:pmc53_H<0}
\end{equation}
\end{itemize}
\normalsize

We find the following solution of  $\text{AMC}[5]\cup \text{PMC}[5]$:
\small
\begin{equation}
\label{eq:Nc=3_Nf=5}
\text{SOL}[N_c=3, N_f=5]=
\left(
\begin{array}{c}
\ell_{b}= \ell_{a}  \\
 \ell_{d}= -\ell_{a}-\ell_{c}  \\
 \ell_{e}= -\ell_{a}-\ell_{c} \\
 \ell_{3}= 2 \ell_{1}-\ell_{a}+\frac{1}{2}\ell_{c}+\frac{1}{6} \\
  \ell_{4}= \ell_{2} \\
  \ell_5 = 0 \\
 \ell_{6}= -\ell_{1}-\ell_{2} \\
 \ell_{7}= -2\ell_{1}-2\ell_{2}+\ell_{a}-\frac{1}{2}\ell_{c}-\frac{1}{6} \\
 \ell_{8}= -\ell_{1}-\ell_{2} \\
 \ell_{9}= 3\ell_{1}+\ell_{2}-\ell_{a}+\frac{1}{2}\ell_{c}+\frac{1}{6} \\
 \ell_{10}= 2 \ell_{2} \\
 \ell_{11}=  2 \ell_{2} \\
 \ell_{12}= 3\ell_{1}+\ell_{2} -\ell_{a}+\frac{1}{2}\ell_{c}+\frac{1}{6} \\
 \ell_{13}=  -\ell_{1}-\ell_{2} \\
 \ell_{14}= -2\ell_{1} - 2 \ell_{2} +\ell_{a}-\frac{1}{2}\ell_{c}-\frac{1}{6} \\
 \ell_{15}=  -\ell_{1}-\ell_{2} \\
 \ell_{16}= \ell_{1} \\
 \ell_{17}=  \ell_{2}  \\
 \ell_{18}= 2 \ell_{1}-\ell_{a}+\frac{1}{2}\ell_{c}+\frac{1}{6} \\
 \ell_{19}=  \ell_{2} \\
 \ell_{20}= \ell_{1} 
\end{array}
\right)\ ,
\end{equation}
\normalsize
where we have chosen the free indices to be $\{ \ell_a, \ell_c, \ell_1, \ell_2 \}$. As noticed earlier, the index $\ell_{5}$ vanishes identically due to parity invariance of the spectrum.

\subsection{Theory with $N_f=4$}

For $N_f=4$, the tensors of~Table~\ref{var_pentaq} are well defined but not all of them correspond to different irreps. In particular, $T_{b},T_5,T_{16}$ are equivalent tensors, and $T_4$ is equivalent to the parity-conjugate of $T_8$. The indices of the corresponding irreps are denoted respectively by $\ell_b$ and $\ell_4$. There 22 irreps in total. The rank of AMC$[4]\cup$PMC$[4]$ is 18 (see Table~\ref{tab:Nc_5}), which means that the family of real solutions has 4 free parameters.

The $[SU(4)_L]^3$ and $[SU(4)_L]^2 U(1)_B$ AMC equations are respectively:
\begin{align}
\label{eq:am_nf=4}
\begin{aligned}
  35 \ell_{a}-\ell_{b}+7 \ell_{c}+22 \ell_{d}-6 \ell_{e}\\
-483 \ell_{1}-237 \ell_{2}-20 \ell_{3}-15 \ell_{4}-525 \ell_{6}-105 \ell_{7}-230 \ell_{9}-42 \ell_{10} \\
-78 \ell_{11}+126 \ell_{12}-42 \ell_{13}-64 \ell_{14}+378 \ell_{15}-21 \ell_{17}-27 \ell_{18}+105\ell_{19}+378 \ell_{20}&=3 \ ,\\
\end{aligned}
\intertext{and}
\label{eq:am_nf=4_2}
\begin{aligned}
  21 \ell_{a}+\ell_{b}+13 \ell_{c}+14 \ell_{d}+2 \ell_{e} \\
-189 \ell_{1}-147 \ell_{2}-44 \ell_{3}-17 \ell_{4}-155 \ell_{6}-35 \ell_{7}+10 \ell_{9}+38 \ell_{10} \\
+114 \ell_{11}+126 \ell_{12}+14 \ell_{13}+192 \ell_{14}+322 \ell_{15}+33 \ell_{17}+71 \ell_{18}+203\ell_{19}+238 \ell_{20}& =1 \ .\\
\end{aligned}
  \end{align}

Next, let us consider PMC equations. For PMC$[4,1]$  we find:
\begin{itemize}
\item \textbf{PMC$[4,1]$ with $H_{1}>0$}
\begin{equation}
\begin{split}
0&= \ell\left(T^{(\eytab{2},\eytab{0})}_{(\eytab{0},\eytab{0})}\,; 1\right) = \ell_{a}+\ell_{c}+\ell_{d}-\ell_{1}-\ell_{2}-\ell_{3}-\ell_{6}-\ell_{7}+\ell_{14}+\ell_{15}+\ell_{18}+\ell_{19}+\ell_{20} \ ,\\
0&= \ell\left(T^{(\eytab{3},\eytab{0})}_{(\eytab{0},\eytab{0})}\,; 1\right)+ \ell\left(T^{(\eytab{6},\eytab{0})}_{(\eytab{1},\eytab{0})}\,; 1\right)+ \ell\left(T^{(\eytab{0},\eytab{6})}_{(\eytab{1},\eytab{0})}\,; 1\right) \\
&= \ell_{b}+\ell_{c}+\ell_{e} -\ell_{2}-\ell_{7}+\ell_{9}-\ell_{12}+\ell_{13}+\ell_{14}+ \ell_{17}+ \ell_{19} \ ,\\
0&= \ell\left(T^{(\eytab{1},\eytab{0})}_{(\eytab{0},\eytab{0})}\,; 2\right)= \ell_{a}+ \ell_{c}+ \ell_{e} -\ell_{1}-\ell_{2}-\ell_{7}+\ell_{9}-\ell_{12}+\ell_{14}+\ell_{19}+\ell_{20}\ ,\\
0&= \ell\left(T^{(\eytab{4},\eytab{0})}_{(\eytab{1},\eytab{0})}\,; 1\right)= \ell_{15}+\ell_{19}+\ell_{20} \ , \quad\quad\quad\quad\quad
0= \ell\left(T^{(\eytab{4},\eytab{0})}_{(\eytab{0},\eytab{1})}\,; 1\right)= -\ell_{1}-\ell_{2}-\ell_{6} \ ,\\
0&= \ell\left(T^{(\eytab{5},\eytab{0})}_{(\eytab{1},\eytab{0})}\,; 1\right)= \ell_{14}+ \ell_{17}+ \ell_{18}+ \ell_{19} \ ,\\
0&= \ell\left(T^{(\eytab{5},\eytab{0})}_{(\eytab{0},\eytab{1})}\,; 1\right)+ \ell\left(T^{(\eytab{1},\eytab{3})}_{(\eytab{1},\eytab{0})}\,; 1\right)=-\ell_{2}-\ell_{3}-\ell_{4}+ \ell_{10}+\ell_{12} \ ,\\
0&= \ell\left(T^{(\eytab{2},\eytab{1})}_{(\eytab{1},\eytab{0})}\,; 1\right)= \ell_{11}+ \ell_{12}+\ell_{14}+\ell_{15} \ , \quad\quad\quad
0= \ell\left(T^{(\eytab{2},\eytab{1})}_{(\eytab{0},\eytab{1})}\,; 1\right)=-\ell_{6}-\ell_{7}-\ell_{9}-\ell_{11} \ ,\\
0&= \ell\left(T^{(\eytab{3},\eytab{1})}_{(\eytab{1},\eytab{0})}\,; 1\right)= \ell_{9}+\ell_{10}+\ell_{13}+\ell_{14} \ , \nonumber
\end{split}
\end{equation}
\begin{equation}
\begin{split}
0&= \ell\left(T^{(\eytab{2},\eytab{0})}_{(\eytab{1},\eytab{0})}\,; 2\right)= \ell_{11}+\ell_{14}+\ell_{15}+\ell_{18}+\ell_{19}+\ell_{20} \ ,\\
0&= \ell\left(T^{(\eytab{2},\eytab{0})}_{(\eytab{0},\eytab{1})}\,; 2\right)= -\ell_{1}-\ell_{2}-\ell_{3}-\ell_{6}-\ell_{7}-\ell_{11} \ , \\
0&= \ell\left(T^{(\eytab{3},\eytab{0})}_{(\eytab{1},\eytab{0})}\,; 2\right)= \ell_{9}+\ell_{13}+\ell_{14}+\ell_{17}+\ell_{19} \ ,\\
0&= \ell\left(T^{(\eytab{1},\eytab{1})}_{(\eytab{1},\eytab{0})}\,; 2\right)= \ell_{6}+\ell_{7}+\ell_{9}+\ell_{10}+\ell_{11}+\ell_{12}+\ell_{14}+\ell_{15} \ ,\\
0&= \ell\left(T^{(\eytab{1},\eytab{0})}_{(\eytab{1},\eytab{0})}\,; 3\right)= \ell_{6}+\ell_{9}+\ell_{11}+\ell_{14}+\ell_{15}+\ell_{19}+\ell_{20} \ ,\\
0&= \ell\left(T^{(\eytab{1},\eytab{0})}_{(\eytab{0},\eytab{1})}\,; 3\right)= -\ell_{1}-\ell_{2}-\ell_{6}-\ell_{7}-\ell_{11}-\ell_{12}-\ell_{15} \ ,\\
0&= \ell\left(T^{(\eytab{0},\eytab{0})}_{(\eytab{1},\eytab{0})}\,; 4\right)= \ell_{1}+\ell_{6}+\ell_{11}+\ell_{15}+\ell_{20} \ ,\\
\end{split}
\label{eq:pmc41_H>0}
\end{equation}
   %
\item \textbf{PMC$[4,1]$ with $H_{1}<0$}
\begin{equation}
\begin{split}
0&= \ell\left(T^{(\eytab{0},\eytab{7})}_{(\eytab{0},\eytab{0})}\,; -1\right)= \ell_{1}- \ell_{20} \ , \quad\quad\quad\quad\quad
0= \ell\left(T^{(\eytab{0},\eytab{8})}_{(\eytab{0},\eytab{0})}\,; -1\right)= \ell_{2}- \ell_{19} \ , \\
 0&= \ell\left(T^{(\eytab{0},\eytab{9})}_{(\eytab{0},\eytab{0})}\,; -1\right)= \ell_{3}- \ell_{18} \ ,  \quad\quad\quad\quad\quad
 0= \ell\left(T^{(\eytab{1},\eytab{4})}_{(\eytab{0},\eytab{0})}\,; -1\right)= \ell_{6}- \ell_{15} \ ,\\
 0&= \ell\left(T^{(\eytab{0},\eytab{10})}_{(\eytab{0},\eytab{0})}\,; -1\right)+ \ell\left(T^{(\eytab{6},\eytab{1})}_{(\eytab{0},\eytab{0})}\,; -1\right)= \ell_{4}+\ell_{13}-\ell_{17} \ ,\\
0&= \ell\left(T^{(\eytab{1},\eytab{5})}_{(\eytab{0},\eytab{0})}\,; -1\right)=\ell_{7}-\ell_{14} \ ,  \quad\quad\quad\quad\quad
 0= \ell\left(T^{(\eytab{3},\eytab{2})}_{(\eytab{0},\eytab{0})}\,; -1\right)=\ell_{9}-\ell_{12} \ , \\
 \end{split}
 \label{eq:pmc41_H<0}
 \end{equation}
 %
\end{itemize}
One can see that PMC$[4,1]$ are in the same form as PMC$[5,2]$, although all the indices in PMC$[4,1]$ are free while those in PMC$[5,2]$ are obtained from the decomposition of the original tensors in the theory with 5 massless flavors.
Furthermore, PMC$[4,1]$ are different from PMC$[5,1]$ or PMC$[6,1]$, hence $N_f$-independence does not hold for $N_f=4$.

By evaluating them on the solutions of PMC$[4,1]$, the two AMC equations become linearly dependent and identical to respectively Eq.~(\ref{Nfeq_Nc=3}) and~(\ref{Nfeq_Nc=3_2}). This proves the $N_f$-independence of the $N_f$-equation for $N_f=4$.

PMC$[4,2]$ equations constrain the indices $ \ell^{(3)}_\alpha$ of irreps of $\mathcal{G}[4,1]$ with $H_1 =0$. The indices $ \ell^{(3)}_\alpha$ are computed by means of Eq.~(\ref{eq:downliftedsol}); we find:
\begin{equation}
\label{eq:ident_4to3_Nf4}
\begin{split}
  \ell^{(3)}_a &= \ell_a-\ell_{1}-\ell_{2}-\ell_{6}+\ell_{15}+\ell_{19}+\ell_{20}\ ,\\
  \ell^{(3)}_b &= 0 \ , \\
  \ell^{(3)}_c&= \ell_c-\ell_{2}-\ell_{3}-\ell_{4}-\ell_{7}+\ell_{14}+\ell_{17}+\ell_{18}+\ell_{19} \ ,\\
  \ell^{(3)}_d &= \ell_d- \ell_{6}- \ell_{7}- \ell_{9}+\ell_{12}+ \ell_{14}+ \ell_{15} \ ,\\
  \ell^{(3)}_e &= \ell_e+\ell_{4}-\ell_{7}+\ell_{9}-\ell_{12}+\ell_{13}+\ell_{14} \ ,\\
  \ell^{(3)}_{2}&= \ell_{2}- \ell_{12}\ ,\\
  \ell^{(3)}_{3} &= \ell_{3}-\ell_{10}\ ,\\
  \ell^{(3)}_7 &= 0 \ , \\
  \ell^{(3)}_\alpha&=\ell_\alpha ,\quad \text{for }  \alpha=1,6,9,11,14,15,18,19,20\ .
\end{split}
\end{equation}
Notice that $T_{5,16}$ are not well defined tensors of $\mathcal{G}[4,1]$, hence $\ell^{(3)}_{5,16}$ do not appear in Eq.~(\ref{eq:ident_4to3_Nf4}). Moreover, the $T_b$ and $T_7$ transform as the same irreps of $\mathcal{G}[4,1]$ as their parity conjugate; as a consequence, the indices $  \ell^{(3)}_{b}$ and $  \ell^{(3)}_{7}$ vanish identically.
Finally, there exist the following groups of equivalent tensors: $T_c$, $T_8$ and $T_{17}$; $T_e$, $T_4$ and $T_{13}$; $T_2$ and the parity conjugate of $T_{12}$; $T_3$ and the parity conjugate of $T_{10}$. The indices of the corresponding irreps are denoted respectively by $  \ell^{(3)}_c$, $  \ell^{(3)}_e$, $  \ell^{(3)}_2$ and $  \ell^{(3)}_3$.

The PMC$[4, 2]$ are:
\begin{itemize}
%
\item \textbf{PMC$[4,2]$ with $H_{2}>0$}
\begin{equation}
\begin{split}
0&= \ell\left(T^{(\eytab{1},\eytab{0})}_{(\eytab{0},\eytab{0})}\,; 2\right)+ \ell\left(T^{(\eytab{3},\eytab{0})}_{(\eytab{1},\eytab{0})}\,; 2\right)+ \ell\left(T^{(\eytab{0},\eytab{3})}_{(\eytab{1},\eytab{0})}\,; 2\right) \\
&= \ell^{(3)}_{a}+ \ell^{(3)}_{c}+ \ell^{(3)}_{e}- \ell^{(3)}_{1}+ \ell^{(3)}_{9}+ \ell^{(3)}_{14}+ \ell^{(3)}_{19}+ \ell^{(3)}_{20}\ ,\\
0&= \ell\left(T^{(\eytab{2},\eytab{0})}_{(\eytab{0},\eytab{0})}\,; 1\right)+ \ell\left(T^{(\eytab{5},\eytab{0})}_{(\eytab{1},\eytab{0})}\,; 1\right)+ \ell\left(T^{(\eytab{1},\eytab{3})}_{(\eytab{1},\eytab{0})}\,; 1\right)\\
&= \ell^{(3)}_{a}+ \ell^{(3)}_{c}+ \ell^{(3)}_{d}- \ell^{(3)}_{1}- \ell^{(3)}_{2}- \ell^{(3)}_{3}- \ell^{(3)}_{6}+ \ell^{(3)}_{14}+ \ell^{(3)}_{15}+ \ell^{(3)}_{18}+ \ell^{(3)}_{19}+ \ell^{(3)}_{20} \ ,\\
0&= \ell\left(T^{(\eytab{4},\eytab{0})}_{(\eytab{1},\eytab{0})}\,; 1\right)=  \ell^{(3)}_{15}+ \ell^{(3)}_{19}+ \ell^{(3)}_{20}\ , \\
0&=  \ell\left(T^{(\eytab{4},\eytab{0})}_{(\eytab{0},\eytab{1})}\,; 1\right)+ \ell\left(T^{(\eytab{2},\eytab{1})}_{(\eytab{1},\eytab{0})}\,; 1\right)= - \ell^{(3)}_{1}- \ell^{(3)}_{2}- \ell^{(3)}_{6}+ \ell^{(3)}_{11}+ \ell^{(3)}_{14}+ \ell^{(3)}_{15} \ , \\
0&=  \ell\left(T^{(\eytab{2},\eytab{0})}_{(\eytab{1},\eytab{0})}\,; 2\right)=  \ell^{(3)}_{11}+ \ell^{(3)}_{14}+ \ell^{(3)}_{15}+ \ell^{(3)}_{18}+ \ell^{(3)}_{19}+ \ell^{(3)}_{20}\ , \\
0&=  \ell\left(T^{(\eytab{2},\eytab{0})}_{(\eytab{0},\eytab{1})}\,; 2\right)+ \ell\left(T^{(\eytab{1},\eytab{1})}_{(\eytab{1},\eytab{0})}\,; 2\right)= - \ell^{(3)}_{1}- \ell^{(3)}_{2}- \ell^{(3)}_{3}+ \ell^{(3)}_{9}+ \ell^{(3)}_{14}+ \ell^{(3)}_{15}\ , \\
0&=  \ell\left(T^{(\eytab{1},\eytab{0})}_{(\eytab{1},\eytab{0})}\,; 3\right)=  \ell^{(3)}_{6}+ \ell^{(3)}_{9}+ \ell^{(3)}_{11}+ \ell^{(3)}_{14}+ \ell^{(3)}_{15}+ \ell^{(3)}_{19}+ \ell^{(3)}_{20}\ , \\
0&=  \ell\left(T^{(\eytab{0},\eytab{0})}_{(\eytab{1},\eytab{0})}\,; 4\right) =  \ell^{(3)}_{1}+ \ell^{(3)}_{6}+ \ell^{(3)}_{11}+ \ell^{(3)}_{15}+ \ell^{(3)}_{20} \ .
\end{split}
\label{eq:pmc42_H>0}
\end{equation}
\item \textbf{PMC$[4,2]$ with $H_{2}<0$}
\begin{equation}
\begin{split}
0&= \ell\left(T^{(\eytab{0},\eytab{7})}_{(\eytab{0},\eytab{0})}\,; -1\right)= \ell^{(3)}_{1}- \ell^{(3)}_{20} \ ,\\
0&= \ell\left(T^{(\eytab{0},\eytab{8})}_{(\eytab{0},\eytab{0})}\,; -1\right)+ \ell\left(T^{(\eytab{3},\eytab{2})}_{(\eytab{0},\eytab{0})}\,; -1\right)= \ell^{(3)}_{2}+ \ell^{(3)}_{9} - \ell^{(3)}_{19}\ ,\\
0&= \ell\left(T^{(\eytab{1},\eytab{4})}_{(\eytab{0},\eytab{0})}\,; -1\right)= \ell^{(3)}_{6}- \ell^{(3)}_{15} \ ,
\end{split}
\label{eq:pmc42_H<0}
\end{equation}
\end{itemize}

The solution of $\text{AMC}[4]\cup \text{PMC}[4]$ is:
\small
\begin{equation}
\label{eq:Nc=3_Nf=4}
\text{SOL}[N_c=3, N_f=4]=
\begin{pmatrix}
  \ell_{b}= \ell_{1}+\ell_{a} \\
 \ell_{d}= -\ell_{a}-\ell_{c} \\ 
 \ell_{e}= -\ell_{a}-\ell_{c} \\
 \ell_{3}= 2 \ell_{1}-  \ell_{a} +\frac{1}{2}\ell_{c}+\frac{1}{6} \\
 \ell_{4}= \ell_{1}+2 \ell_2 \\
\ell_{6}= -\ell_{1}-\ell_{2} \\
 \ell_{7}= -2\ell_{1}-2\ell_{2}+\ell_{a}-\frac{1}{2}\ell_{c}-\frac{1}{6} \\
 \ell_{9}= 3 \ell_{1}+\ell_{2} - \ell_{a}+\frac{1}{2}\ell_{c}+\frac{1}{6}\\
 \ell_{10}= 2\ell_{2} \\
 \ell_{11}= 2\ell_{2} \\
 \ell_{12}= 3\ell_{1}+\ell_{2} -\ell_{a}+\frac{1}{2}\ell_{c}+\frac{1}{6} \\
 \ell_{13}= -\ell_{1}-\ell_{2} \\
 \ell_{14}= -2\ell_{1}-2\ell_{2} + \ell_{a}-\frac{1}{2}\ell_{c}-\frac{1}{6} \\
 \ell_{15}= -\ell_{1}-\ell_{2} \\
 \ell_{17}= \ell_{2} \\
 \ell_{18}= 2 \ell_{1}- \ell_{a}+\frac{1}{2}\ell_{c}+\frac{1}{6} \\
 \ell_{19}= \ell_{2} \\
 \ell_{20}= \ell_{1} 
\end{pmatrix}\, ,
\end{equation}
\normalsize
where we have chosen the free indices to be $\{ \ell_a, \ell_c, \ell_1, \ell_2\}$.

\subsection{Theory with $N_f=3$}

For $N_f=3$, $T_{5,16}$ are not well defined tensors for ${\cal G}[3]$. Among the remaining tensors of~Table~\ref{var_pentaq}, the following groups are equivalent: $T_c$, $T_8$ and $T_{17}$; $T_e$, $T_4$ and $T_{13}$; $T_2$ and the parity conjugate of $T_{12}$; $T_3$ and the parity conjugate of $T_{10}$. The indices of the corresponding irreps are denoted respectively by $\ell_c$, $\ell_e$, $\ell_2$ and $\ell_3$. Furthermore, $T_b$ and $T_7$ transform as the same irreps of ${\cal G}[3]$ as their parity conjugate;  as a consequence of parity invariance of the spectrum, the indices $\ell_b$ and $\ell_7$ vanish identically and do not appear in AMN and PMC equations. There exist 17 irreps in total. The rank of the of AMC$[5]\cup$PMC$[5]$ is 11 (see Table~\ref{tab:Nc_5}), and we conclude that the family of real solutions has 4 free parameters.

The $[SU(3)_L]^3$ and $[SU(3)_L]^2 U(1)_B$ AMC equations are respectively
\begin{align}
  \label{eq:am_nf=3}
  \begin{aligned}
27 \ell_{a}+15 \ell_{d}-6\ell_{e}-246 \ell_{1} -57 \ell_{2} +15 \ell_{3} -216 \ell_{6}-84 \ell_{9} \\
-21 \ell_{11}-57 \ell_{14}+168 \ell_{15}-27 \ell_{18}+189 \ell_{20}&=3 \ ,
\end{aligned}
\intertext{and}
\label{eq:am_nf=3_2}
\begin{aligned}
15 \ell_{a}+6 \ell_{c}+9\ell_{d}-90 \ell_{1}-45 \ell_{2}-9 \ell_{3}-60\ell_{6}\\
+45 \ell_{11}+45 \ell_{14}+126 \ell_{15}+15 \ell_{18}+54 \ell_{19}+105 \ell_{20} & =1 \ .
\end{aligned}
\end{align}
All the anomaly coefficients in Eq.~(\ref{eq:am_nf=3_2}) are multiples of $3$, so there is no integral solution for this equation and more in general for the entire system. This is an explicit check of the `prime factor' argument~\cite{Preskill:1981sr, Weinberg:1996kr, CLRX2}, and is enough to prove $\chi$SB in this case.

Next, we consider PMC.
The PMC$[3,1]$ equations are:
\begin{itemize}
%
\item \textbf{PMC$[3,1]$ with $H_{1}>0$}
\begin{equation}
\begin{split}
0&= \ell\left(T^{(\eytab{1},\eytab{0})}_{(\eytab{0},\eytab{0})}\,; 2\right)+ \ell\left(T^{(\eytab{3},\eytab{0})}_{(\eytab{1},\eytab{0})}\,; 2\right)+ \ell\left(T^{(\eytab{0},\eytab{3})}_{(\eytab{1},\eytab{0})}\,; 2\right) \\
&=\ell_{a}+\ell_{c}+\ell_{e}-\ell_{1}+\ell_{9}+\ell_{14}+\ell_{19}+\ell_{20}\ ,\\
0&= \ell\left(T^{(\eytab{2},\eytab{0})}_{(\eytab{0},\eytab{0})}\,; 1\right)+ \ell\left(T^{(\eytab{5},\eytab{0})}_{(\eytab{1},\eytab{0})}\,; 1\right)+ \ell\left(T^{(\eytab{1},\eytab{3})}_{(\eytab{1},\eytab{0})}\,; 1\right)\\
&=\ell_{a}+\ell_{c}+\ell_{d}-\ell_{1}-\ell_{2}-\ell_{3}-\ell_{6}+\ell_{14}+\ell_{15}+\ell_{18}+\ell_{19}+\ell_{20} \ ,\\
0&= \ell\left(T^{(\eytab{4},\eytab{0})}_{(\eytab{1},\eytab{0})}\,; 1\right)= \ell_{15}+ \ell_{19}+ \ell_{20}\ , \\
0&=  \ell\left(T^{(\eytab{4},\eytab{0})}_{(\eytab{0},\eytab{1})}\,; 1\right)+ \ell\left(T^{(\eytab{2},\eytab{1})}_{(\eytab{1},\eytab{0})}\,; 1\right)= -\ell_{1}-\ell_{2}-\ell_{6}+\ell_{11}+\ell_{14}+\ell_{15} \ , \\
0&=  \ell\left(T^{(\eytab{2},\eytab{0})}_{(\eytab{1},\eytab{0})}\,; 2\right)= \ell_{11}+\ell_{14}+\ell_{15}+\ell_{18}+\ell_{19}+\ell_{20}\ , \\
0&=  \ell\left(T^{(\eytab{2},\eytab{0})}_{(\eytab{0},\eytab{1})}\,; 2\right)+ \ell\left(T^{(\eytab{1},\eytab{1})}_{(\eytab{1},\eytab{0})}\,; 2\right)= -\ell_{1}-\ell_{2}-\ell_{3}+\ell_{9}+\ell_{14}+\ell_{15}\ , \\
0&=  \ell\left(T^{(\eytab{1},\eytab{0})}_{(\eytab{1},\eytab{0})}\,; 3\right)= \ell_{6}+\ell_{9}+\ell_{11}+\ell_{14}+\ell_{15}+\ell_{19}+\ell_{20}\ , \\
0&=  \ell\left(T^{(\eytab{0},\eytab{0})}_{(\eytab{1},\eytab{0})}\,; 4\right) = \ell_{1}+ \ell_{6}+ \ell_{11}+ \ell_{15}+ \ell_{20} \ ,
\end{split}
\label{eq:pmc31_H>0}
\end{equation}
\item \textbf{PMC$[3,1]$ with $H_{1}<0$}
\begin{equation}
\begin{split}
0&= \ell\left(T^{(\eytab{0},\eytab{7})}_{(\eytab{0},\eytab{0})}\,; -1\right)= \ell_{1}- \ell_{20} \ ,\\
0&= \ell\left(T^{(\eytab{0},\eytab{8})}_{(\eytab{0},\eytab{0})}\,; -1\right)+ \ell\left(T^{(\eytab{3},\eytab{2})}_{(\eytab{0},\eytab{0})}\,; -1\right)=\ell_{2}+\ell_{9} -\ell_{19}\ ,\\
0&= \ell\left(T^{(\eytab{1},\eytab{4})}_{(\eytab{0},\eytab{0})}\,; -1\right)=\ell_{6}-\ell_{15} \ .
\end{split}
\label{eq:pmc31_H<0}
\end{equation}
\end{itemize}
It is easy to see that PMC$[3,1]$ are in the same form as PMC$[4,2]$, although all the indices in PMC$[3,1]$ are free while those in PMC$[4,2]$ are obtained from the decomposition of the original tensors in the theory with 4 massless flavors.
Furthermore, PMC$[3,1]$ are different from PMC$[4,1]$, PMC$[5,1]$ or PMC$[6,1]$, hence $N_f$-independence does not hold for $N_f=3$.

Once evaluated on the solution of PMC$[3,1]$, the two AMC equations become linearly dependent and can be rewritten as in Eqs.~(\ref{Nfeq_Nc=3}),~(\ref{Nfeq_Nc=3_2}). We thus find that the $N_f$-equation is $N_f$-independent for $N_f=3$.

The solution of $\text{AMC}[3]\cup \text{PMC}[3]$ is:
\small
\begin{equation}
\label{eq:Nc=3_Nf=3}
\text{SOL}[N_c=3, N_f=3]=
\left(
  \begin{array}{c}
    \ell_b = 0 \\
  \ell_{c}= \ell_6-\ell_a-\ell_d \\
  \ell_{e}=-\ell_{1}-\ell_6+\ell_d\\
  \ell_{2}= -3 \ell_{1}+\frac{3}{2}\ell_{a}+\frac{1}{2}\ell_{d}-\frac{1}{6} \\
    \ell_{3}= 4 \ell_{1}-\frac{1}{6} (-12 \ell_6+9\ell_a+3\ell_d-1) \\
    \ell_7 = 0 \\
  \ell_{9}= \frac{1}{6} (12 \ell_1-6 \ell_6 -9 \ell_a-3\ell_d+1) \\
 \ell_{11}= -2( \ell_{1}+\ell_{6}) \\
 \ell_{14}= \frac{1}{6} (12 \ell_6+9 \ell_a+3 \ell_d-1)\\
 \ell_{15}= \ell_{6} \\
 \ell_{18}= \frac{1}{6} (12 \ell_1-9 \ell_a-3\ell_d+1) \\
 \ell_{19}= -\ell_{1}-\ell_{6} \\
 \ell_{20}= \ell_{1}
\end{array}
\right)\ ,
\end{equation}
\normalsize
where the free indices have been chosen to be $\{ \ell_a, \ell_d, \ell_1, \ell_6\}$. Notice that, as previously discussed, $\ell_b$ and $\ell_7$ vanish identically due to parity invariance of the spectrum.

\subsection{Theory with $N_f=2$}

Tensors $T_b$, $T_4$, $T_5$, $T_8$, $T_{13}$, $T_{16}$ and $T_{17}$ are not well defined tensors for $N_f=3$.
Among the other tensors of Table~\ref{var_pentaq}, we identify the following groups of equivalent tensors: $T_a$, $T_{12}$, and $T_{19}$; $T_c$, $T_3$, $T_{10}$, $T_{18}$ and the parity conjugate of $T_e$; $T_d$, $T_7$, $T_{14}$ and the parity conjugates of $T_2$ and $T_9$; $T_1$ and the parity conjugate of $T_{15}$; $T_6$ and the parity conjugate of $T_{11}$. The indices of the corresponding irreps are denoted respectively by $\ell_a$, $\ell_c$, $\ell_d$, $\ell_1$ and $\ell_6$. There are in total 6 irreps. The $[SU(2)_L]^2 U(1)_B$ AMC equation is~\footnote{The 6 irreps are the same as those in the theory with $N_f=2$ and only massless baryons, see Table~\ref{var_nc_5}. As a consequence, the AMC equation (\ref{eq:Nc=3_Nf=2}) is the same as Eq.~(\ref{eq:Nf=2}) up to a relabeling of indices.}
\begin{equation}
  \label{eq:Nc=3_Nf=2}
10 \ell_a+ \ell_c+5 \ell_d-35 \ell_{1}  -14 \ell_{6}   + 35 \ell_{20} =1\ ,
\end{equation}
while there is no $[SU(2)_L]^3$ AMC equation since all the representations of $SU(2)$ are real or pseudo-real. There are no PMC as well for a parity-invariant spectrum.
It is simple to see Eq.~(\ref{eq:Nc=3_Nf=2}) admits integral solutions.

\subsection{Downlifting: from $N_f=6$ to $N_f=2$}
\label{downlifting_baryons_pentaquarks}

We have shown that solutions of AMC$[N_f]\cup$PMC$[N_f]$ are $N_f$-independent in our second example only for $N_f\geq 6$. This result agrees with the general condition stated in~\cite{Ciambriello:2022wmh}.  Irrespectively of the validity of $N_f$-independence, solutions can be downlifted by means of Eq.~(\ref{eq:downliftedsol}), i.e. one can obtain a solution for a theory with $N_f-1$ massless flavors from the solution of a theory with $N_f$ massless flavors. 
By iteration, one can construct downlifted solutions down to $N_f=2$.

For $N_f \geq 6$, all massless baryons and pentaquarks in the spectrum are class A. Equation~(\ref{eq:uplift}) then holds, and the correspondence between irreps of ${\cal G}[N_f+1]$ and ${\cal G}[N_f]$ is one-to-one. Hence, the downlifted solution implied by Eqs.~(\ref{eq:downliftedsol}) and~(\ref{eq:uplift}) is simply~\footnote{As for Sec.~\ref{downlifting_baryons}, in this section indices are equipped with an upper label to indicate that they refer to the theory with the corresponding number of flavors.} 
\begin{equation}
\tilde\ell^{(N_f)}_\alpha\equiv \ell^{(N_f+1)}_\alpha ,\quad \alpha=a,\cdots,e,1,\cdots,20\ .
\end{equation}

Starting from $N_f=6$, the downlifted solution of AMC$[5]\cup$PMC$[5]$ obtained through Eq.~(\ref{eq:downliftedsol}) is the following:
\begin{equation}
\label{eq:identification_6to5}
\begin{split}
\tilde \ell^{(5)}_a&\equiv \ell^{(6)}_a-\ell^{(6)}_{1}-\ell^{(6)}_{2}-\ell^{(6)}_{6}+\ell^{(6)}_{15}+\ell^{(6)}_{19}+\ell^{(6)}_{20}\ ,\\
\tilde \ell^{(5)}_b&\equiv \ell^{(6)}_b-\ell^{(6)}_{4}-\ell^{(6)}_{5}-\ell^{(6)}_{8}+\ell^{(6)}_{13}+\ell^{(6)}_{16}+\ell^{(6)}_{17} \ ,\\
\tilde \ell^{(5)}_c&\equiv \ell^{(6)}_c-\ell^{(6)}_{2}-\ell^{(6)}_{3}-\ell^{(6)}_{4}-\ell^{(6)}_{7}+\ell^{(6)}_{14}+\ell^{(6)}_{17}+\ell^{(6)}_{18}+\ell^{(6)}_{19} \ ,\\
\tilde \ell^{(5)}_d&\equiv \ell^{(6)}_d-\ell^{(6)}_{6}-\ell^{(6)}_{7}-\ell^{(6)}_{9}+\ell^{(6)}_{12}+\ell^{(6)}_{14}+\ell^{(6)}_{15} \ ,\\
\tilde \ell^{(5)}_e&\equiv \ell^{(6)}_e-\ell^{(6)}_{7}-\ell^{(6)}_{8}+\ell^{(6)}_{9}-\ell^{(6)}_{12}+\ell^{(6)}_{13}+\ell^{(6)}_{14} \ ,\\
\tilde \ell^{(5)}_5 &\equiv 0 \ ,\\
\tilde \ell^{(5)}_\alpha&\equiv \ell^{(6)}_\alpha\ ,\ \   (\alpha=1,\cdots, 4, 6, \cdots, 20) \ .
\end{split}
\end{equation}
The downlifted index $\tilde \ell^{(5)}_5$ vanishes identically due to parity invariance of the spectrum and consistently with Eq.~(\ref{eq:Nc=3_Nf=5}).
Notice that, differently from pentaquarks, baryons are interpolated by class~A tensors for $N_f=5$. Consistently with this fact, one can verify that by focusing on the subfamily of purely baryonic solutions of Eq.~(\ref{eq:Nc=3_Nf>=6}), i.e. setting the pentaquark indices to zero ($\ell^{(6)}_i = 0$ for $i=1, \cdots, 20$), the downlifted solution is simply $\tilde\ell^{(5)}_\alpha\equiv \ell^{(6)}_\alpha$, for $\alpha = a, \cdots , e$. This agrees with the implications of Eq.~(\ref{eq:uplift}) given the one-to-one correspondence that exists between baryons in theories with $N_f=6$ and $N_f=5$.

By downlifting from $N_f=5$ to $N_f=4$ we find the following solution:
\begin{equation}
\label{eq:identification_5to4}
\begin{split}
\tilde \ell^{(4)}_a&\equiv  \ell^{(5)}_a- \ell^{(5)}_{1}-\ell^{(5)}_{2}- \ell^{(5)}_{6}+ \ell^{(5)}_{15}+ \ell^{(5)}_{19}+ \ell^{(5)}_{20}\ ,\\
\tilde \ell^{(4)}_b&\equiv \ell^{(5)}_b- \ell^{(5)}_{4}-\ell^{(5)}_{8}+\ell^{(5)}_{13}+ \ell^{(5)}_{16}+ \ell^{(5)}_{17} \ ,\\
\tilde \ell^{(4)}_c&\equiv \ell^{(5)}_c-\ell^{(5)}_{2}-\ell^{(5)}_{3}-\ell^{(5)}_{4}-\ell^{(5)}_{7}+ \ell^{(5)}_{14}+ \ell^{(5)}_{17}+ \ell^{(5)}_{18}+ \ell^{(5)}_{19} \ ,\\
\tilde \ell^{(4)}_d&\equiv \ell^{(5)}_d- \ell^{(5)}_{6}- \ell^{(5)}_{7}- \ell^{(5)}_{9}+ \ell^{(5)}_{12}+ \ell^{(5)}_{14}+ \ell^{(5)}_{15} \ ,\\
\tilde \ell^{(4)}_e&\equiv \ell^{(5)}_e- \ell^{(5)}_{7}- \ell^{(5)}_{8}+ \ell^{(5)}_{9}- \ell^{(5)}_{12}+ \ell^{(5)}_{13}+ \ell^{(5)}_{14} \ ,\\
\tilde \ell^{(4)}_4&\equiv  \ell^{(5)}_4- \ell^{(5)}_8\ ,\\
\tilde \ell^{(4)}_\alpha&\equiv \ell^{(5)}_\alpha\ ,\ (\alpha=1,2,3,6,7,9,\cdots ,15, 17,\cdots, 20) \ .
\end{split}
\end{equation}
%
As before, the downlift of a purely baryonic solution is simply $\tilde\ell^{(4)}_\alpha\equiv \ell^{(5)}_\alpha$, for $\alpha = a, \cdots , e$. This follows because baryons are interpolated by class A tensors for $N_f=4$, and agrees with Eq.~(\ref{eq:uplift}).

Going one step further, i.e. downlifting from $N_f=4$ to $N_f=3$ we find:
\begin{equation}
\label{eq:identification_4to3}
\begin{split}
  \tilde \ell^{(3)}_a & \equiv \ell^{(4)}_a-\ell^{(4)}_{1}-\ell^{(4)}_{2}-\ell^{(4)}_{6}+\ell^{(4)}_{15}+\ell^{(4)}_{19}+\ell^{(4)}_{20}\ ,\\
  \tilde \ell^{(3)}_b & \equiv 0 \ , \\
\tilde \ell^{(3)}_c&\equiv \ell^{(4)}_c-\ell^{(4)}_{2}-\ell^{(4)}_{3}-\ell^{(4)}_{4}-\ell^{(4)}_{7}+\ell^{(4)}_{14}+\ell^{(4)}_{17}+\ell^{(4)}_{18}+\ell^{(4)}_{19} \ ,\\
\tilde \ell^{(3)}_d &\equiv \ell^{(4)}_d- \ell^{(4)}_{6}- \ell^{(4)}_{7}- \ell^{(4)}_{9}+\ell^{(4)}_{12}+ \ell^{(4)}_{14}+ \ell^{(4)}_{15} \ ,\\
\tilde \ell^{(3)}_e&\equiv \ell^{(4)}_e+\ell^{(4)}_4 -\ell^{(4)}_{7}+\ell^{(4)}_{9}-\ell^{(4)}_{12}+\ell^{(4)}_{13}+\ell^{(4)}_{14} \ ,\\
\tilde \ell^{(3)}_{2}&\equiv \ell^{(4)}_{2}- \ell^{(4)}_{12}\ ,\\
\tilde \ell^{(3)}_{3}&\equiv \ell^{(4)}_{3}-\ell^{(4)}_{10}\ ,\\
 \tilde \ell^{(3)}_7 & \equiv 0 \ , \\
\tilde \ell^{(3)}_\alpha&\equiv \ell^{(4)}_\alpha\ ,\  (\alpha=1,6,9,11,14,15,18,19,20) \ .
\end{split}
\end{equation}
The downlifted indices $\tilde \ell^{(3)}_b$ and $\tilde\ell^{(3)}_{7}$ vanish identically due to parity invariance of the spectrum and consistently with Eq.~(\ref{eq:Nc=3_Nf=3}).

Finally, downlifting from $N_f=3$ to $N_f=2$ gives:
\begin{equation}
\label{eq:identification_3to2}
\begin{split}
  \tilde\ell^{(2)}_a &\equiv \ell^{(3)}_a-\ell^{(3)}_1-\ell^{(3)}_2-\ell^{(3)}_6+\ell^{(3)}_{15}+\ell^{(3)}_{19}+\ell^{(3)}_{20}\ ,\\
  \tilde\ell^{(2)}_{c} &\equiv \ell^{(3)}_{c}-\ell^{(3)}_{e}-\ell^{(3)}_{2}-\ell^{(3)}_{9}+\ell^{(3)}_{18}+\ell^{(3)}_{19}\ ,\\
\tilde\ell^{(2)}_{d}&\equiv \ell^{(3)}_{d}-\ell^{(3)}_{2}-\ell^{(3)}_{6}-\ell^{(3)}_{9}+\ell^{(3)}_{14}+\ell^{(3)}_{15}\ ,\\
\tilde\ell^{(2)}_{1} &\equiv \ell^{(3)}_{1}-\ell^{(3)}_{15}\ ,\\
\tilde\ell^{(2)}_{6} &\equiv \ell^{(3)}_{6}-\ell^{(3)}_{11}\ ,\\
\tilde\ell^{(2)}_{20} &\equiv \ell^{(3)}_{20} \ .
\end{split}
\end{equation}

In all the cases except for $N_f=2$, the downlifted solution is  the most general solution of AMC and PMC equations.

\vspace{1cm}
\noindent
{\bf Acknowledgements.}
\hspace{0.04cm}
We are grateful to Andrea Luzio and Marcello Romano for collaboration in companion papers. 
We would like to thank Ethan Neil, Slava Rychkov, Yael Shadmi, and Giovanni Villadoro for useful discussions and comments.
This research was supported in part by the MIUR under contract 2017FMJFMW (PRIN2017). The work of R.C. was partly supported by the Munich Institute for Astro- and Particle Physics (MIAPP), which is funded by the Deutsche Forschungsgemeinschaft (DFG, German Research Foundation) under Germany Excellence Strategy - EXC-2094 - 390783311.
L.X.X. would like to thank Scuola Normale Superiore, where this project was initiated, for its warm hospitality. The work of L.X.X. was partially supported by ERC grant n.101039756. 

\appendix

\newpage
\section{Representations, dimensionalities, and anomalies}
\label{app:table}

\begingroup
\renewcommand*{\arraystretch}{1.8}
\begin{longtable}{||c|c|c|c|c||}
	\hline\hline
	Label&Young Tableau & Dimensionality $d$ & Anomaly $D_2$ & Anomaly $D_3$\\
	\hline\hline
	$R_{0}$&.&1&$0$&$0$\\
  \hline
	$R_{1}$&$\includegraphics[scale=0.05]{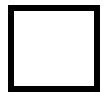}$&$n$&$1$&$1$\\
  \hline
	$R_{2}$&$\includegraphics[scale=0.05]{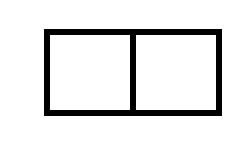}$&$\frac{(n+1)n}{2}$&$n+2$&$n+4$\\
  \hline
	$R_{3}$&$\includegraphics[scale=0.05]{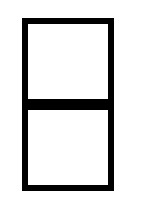}$&$\frac{n(n-1)}{2}$&$n-2$&$n-4$\\
  \hline
	$R_{4}$&$\includegraphics[scale=0.05]{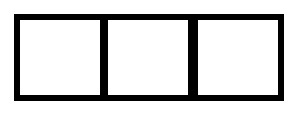}$&$\frac{(n+2)(n+1)n}{6}$&$\frac{(n+3)(n+2)}{2}$&$\frac{(n+6)(n+3)}{2}$\\
  \hline
	$R_{5}$&$\includegraphics[scale=0.05]{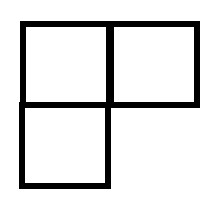}$&$\frac{(n+1)n(n-1)}{3}$&$n^2-3$&$n^2-9$\\
  \hline
	$R_{6}$&$\includegraphics[scale=0.05]{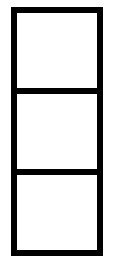}$&$\frac{n(n-1)(n-2)}{6}$&$\frac{(n-2)(n-3)}{2}$&$\frac{(n-3)(n-6)}{2}$\\
  \hline
	$R_{7}$&$\includegraphics[scale=0.05]{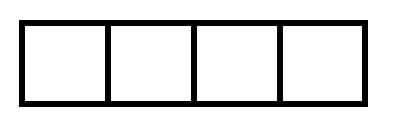}$&$\frac{(n+3)(n+2)(n+1)n}{24}$&$\frac{(n+4)(n+3)(n+2)}{6}$&$\frac{(n+8)(n+4)(n+3)}{6}$\\
  \hline
	$R_{8}$&$\includegraphics[scale=0.05]{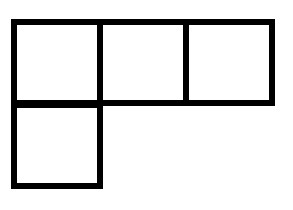}$&$\frac{(n+2)(n+1)n(n-1)}{8}$&$\frac{(n^2+n-4)(n+2)}{2}$&$\frac{(n^2+n-8)(n+4)}{2}$\\
  \hline
	$R_{9}$&$\includegraphics[scale=0.05]{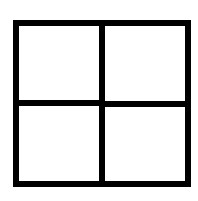}$&$\frac{(n+1)n^2(n-1)}{12}$&$\frac{(n^2-4)n}{3}$&$\frac{(n^2-16)n}{3}$\\
  \hline
	$R_{10}$&$\includegraphics[scale=0.05]{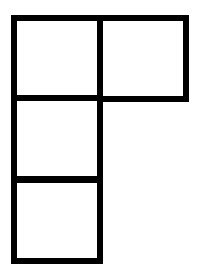}$&$\frac{(n+1)n(n-1)(n-2)}{8}$&$\frac{(n-2)[n(n-1)-4]}{2}$&$\frac{(n-4)[n(n-1)-8]}{2}$\\
  \hline
	$R_{11}$&$\includegraphics[scale=0.05]{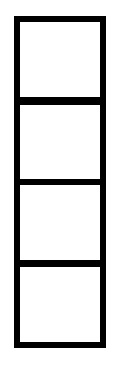}$&$\frac{n(n-1)(n-2)(n-3)}{24}$&$\frac{(n-2)(n-3)(n-4)}{6}$&$\frac{(n-3)(n-4)(n-8)}{6}$\\
  \hline
	$R_{12}$&$\includegraphics[scale=0.05]{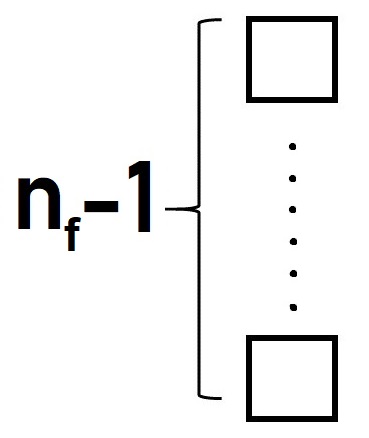}$&$n$&$1$&$-1$\\
  \hline
	$R_{13}$&$\includegraphics[scale=0.05]{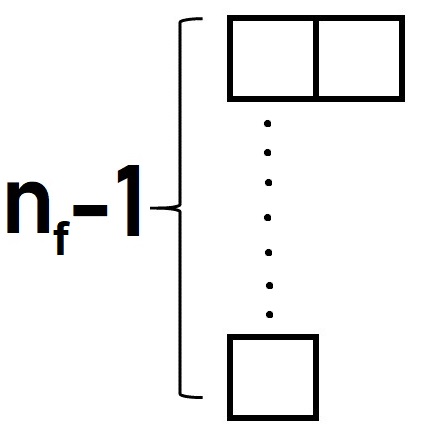}$&$n^2-1$&$2n$&$0$\\
  \hline
	$R_{14}$&$\includegraphics[scale=0.05]{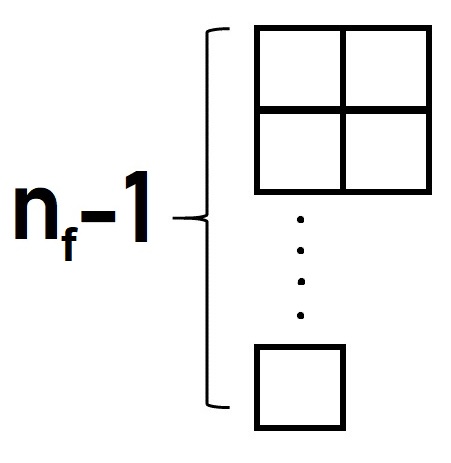}$&$\frac{(n+1)n(n-2)}{2}$&$\frac{(3n+1)(n-2)}{2}$&$\frac{n(n-7)-2}{2}$\\
  \hline
	$R_{15}$&$\includegraphics[scale=0.05]{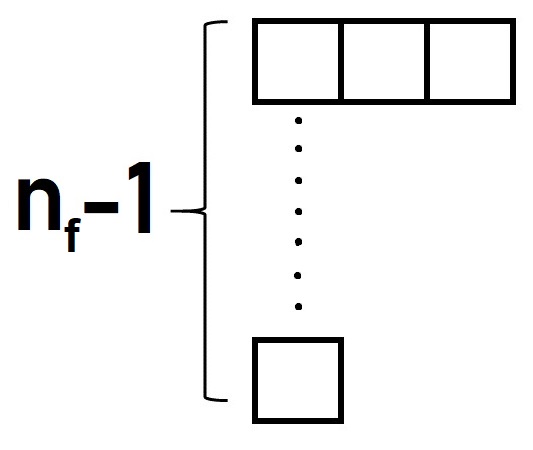}$&$\frac{(n+2)n(n-1)}{2}$&$\frac{(3n-1)(n+2)}{2}$&$\frac{n(n+7)-2}{2}$\\
  \hline
	$R_{16}$&$\includegraphics[scale=0.05]{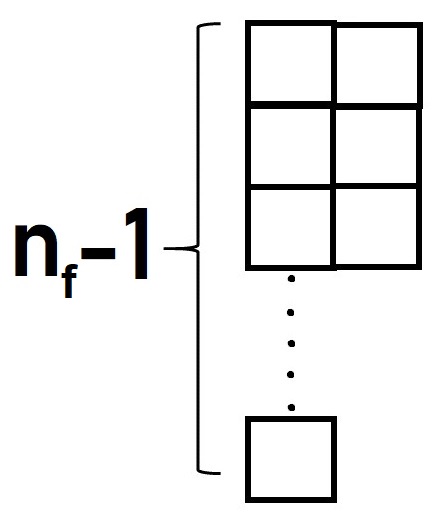}$&$\frac{(n+1)n(n-1)(n-3)}{6}$&$\frac{(2n+1)(n-2)(n-3)}{3}$&$\frac{(n-3)[n(n-9)-4]}{3}$\\
  \hline
	$R_{17}$&$\includegraphics[scale=0.05]{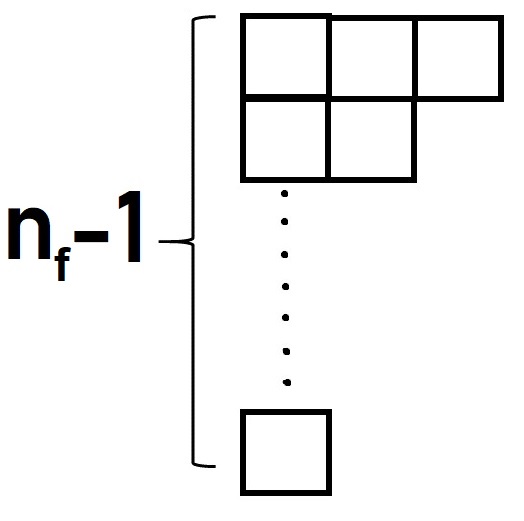}$&$\frac{(n+2)n^2(n-2)}{3}$&$\frac{4n(n^2-4)}{3}$&$\frac{2n(n^2-16)}{3}$\\
  \hline
	$R_{18}$&$\includegraphics[scale=0.05]{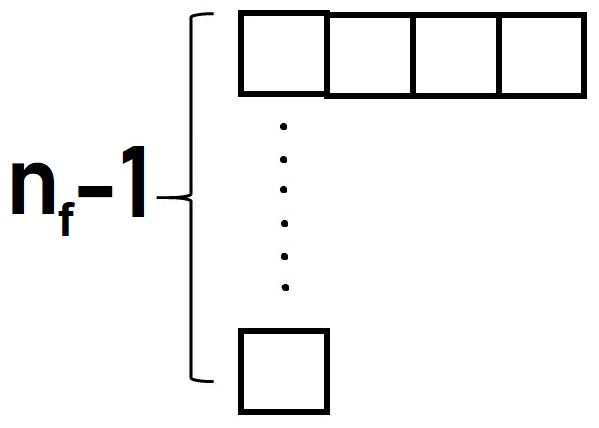}$&$\frac{(n+3)(n+1)n(n-1)}{6}$&$\frac{(2n-1)(n+3)(n+2)}{3}$&$\frac{(n+3)[n(n+9)-4]}{3}$\\
  \hline
$R_{19}$&$\includegraphics[scale=0.05]{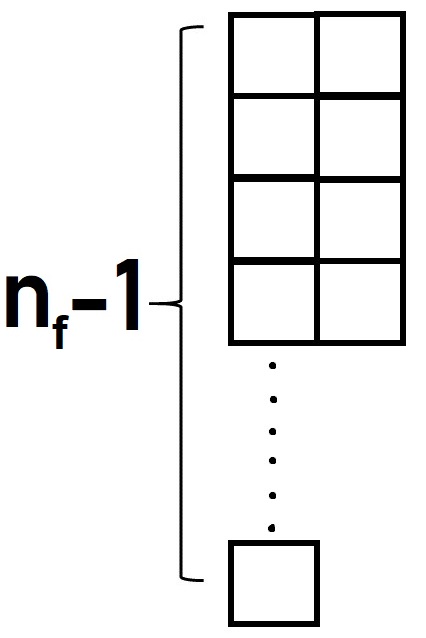}$&$\frac{(n+1)n(n-1)(n-2)(n-4)}{24}$&$\frac{(5n+3)(n-2)(n-3)(n-4)}{24}$&$\frac{[n(n-11)-6](n-3)(n-4)}{8}$\\
  \hline
	$R_{20}$&$\includegraphics[scale=0.05]{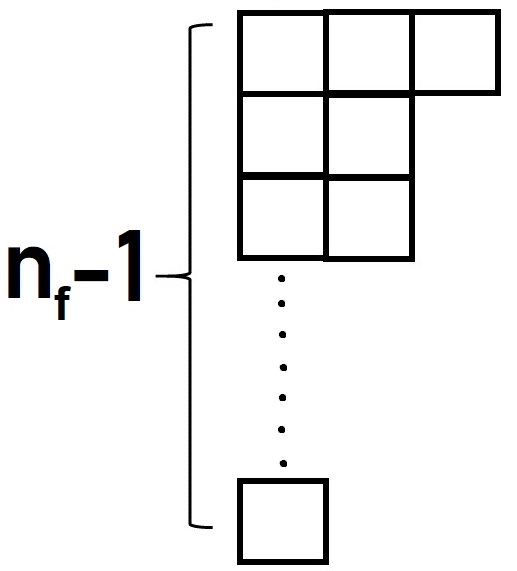}$&$\frac{(n+2)n^2(n-1)(n-3)}{8}$&$\frac{(5n-9)(n+2)n(n-3)}{8}$&$\frac{3(n+3)n(n-3)(n-6)}{8}$\\
  \hline
	$R_{21}$&$\includegraphics[scale=0.05]{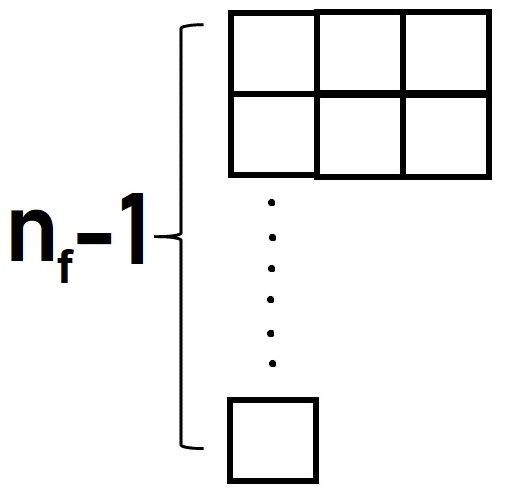}$&$\frac{(n+2)(n+1)n(n-1)(n-2)}{12}$&$\frac{5n^4-29n^2+36}{12}$&$\frac{n^4-25n^2+36}{4}$\\
  \hline
	$R_{22}$&$\includegraphics[scale=0.05]{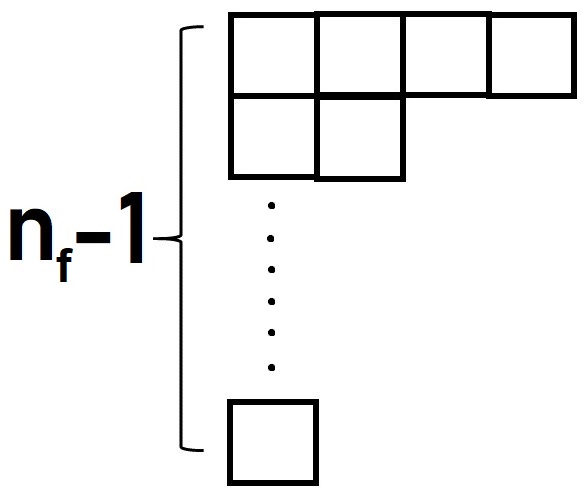}$&$\frac{(n+3)(n+1)n^2(n-2)}{8}$&$\frac{(5n+9)(n+3)n(n-2)}{8}$&$\frac{3(n+6)(n+3)n(n-3)}{8}$\\
  \hline
$R_{23}$&$\includegraphics[scale=0.05]{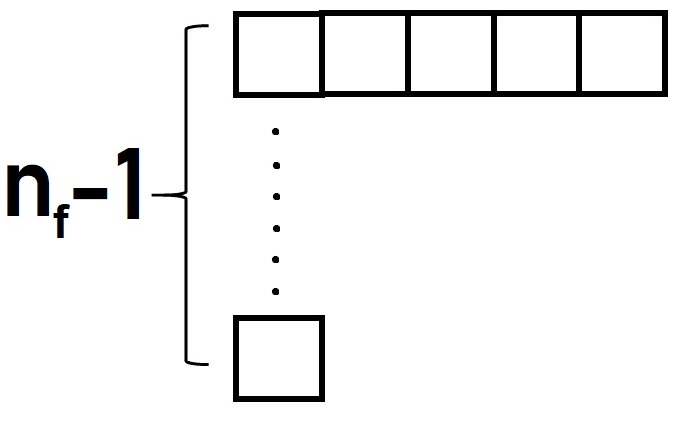}$&$\frac{(n+4)(n+2)(n+1)n(n-1)}{24}$&$\frac{(5n-3)(n+4)(n+3)(n+2)}{24}$&$\frac{[n(n+11)-6](n+4)(n+3)}{8}$\\	
  \hline
	$R_{24}$&$\includegraphics[scale=0.05]{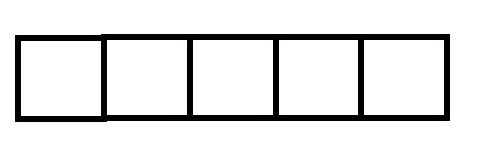}$&$\frac{n (1 + n) (2 + n) (3 + n) (4 + n)}{120}$&$\frac{(2 + n) (3 + n) (4 + n) (5 + n)}{24}$&$\frac{(3 + n) (4 + n) (5 + n) (10 + n)}{24}$\\
  \hline
	$R_{25}$&$\includegraphics[scale=0.05]{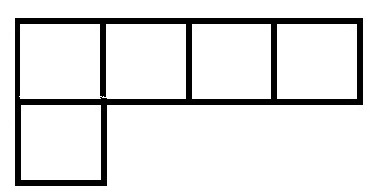}$&$\frac{(-1 + n) n (1 + n) (2 + n) (3 + n)}{30}$&$\frac{(2 + n) (3 + n) (-5 + n (2 + n))}{6}$&$\frac{(5 + n)^2 (-6 + n + n^2)}{6}$\\
  \hline
	$R_{26}$&$\includegraphics[scale=0.05]{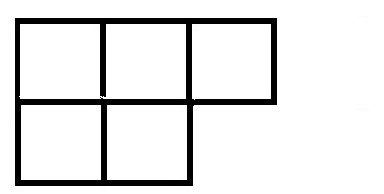}$&$\frac{(-1 + n) n^2 (1 + n) (2 + n)}{24}$&$\frac{n (2 + n) (-25 + n (4 + 5 n))}{24}$&$\frac{n (5 + n) (-50 + n (-3 + 5 n))}{24}$\\
  \hline
	$R_{27}$&$\includegraphics[scale=0.05]{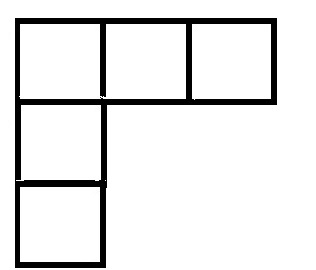}$&$\frac{(-2 + n) (-1 + n) n (1 + n)(2+n)}{20}$&$\frac{(20 - 9 n^2 + n^4)}{4}$&$\frac{(100 - 17 n^2 + n^4)}{4}$\\
  \hline
	$R_{28}$&$\includegraphics[scale=0.05]{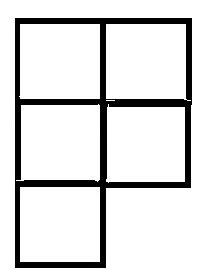}$&$\frac{(-2 + n) (-1 + n) n^2 (1 + n)}{24}$&$\frac{(-2 + n) n (-25 + n (-4 + 5 n))}{24}$&$\frac{(-5 + n) n (-50 + n (3 + 5 n))}{24}$\\
  \hline
	$R_{29}$&$\includegraphics[scale=0.05]{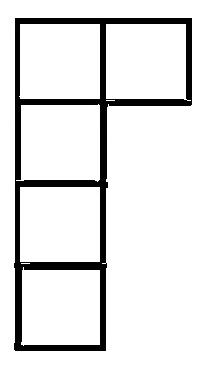}$&$\frac{(-3 + n) (-2 + n) (-1 + n) n (1 + n)}{30}$&$\frac{(-3 + n) (-2 + n) (-5 + (-2 + n) n)}{6}$&$\frac{(-5 + n)^2 (-3 + n) (2 + n)}{6}$\\
  \hline
	$R_{30}$&$\includegraphics[scale=0.05]{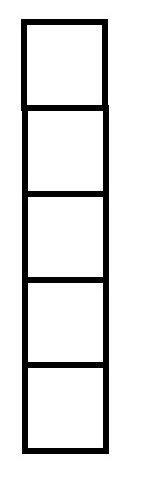}$&$\frac{(-4 + n) (-3 + n) (-2 + n) (-1 + n) n}{120}$&$\frac{(-5 + n) (-4 + n) (-3 + n) (-2 + n)}{24}$&$\frac{(-10 + n) (-5 + n) (-4 + n) (-3 + n)}{24}$\\
	\hline\hline
  \caption{List of the irreps of $SU(n)$ encountered in the calculation, together with their dimensionalities $d$ and their anomaly constants $D_2$ and $D_3$, cf. Eq.~(\ref{eq:D2_D3}). Notice that $D_3$ is only well-defined for $n\geq 3$.}
	\label{list}
\end{longtable}
\endgroup

\bibliography{ChSB_refs}

\end{document}